# $(\vec{p}, \vec{p}')$ Polarization-transfer observables with polarized beams[*]

A.V. Plavko[†]

*Doctoral Dissertation Council on Physics,*
*St. Petersburg State Polytechnic University,*
*St. Petersburg, 195251, Russia*
Email: plavko_av@spbstu.ru

A review is given of our present knowledge of different parametrization of experimental results in inelastic scattering process with polarized proton beams. Spin observables in ($\vec{p}, p'$) and ($\vec{p}, \vec{p}'$) reactions for a set of lightweight nuclei are studied at intermediate energies in the region of 100–500 MeV. Two important types of DWIA calculations are used in the analysis of experiments. Important spin observables representing the difference functions $(P–A_y)$ and $(P–A_y)\sigma$ are examined along with the polarization-transfer (spin-transfer) coefficients $D_{ij}$ ($D_{NN}$, $D_{SS}$, $D_{LL}$, $D_{LS}$, and $D_{SL}$) in inelastic scattering of polarized protons. The above indicated value of $\sigma$ is the differential cross section. These spin observables are treated in the context of the exchange features of the effective interaction. The linear combinations $D_K$ (or $D_i$) of the complete polarization-transfer observables $D_{ij}$ in the ($\vec{p}, \vec{p}'$) reaction at intermediate energies are extensively demonstrated. A comparison between systematized, measured and calculated quantities of the combinations $D_K$ and, partially, $\sigma D_K$ for the $1^+$ ($T = 0$ and $T = 1$) levels in $^{12}C$, for the $4^-$ ($T = 0$ and $T = 1$) states in $^{16}O$, and for the $6^-$ ($T = 0$ and $T = 1$) levels in $^{28}Si$ are reported. Particularities in angular distributions of transverse- and longitudinal-spin-transfer probabilities $D_K$ for $T = 0$ and $T = 1$ unnatural-parity states in some nuclei are discussed. The spin-observable combinations $D_{ls}$ allowed to differentiate reliably the strength of isoscalar and isovector spin-orbit interactions. The comparison of the experimental and calculated $D_K$ for the $1^+$, $T = 0$ state in $^{12}C$ with the use of zero-range treatment (LEA code) and exact finite-range calculations (DWBA-91 program) made it possible to identify the role of exchange contributions. Other observables are also considered for the same purposes.

## Introduction, Motivation, and Sum Rules

The performed polarization-transfer (spin-transfer) experiments and their analysis proved to be important in establishing new relations for nucleon–nucleus scattering. In particular, these studies demonstrated that spin-transfer observables could be used to isolate certain parts of the nucleon–nucleon (*NN*) amplitude and specific nuclear form factors.

Some distinguished theoreticians (Petrovich, Carr, McManus, etc.) believe that efforts in this area are certain to increase in the near future. However, recently these efforts have slowed down due to their complexity, high cost, and the lack of adequate financing. Nevertheless, work goes on, though exclusively due to the enthusiasm of individual researchers. Sammarruca, Stephenson, etc. assert that such studies will continue to be a "powerful diagnostic tool".

The present review was completed by the author in the summer and fall of 2016, based mainly on the contributions from A.V. Plavko, M.S. Onegin and V.I. Kudriashov, presented at the following International Conferences: 62[nd] Int. Conf. on Nuclear Spectroscopy and Structure of Atomic Nucleus. – "Nucleus 2012", Voronezh, 2012 (see Sec. 1 in the present review); 63[rd] Int. Conf. on Nuclear Spectroscopy and Nuclear Structure. – "Nucleus 2013", Moscow, 2013 (see Sec. 2 in this review); 64[th] Int. Conf. on Nuclear Physics. – "Nucleus 2014", Minsk, 2014 (see Sec. 3 in this review), and 65[th] Int. Conf. on Nuclear Physics. – "Nucleus 2015", St. Petersburg, 2015 (see Sec. 4 in this review, and a report by A.V. Plavko, M.S. Onegin, and V.I. Kudriashov of Jul

---

[*] In memory of the warm hospitality extended to the author during his several long stays at IUCF, Bloomington, Indiana, and LANL (LAMPF), Los Alamos, U.S.A.
[†] This work represents a combined effort together with M.S. Onegin and V.I. Kudriashov

03 2015 nucl-th arχiv: 1507, 00468V1). For the present review, the above-mentioned contributions were complemented by additional commentaries from the author. Final section 5 is written by the author for the purpose of combining and significantly complementing the ideas and results presented in the previous sections. For the present review, the above-mentioned contributions were complemented by additional commentaries from the author. Final section 5 is written by the author for the purpose of combining and significantly complementing the ideas and results presented in the previous sections.

The primary purpose of the review is to demonstrate that polarization-transfer coefficients and their combinations can provide one of the best ways to test spin-dependent $NN$ amplitudes and evaluate the treatment of the observed effects in order to provide tentative explanations of the data.

In theoretical methods, a matrix of amplitudes is usually obtained, for example, for the scattering of a beam of particles with spin 1/2 on a scatterer with spin 0. These can be expressed via decomposition into a certain number of matrices. If the principles of invariance are observed, such decomposition yields an expression with a set of independent scalar amplitudes. The latter are frequently referred to as Wolfenstein parameters, after the author of a triple scattering analysis (L. Wolfenstein. Annu. Rev. Nucl.Sci. 1956. Vol. 6, p. 43).

Following the above-mentioned theoretical works, the experimental techniques were then radically improved. For example, the initial beam is now not only polarized, but also has its particles alternately polarized in three different directions. First, there is the traditional direction of polarization, oriented normal to the reaction plane ($\hat{N}$) and frequently perpendicular to the horizontal plane. Second, a longitudinally polarized beam ($\hat{L}$) is formed, whose polarization vector is determined by the direction of the incident beam. And third, a beam is formed with the so-called side polarization vector ($\hat{S}$), characterized by its direction in the scattering plane but orthogonal to the direction $\hat{N}$, i.e., $\hat{S} = \hat{N} \times \hat{L}$. A similar experiment was performed for the first time at $E_p$ = 500 MeV (LANL, Los Alamos), then at $E_p$ = 200 MeV (IUCF, Bloomington, Indiana, USA), at $E_p$ = 350 MeV (TRIUMF, Vancouver, Canada), and, finally, at $E_p \simeq$ 400 MeV in the range of zero degrees scattering angles (RCNP, Osaka, Japan).

As a result, the spin (polarization)-transfer coefficients $D_{NN}$, $D_{LL}$, $D_{SS}$, $D_{LS}$, and $D_{SL}$ were measured in inelastic scattering of polarized protons with the excitation of levels of abnormal parity (as the most interesting) in light nuclei. All these coefficients proved to be identical to the respective Wolfenstein parameters from the triple scattering theory. Traditionally, azimuthal asymmetries, such as the scattering-induced polarization $P$ of particles and the analyzing power $A_y$ that arises in scattering of polarized protons, play an important role in the scattering of protons. Difference spin functions based on combinations of a ($P$–$A_y$) and other types are also of great importance in understanding scattering processes.

Let us designate the polarization of incident protons as $P_N^0, P_S^0$ and $P_L^0$, where the subscripts denote three possible directions of the polarization vector of the initial incident beam noted above. The polarizations of the respective scattered protons are, in turn, presented as $P_N$, $P_S$, and $P_L$. Having somewhat generalized these expressions, we can say that the polarization components of the incident beam are related to the components of the scattered beam through the intermediary of Wolfenstein parameters, based on the well-known system of equations. If the spins of the target are not analyzed, the polarization vector components of the scattered beam can be expressed as

$$P_S = (D_{SS} P_S^0 + D_{LS} P_L^0) / (1 + D_{N0} P_N^0),$$
$$P_N = (D_{NN} P_N^0 + D_{0N}) / (1 + D_{N0} P_N^0), \qquad (1)$$
$$P_L = (D_{LL} P_L^0 + D_{SL} P_S^0) / (1 + D_{N0} P_N^0).$$

For inelastic proton scattering, we can measure eight independent observables. These include the differential scattering cross sections $\sigma$ and the seven quantities contained in Equations (1). Among these seven quantities are five spin transfer coefficients. For these coefficients, we also indicate, in parentheses, the respective alternative designations proposed earlier by Wolfenstein: $D_{SS}$ (R), $D_{NN}$ (D), $D_{LL}$ (A'), $D_{LS}$ (A), and $D_{SL}$ (R'). Equations (1) also include the observable $D_{0N}$, equivalent to the traditional polarization of particles during scattering (P), along with the observable $D_{N0}$, which is the ordinary analyzing power ($A_y$), i.e., the asymmetry of scattering of particles, polarized normal to the scattering plane.

All of the seven spin observables are components of the proton depolarization tensor. Relationships (1) between the polarization components of the incident and scattered nucleons can be combined into a single general matrix form, in which the parameters $D_{SS}$, $D_{NN}$, and $D_{LL}$ are diagonal matrix elements, while $D_{LS}$ and $D_{SL}$ are off-diagonal. From this we derive the names diagonal and off-diagonal spin (polarization)-transfer coefficients.

If we write the spin-transfer coefficients in a general form, i.e. $D_{ij}$, this will mean that the spin of a proton prior to its scattering is oriented along the $i$-axis, and then along the $j$-axis after being scattered. Let the wave vector of a proton, prior to its scattering, be $\vec{k}$; then after scattering it becomes $\vec{k}'$. It is convenient to use a rectangular coordinate system in which the axis $L$ ($L'$) is parallel to the vector $\vec{k}(\vec{k}')$, the axis $N(N')$ is parallel to the vector $\vec{k} \times \vec{k}'$, the axis $S(S')$ is parallel to the direction $N \times L$ ($N' \times L'$), and the axes $N$ and $N'$ being coincident; i.e., $N = N'$. This coordinate system is essentially the same laboratory system of the beam direction described above. The respective indices on the measured and calculated spin transfer coefficients $D_{ij}$ are then

$$(i, j) = N, L, S;\ N', L', S'. \qquad (2)$$

The primes on the axis designations are usually omitted in the current literature for purposes of simplification, as is done in Eq. (1).

Expressions (1) followed from the most general forms. Then the general results were simplified, since certain particular conditions, imposed by parity conservation and time reversal invariance, were applied. Indeed, all spin observables should change according to specific rules under the appropriate transformations. The first rule can be derived by applying parity conservation. The second rule can be derived via applying rotation invariance, and the third – by applying time reversal invariance. All the findings are summarized and written as Forms (1).

In fact, the above-indicated rules are a mathematical expression of symmetries. When these symmetries are observed, the corresponding interactions cannot mix polarization components, normal to the scattering plane $(\hat{N})$, with those that lie in this plane ($\hat{L}$ and $\hat{S}$). Therefore, the remaining off-diagonal polarization-transfer observables ($D_{NL}$, $D_{LN}$, $D_{NS}$ and $D_{SN}$) are zero.

Certainly, these observables for parity-conserving reactions with a polarized spin ½ particle 'in' and a polarized spin ½ particle 'out' were not included in Forms (1). Indeed, the polarization-transfer coefficients $D_{NL}$, etc., were initially included in fully expanded forms, obtained in mathematical terms of the density matrix. General expressions of type (1) were established within the framework of the relation between the initial and final state of the density matrix. Subsequently, parameters that did not respect the requirements of the rules, based on symmetries, were excluded from the fully expanded forms.

All these rules, as well as some other regulations, are described in detail and in comprehensible form in G.G. Ohlsen's review (Rep. Prog. Phys., 1972, vol. 35, p. 717).

Above-mentioned Section 5, which the author wrote specifically for the present review, is devoted to the analysis of polarization observables, introduced by Bleszynski et el. (E. Bleszynski, M. Bleszynski, and C.A. Whitten, Jr., Phys. Rev., 1982, vol.25, p. 2053). In the review, $(\vec{p}, p'\gamma)$ observables and analogous coincident combinations of polarization data are also studied. And the final analysis is mainly devoted to the treatment of knock-on exchange in DWIA.

Different treatments of knock-on exchange were achieved through using successively the DWBA 91 program (personally provided by J. Raynal), and the LEA program (personally provided by J. Kelly). Calculations, based on these programs, were then complemented in important cases by the results obtained within the framework of the DBHF (Dirac–Brueckner Hartree–Fock) model, personally provided by E.J. Stephenson. In Section 5, the author of the review demonstrates excellent compatibility of these three calculation techniques.

The bulk of the data used in the present review is available in the following works by A.V. Plavko, M.S. Onegin, and V.I. Kudriashov published in:
1. ISSN 1062–8738, Bulletin of the Russian Academy of Science, Physics, 2013, vol. 77, No. 7, pp. 871–879. © Allerton Press, Inc., 2013.
2. ISSN 1062–8738, Bulletin of the Russian Academy of Science, Physics, 2014, vol. 78, No. 7, pp. 663–671. © Allerton Press, Inc., 2014.
3. ISSN 1062–8738, Bulletin of the Russian Academy of Science, Physics, 2015, vol. 79, No. 7, pp. 838–847. © Allerton Press, Inc., 2015.

A more detailed description of the relevant materials was presented at the conferences mentioned at the beginning of the Introduction to the present review. Now these materials become available to a much wider range of potential readers – and, what is new, in a more generalized and expanded form with detailed commentary. The author of the review takes responsibility for all the additional comments and notes relating to the study.

A schematic description of the equipment used at the Los Alamos Meson Physics Facility (LAMPF) for $(\vec{p}, \vec{p}\,')$, $(\vec{p}, n)$ and other studies is available in publications by J.B. MacClelland (e.g. in Can. J. Phys., 1987, vol. 65, p. 633) and other researches.

The author of the review acknowledges Dr. J.B. MacClelland, Head of the experimental team at LAMPF, for the informative conversations, regarding $(\vec{p}, \vec{p}\,')$ operations, at the location where the experiments were carried out.

The author is also grateful to the entire staff of the Indiana Cyclotron Facility for the opportunity given to him to get unlimited information on methods for measuring polarization observables and participate in some of the experiments with the polarized proton beam. These experiments dealt with the primary scattering reaction and provided a measure of the second (analyzer) scattering reaction.

At this stage, the review covers only the study of polarized transfer coefficients and other observables in the case of excitations of abnormal parity in nuclei. We intend to expand our research to natural-parity transitions in order to investigate how the localization of transition form factors (at the surface or in the interior probes) influences $(\vec{p}, \vec{p}\,')$ processes and the corresponding observables. The list below includes relevant works in this direction:
1. E.J. Stephenson, Colloque de physique. Colloque C6, Suppl. au nº 22, T. 51, 1990, p. C6–85.
2. B.S. Flanders, J.J. Kelly, H. Seiferet et al., Phys. Rev. C, V. 43, 1991, p. 2103.
3. Jian Liu, E.J. Stephenson, A.D. Bacher et al., Phys. Rev. C., V. 53, 1996, p. 1711.
4. M.S. Onegin, A.V. Plavko. 15[th] Natioanl Conf. on Nucl. Phys. Book of Abstracts. St. Petersburg, 2005, p. 270.

# SECTION 1
# Relations Between Measured and Calculated Spin Observables in Inelastic Scattering of Polarized Protons for the Excitation of the $1^+$, $T = 0$ State in $^{12}$C

## Difference Function ($P$–$A_y$)

In this section we investigate an exchange treatment of energy dependence of spin observables in proton scattering for the excited $1^+$, $T = 0$ state in $^{12}$C. Our primary aim is to analyze spin observables from the point of view of the difference between polarization and the analyzing power ($P$–$A_y$). Here we restrict our research to the treatment of exchange required for the excited $1^+$, $T = 0$ state in $^{12}$C, and expand our study of the $1^+$, $T = 1$ state in $^{12}$C [1, 2].

We use a similar approach to exchange approximations in our calculations of proton scattering in order to advance our research [3] of energy dependence for the spin-transfer parameters $D_{NN}$ from inelastic scattering of polarized protons in the excitation of the $1^+$, $T = 0$ state in the $^{12}$C nucleus. We also consider the emergent situation for the case of the spin-transfer parameters $D_{SS}$ for the same nuclear excitation.

All the calculations of proton scattering were made using two computer programs: the DWBA 91 of Raynal [4] and the LEA (linear expansion analysis) of Kelly [5].

The same as it was performed for $T = 1$ [1, 2], here we conduct our analysis mainly in the region of the so-called energy window for the nuclear structure, ranging from ~100 MeV to ~500 MeV. In this energy interval researchers find the nucleus most transparent [6]. It is the studies of the energy dependence of the observable that may provide insight into the source of the contributions that are present in the interaction [7].

The difference function ($P$–$A_y$) for the $1^+$, $T = 0$ transition in $^{12}$C has been a long-standing problem. It is interesting to note that historically the basic quality understanding of the function behavior had been achieved [8] before a sufficient amount of experimental data were accumulated [9–11].

Consequently, the time-reversal invariance requires, for elastic scattering of protons, the polarization $P$ to be identical to the reaction analyzing power $A_y$, which is known from numerous papers. For inelastic scattering, various nonrelativistic and relativistic models predict nonzero quantities for ($P$–$A_y$); however, in some cases these quantities can be rather significant.

It is shown in Amado's work (see Ref. [8]), based on invariance principles only, that for inelastic scattering at the energy of approximately 500 MeV the ($P$–$A_y$) quantity in a direct PWIA treatment vanishes, except for very small effects related to the $Q$ reaction quantity. At the same time, in research [9] at a somewhat lower energy (100–200 MeV), exchange processes permitted a larger ($P$–$A_y$). Moreover, even in an early research by Satchler that could be considered a study of the difference ($P$–$A_y$) in an adiabatic limit [12], the problem of the model absence of the exchange interaction as a limited version of approximation was put forward. A number of subsequent studies showed that if nonlocal terms appear explicitly in the $NN$ effective interaction [7], or if the nonlocality (or velocity dependence) of an effective coupling between projectile and target nucleons takes place, then the model gives ($P$–$A_y$) $\neq 0$, even for the quantity $Q = 0$, and even in the absence of distortion [13]. In particular, such nonlocal effects can arise through an explicit treatment of knock-on exchange processes [9]. Especially important in these processes is a tensor exchange contribution. Therefore, the exchange terms arising from the tensor force are a more important source of ($P$–$A_y$) $\neq 0$ than those associated with the central force [13]. Based on the assumption that ($P$–$A_y$) is driven primarily by tensor exchange contributions and knowing the behavior of the latter at different energies, we can observe quality energy changes in the ($P$–$A_y$) quantity [7]. This is what we demonstrated in our papers [1, 2] for the excitations of the $1^+$, $T = 1$ state in $^{12}$C.

Even the first experimental data for inelastic-proton transitions to $1^+$ states in $^{12}$C ($T = 0$ and $T = 1$) at 150 MeV showed large differences between $P$ and $A_y$ [9]. Furthermore, the theoretical analysis of the data [9] reveals that these differences arise mainly from the nonlocal (exchange) character of the effective $NN$ interaction, primarily from the tensor component. Although, it is possible to reproduce to a certain extent the shape of the distribution ($P$–$A_y$) in the case of $T = 1$; for $T = 0$ there is a strong discrepancy in phase between description and experiment. A further theoretical analysis [14] failed to improve the situation for $T = 0$.

The results got better later on, and the description of spin-difference functions for $T = 0$ appeared to be, on the whole, not worse than for the $T = 1$ excited state, in both nonrelativistic [13, 15] and relativistic [15] approaches. The nonrelativistic calculations presented in Ref. [15] are consistent with those presented in Ref. [13], but significantly differ from those reported in Refs. [9, 14]. It is due to a different version of the tensor force in two types of treatment used in Refs. [9, 14], and in Refs. [13, 15].

In a nonrelativistic impulse approximation, the nonzero quantities ($P$–$A_y$) were obtained through inclusion of nonlocal effects, such as knock-on exchange, and found partial success at 150 MeV for the $1^+$, $T = 0$ state in $^{12}$C [13, 15].

The relativistic impulse treatment possesses a structure automatically having those terms that are necessary for the reproduction of a similar nonzero spin-difference function even in a plane-wave approximation [15]. When distortions are added, the relativistic theory gives a good description of the spin-difference data ($P$–$A_y$) for the $1^+$, $T = 0$ state in $^{12}$C at 150 MeV [15]. As is pointed out in Ref. [15] for $T = 1$, these relativistic approaches proved to be less successful. However, there are other hidden possibilities for relativistic calculations, which include exchange and other nonlocalities [15].

## Various Treatments of Knock-on Exchange for the Function ($P$–$A_y$)

Thus, the accumulated data make it clear that in order to describe the spin-difference function ($P$–$A_y$) for the $1^+$, $T = 0$ state in $^{12}$C we must choose a program which would contain a finite-range exchange needed to handle the tensor exchange amplitude. Lately the role of such a program has been assigned to the program DWBA 91 [4], primarily used in the case of unnatural parity transitions. We show the results of using this program in Fig. 1, as well as in the subsequent figures. For the purpose of comparison, we also show our calculations made using the LEA code [5], which is normally employed for natural parity transitions. LEA offers a better way to incorporate electric and magnetic form factors [5]. Strictly speaking, such a comparison is not entirely correct. However, it is aimed at singling out the basic approximations of these two programs and examining the role of additional degrees of freedom in the baseline interaction.

Certainly, it would be more appropriate to compare the results of the calculations made, for example, with the use of the standard LEA program from Kelly [5], on the one hand, and with the LEA program, modified by Stephenson's group [16–18], on the other. A zero-range treatment of knock-on exchange in the standard DWIA program LEA [5] was modified to follow nucleon-nucleus kinematics. This modification agrees well with exact finite-range calculations made using a series of finite-range DWIA programs based on the work by Schaeffer and Raynal (see e.g. Ref. [4]). Unlike prior applications of LEA [5], the kinematics for the exchange amplitude added here used momentum transfer calculations based on certain assumptions [4, 16–18]. Normally, direct and exchange pieces are added to get a total interaction for input into LEA. A full treatment of exchange is important for the tensor component of the interaction, as the direct and exchange parts of the interaction have different spin operators.

This scheme was adopted after some trial and error, but it appears to work well when the angular momentum transfer $L \geq 3$ [16–18]. Lower spin unnatural-parity states pick up sensitivities to nuclear current and finite-range exchange, which complicates the analysis. Moreover, lower spin states are sensitive to the choice of amplitudes for many particle-hole configurations [13, 19].

Taking into account the above-stated information, we chose the simplest method available. So in Fig.1, we compared the results of calculations made with the full treatment of exchange, using the program DWBA 91 [4], with those available in LEA [5], based upon a zero-range exchange approximation [5] with the Geramb (Paris density-dependent) interaction [6].

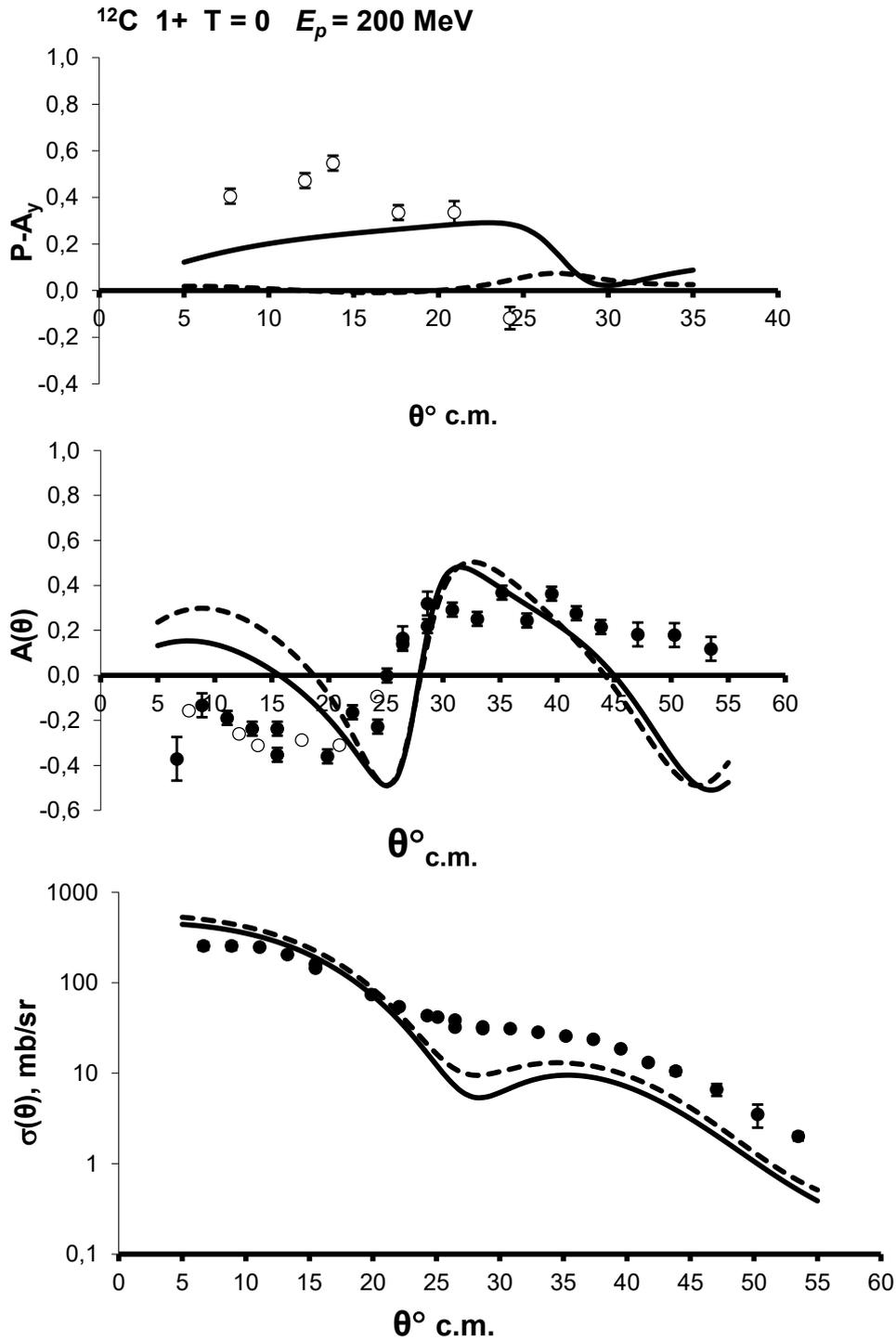

**Fig. 1.** Typical relations between experimental and calculated observables in polarized-proton scattering at the excitation of the $1^+$ ($T = 0$) state in $^{12}C$ in the region of 200 MeV. Measurements of the angular distributions of the differential cross section ($\sigma$) [20], of the analyzing power ($A_y$): [20] – solid points, [7] – open points, and of the difference function ($P$–$A_y$) [7]. The solid lines are DWBA 91 calculations, using the Paris density-dependent (PD) interaction of von Geramb [6], while the dashed lines use the same interaction but in the LEA code.

It is clear that against the background of LEA results, close to $(P–A_y) = 0$, an important source of $P \neq A_y$ is the treatment associated with nonlocality (finite-range) terms in DWBA 91. Such significant differences in calculations (with exact finite-range and zero-range) of the function $(P–A_y)$ for the $^{12}$C $1^+$, $T = 0$ state are demonstrated here for the first time. The difference between a zero-range and a finite-range treatment in this case appears to be very significant. It is true that the agreement between the calculated and experimental data with finite-range amplitudes is achieved only schematically. However, as we will see further on, the results of calculations also depend on the type of the *NN* interaction used. Eventually, we managed to reproduce in our other calculations the basic shape of the difference function $(P–A_y)$, obtained in an experiment at 200 MeV for the $1^+$, $T = 0$ state in $^{12}$C (see Fig. 2 and Ref. [7]).

Just as the $(P–A_y)$ theoretical quantities shown in Fig.1, the $(P–A_y)$, measurements made at various beam energies can be contrasted, too (Fig. 2). The decrease in quantities $(P–A_y)$ with increasing energy $E_p$ is consistent with the assumption that the $(P–A_y)$ quantity is driven primarily by exchange terms arising from the tensor force. Indeed, tensor exchange contributions are expected to become less important at higher energies [7, 13]. These contrasts of $(P–A_y)$ were clearly demonstrated for $1^+$ states in $^{12}$C in the case of $T = 1$ in Refs. [1, 2]. Now a similar pattern is observed for $T = 0$ (Fig. 2).

The discrepancy between two experimental distributions of the $(P–A_y)$ function, indicated in Fig. 2 by open and solid points, respectively, is also observed, though within reasonable limits, in both calculations: those made using the program with a full treatment of exchange (dashed curve), on the one hand, and those with a zero-range treatment (solid curve), on the other hand. The discrepancy between these curves in the case of the $(P–A_y)$ function is significant. However, the discrepancy between the curves, corresponding to similar calculations for $A_y$ and $d\sigma/d\Omega$ distributions in the region of the first (main) diffraction maximum, is not of great importance. This clearly indicates that the $(P–A_y)$ quantities are driven primarily by tensor exchange contributions. At the same time, these parameters do not exert strong influence upon the description of the $A_y$ and the cross section.

As is shown in Ref. [13], in the case of $(P–A_y)\sigma$ observables, the comparison of DWIA calculations with the data for the excitation of the two lowest $1^+$ states in $^{12}$C by 150 MeV protons should prove quite helpful in discriminating between different *NN* interactions having different nonlocal behaviors.

The type of a nucleon-nucleus amplitude, suggested in Ref. [13] and used in the corresponding computer programs, for example, in Ref. [4], arises when we explicitly account for the nonlocality present in exchange terms. In the context of Ref. [13], the nonlocality corresponds to different coordinates for incident and scattered nucleons. As was noted earlier, the exchange terms arise primarily from the tensor force [9, 13].

According to works by J.R Comfort et al. (see [20], etc.), for the $1^+$ state at 12.71 MeV most of the direct contributions are intrinsically weak, but in this case the above state is strongly influenced by the exchange coupling. Here the isoscalar tensor interaction is also very weak, however, there are significant contributions from the isovector interaction through knock-on exchange.

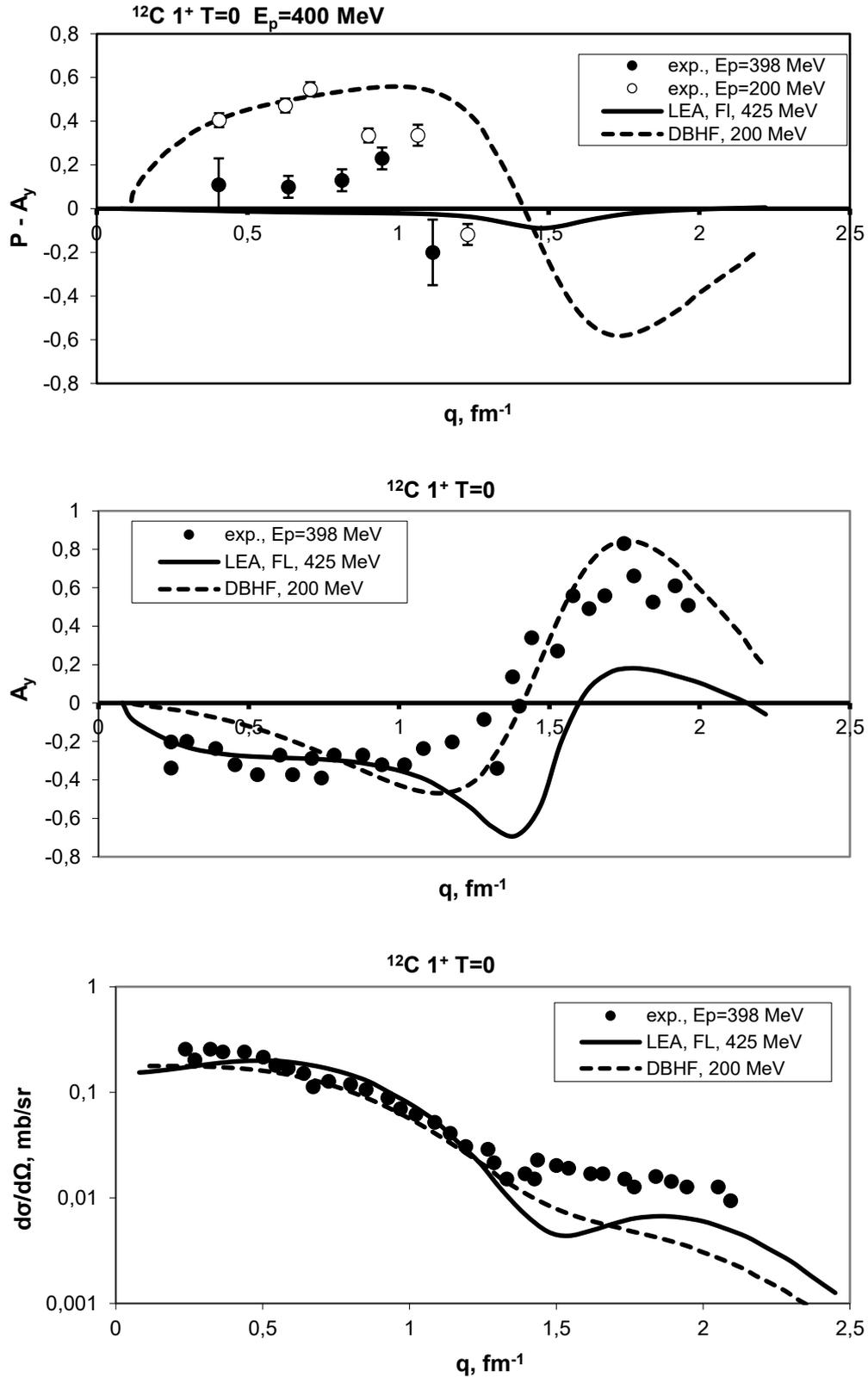

**Fig. 2.** A comparison of momentum-transfer dependence calculations made using the LEA code with the Franey and Love interaction (425 MeV – solid lines) and momentum-transfer dependence calculations made using the DWBA 86 program [21], with the DBHF (Dirac-Brueckner Hartree-Fock) interaction (200 MeV – dashed curves) for the $1^+$, $T = 0$ transition in $^{12}C$ $(\vec{p}, p')$ $^{12}C$. Measurements are represented by solid points for the differential cross-section $d\sigma/d\Omega$, the analyzing power $A_y$ and the difference function $(P–A_y)$ at 398 MeV [10, 22], and by open points for the difference function $(P–A_y)$ at 200 MeV [7].

The $(P–A_y)\sigma$ quantity is usually understood [9, 23] as the product of $(P–A_y)$ and the differential cross section $d\sigma/d\Omega$. These very quantities are represented in Fig. 3 in our model calculations at the indicated energies $E_p$. It is obvious that just as it was in the case of polarization-analyzing-power differences (Fig. 1), a noticeable effect is observed here in approximations in which the *NN* coupling is a local or a nonlocal exchange term, respectively.

In Fig. 3, as we did earlier in Figs. 1 and 2, we compared our calculations for the exact finite-range program DWBA 91 [4] with the exchange part of scattering, on the one hand, and with the calculations made using the program LEA [5] with a zero-range approximation, on the other. The difference between these two approaches appeared to be rather significant.

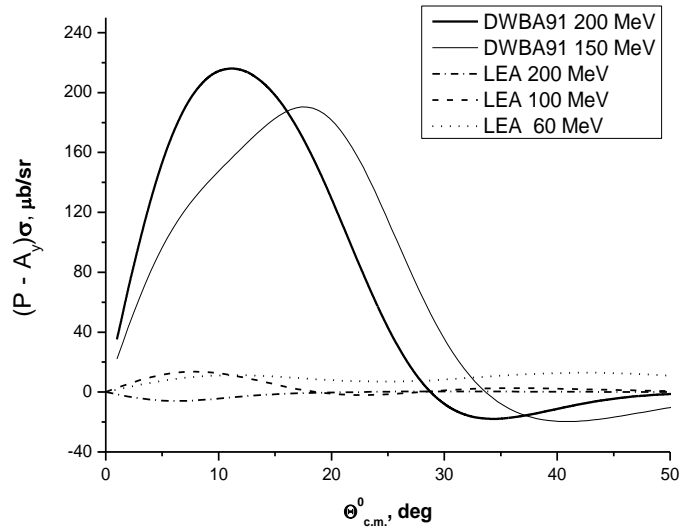

**Fig. 3.** Products of the difference functions $(P–A_y)$ and the differential cross sections $\sigma$, and their variations, arising in the indicated DWIA calculations made using the programs DWBA 91 (solid curves) and LEA (dashed lines), respectively. A significant difference between the calculations arises from the nonlocal and local exchange nature of the *NN* interaction of a Franey and Love type [24]. The corresponding values of proton energies are indicated in the box at the top of the figure.

The fact that a zero-range treatment of knock-on exchange in LEA is also absolutely improper for the description of the experimental data $(P–A_y)\sigma$ is obvious from Fig. 4.

As indicated above, many distinguishing aspects of the two approaches became much clearer in the process of modernizing the standard program LEA by E.J. Stephenson et al. [17, 18]. The scheme that the authors adopted was described in Appendix B in Ref. [17]. According to Refs. [17, 19], there are at least two reasons why LEA calculations could not be reliable, the principal one being that these calculations do not contain a full treatment of exchange. It is most important for such a transition, where there is only a small contribution from the spin-orbit term and a large contribution of the tensor component of the *NN* interaction. According to Ref. [17], this requires some prescription for the calculation of the crossed momentum transfer ($Q$) in terms of the momentum transfer ($q$).

All the above data clearly demonstrate the weakness of using a zero-range exchange integral of the LEA program and the strength of using finite-range DWIA programs, DWBA 91 [4] and DWBA 86 [25], for the description of the $(P–A_y)\sigma$ data in the case of the $1^+$, $T = 0$ state in $^{12}$C at $E_p$=200 MeV. The variation of the *NN* interaction has of course the same meaning, but it is of a secondary importance as less significant (see Fig. 4).

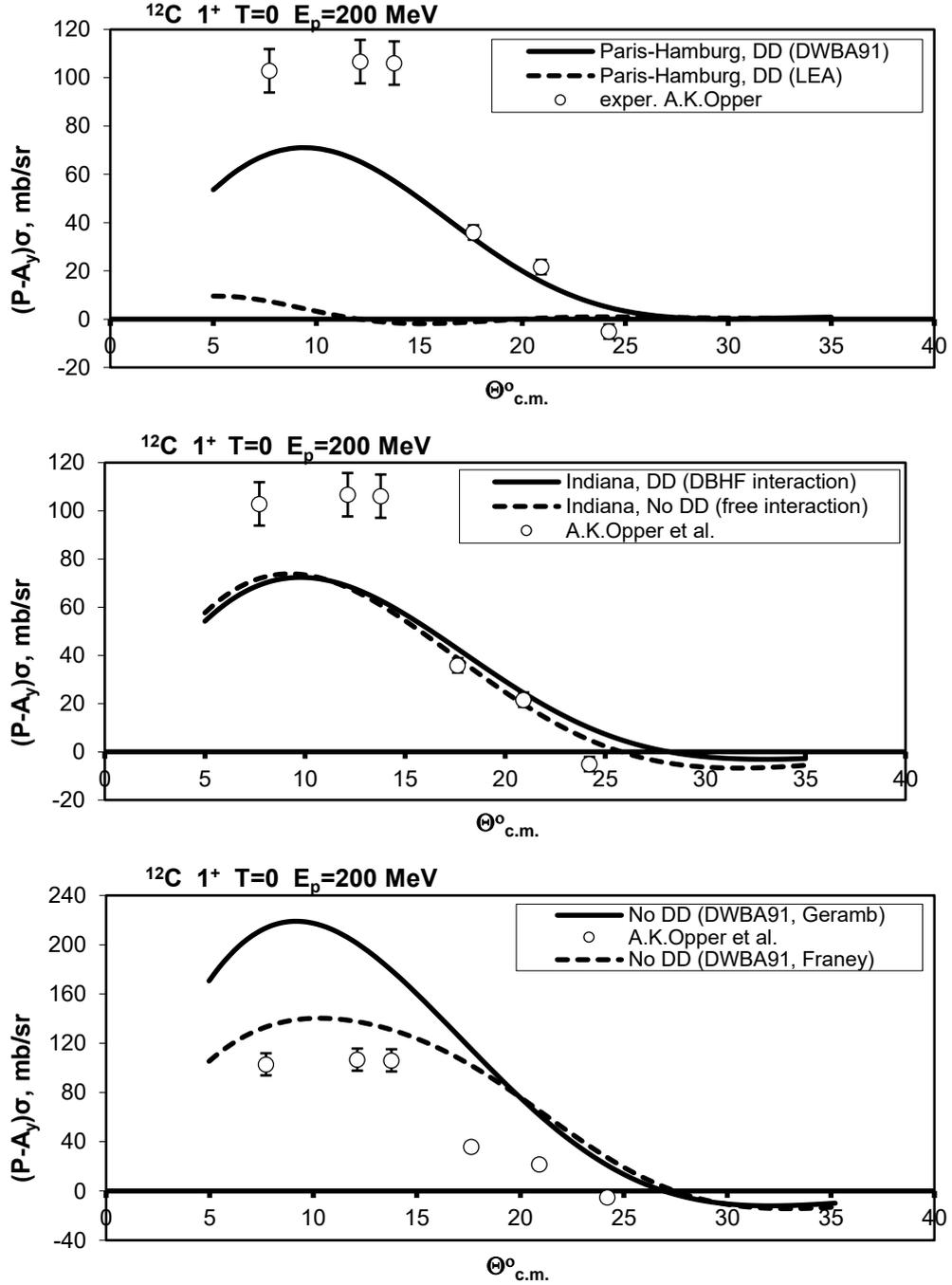

**Fig. 4.** A comparison of DWIA calculations (curves) for the observables $(P–A_y)\sigma$ with the corresponding experimental data (points), based on measurements from Ref. [7] (in the figure designated as A.K. Opper et al.), for the $1^+$, $T = 0$ state in $^{12}$C at $E_p = 200$ MeV. In the top panel, shown are solid (dashed) curves of the calculated observable, obtained with the DWBA 91 (LEA) programs, respectively, using the Paris density-dependent (*DD*) interaction. In the bottom panel, both curves were obtained using the DWBA 91 program only, but with different *NN* interactions. Thus, the dashed curve corresponds to the Franey and Love interaction (designated as Franey), and the solid curve corresponds to the Paris group interaction (designated as Geramb), but without the inclusion of the density-dependent interaction (No DD). The middle panel shows the resulting calculated observables (designated as Indiana), based on individual calculated data from E.J. Stephenson, who used the program DWBA 86. In the latter case both calculated observables were obtained using the DBHF (Dirac–Brueckner Hartree–Fock) interaction of Sammarruca and Stephenson [17]. The solid lines are such calculations with the density-dependent (DD) DBHF interaction, while the dashed line is the free interaction of the same group (No DD).

## Some Aspects of the Description of Polarization (Spin) Transfer Coefficients

It is suggested in Ref. [26] that the addition of exchange has a large effect on the spin-transfer observables $D_{NN}$ and $D_{SS}$. In our previous study [3], we established that for the $1^+$, $T = 0$ state in $^{12}C$ at $E_p \sim 425$ MeV, the description of the $D_{NN}$ data was quite satisfactory when the program LEA was used. At the same time, we showed that in studies of $D_{NN}$ at 150 and 200 MeV the program DWBA 91 was necessary. Here to analyze the $D_{NN}$ results at one of these energies (200 MeV), we will take a comprehensive approach: namely, we will use both programs (LEA and DWBA 91) at a similar effective interaction. So in Fig. 5 (left), in the cases of these two programs, we use the Paris density-dependent (PD) interaction derived from the Paris potential, represented by H.V. von Geramb in Ref. [6]. In addition, in Fig. 5 (right) we show our calculations made with the use of the same programs, employing the Franey and Love parameterization of the *NN t*-matrix [24].

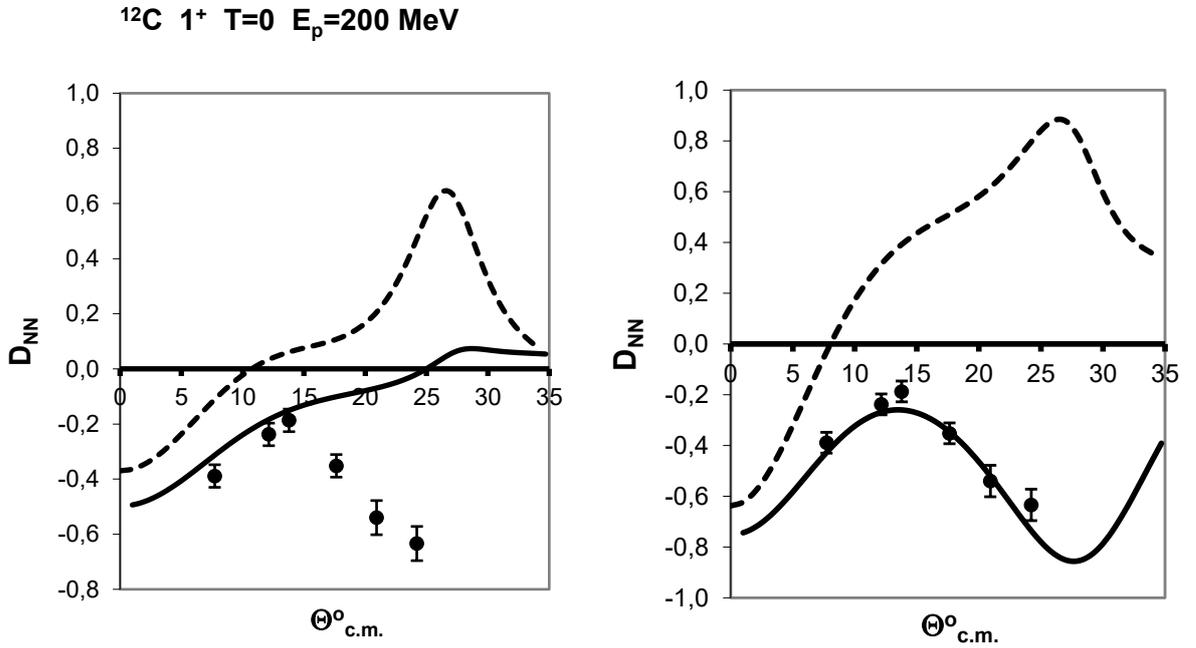

**Fig. 5.** The polarization transfer observables $D_{NN}$ versus c.m. scattering angles for the excitation of the $1^+$, $T = 0$ state in $^{12}C$ at 200 MeV. The results of the experiment [7] are plotted by solid points with statistical errors. The solid and dashed curves are the results of DWIA calculations made using the DWBA 91 and LEA programs, respectively. On the left are calculations with the Paris density-dependent (DD) interaction by von Geramb, and on the right are calculations with the Franey and Love effective interaction (FL).

As is seen in Fig. 5, the advantage of using DWBA 91 is obvious, as this program allows for a finite-range DWIA for exchange contributions. Certainly, the type of the effective *NN* interaction also influences the description of the $D_{NN}$ data. However, the overall interaction impact is weaker than the above-indicated exchange effect.

A similar situation occurs in the case of $D_{SS}$ polarization transfer observables. In Fig. 6 (top panel) we compare the results of LEA calculations with $D_{SS}$ experimental quantities. Almost to the same extent, the measured $D_{SS}$ observables for the excitation of the $1^+$, $T = 0$ state in $^{12}C$ are not consistent with DWIA predictions, regardless of the type of a *NN* interaction: the effective interaction by Franey and Love (FL), or the density-dependent (DD) interaction by von Geramb. At the same time, calculations made using the FL and von Geramb DD interactions describe $D_{SS}$ much better when the program DWBA 91 is used, which is shown in Fig. 6 (bottom panel). Despite the fact that in this case the calculations do not reproduce the $D_{SS}$ data well enough, the overall advantage of using the program DWBA 91 is quite obvious.

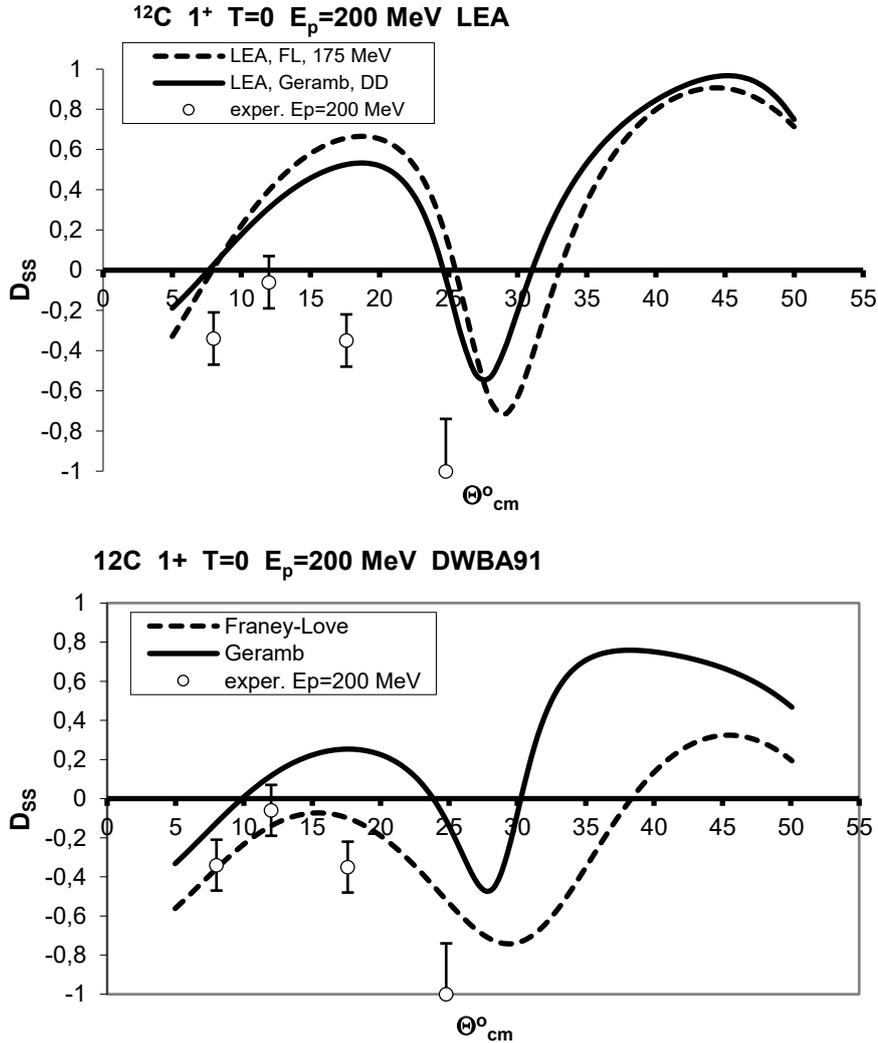

**Fig. 6.** $D_{SS}$ polarization transfer observables versus c.m. scattering angles for the excitation of the $1^+$, $T = 0$ state in $^{12}$C at 200 MeV. The results of the experiment [27] are plotted by open points with statistical errors. The solid and dashed curves are the results of DWIA calculations with the Paris density-dependent interaction by von Geramb (designated as Geramb DD, or simply Geramb) and with the Franey and Love interaction (FL), respectively. In the top panel are calculations made with the use of LEA program, and in the bottom panel are calculations made using the DWBA 91 program.

Consequently, the description of $(p, p')$-scattering in the region of $E_p = 200$ MeV for the $1^+$, $T = 0$ excited state in $^{12}$C becomes rather critical in the case of the following data: $(P-A_y)$ (Fig. 1), $(P-A_y)\sigma$ (Fig. 4), $D_{NN}$ (Fig. 5), and $D_{SS}$ (Fig. 6), from the point of view of the use of the programs DWBA 91 or LEA. It is obvious that such cases require the use of finite-range programs for the exchange part of the scattering.*

A contrasting situation is observed for the cross section (Fig. 1), where only an insignificant difference between calculations using both programs can be seen. Although the treatment of exchange in a local approximation is somewhat similar to the case of $(P-A_y)$ data at 425 MeV, is not appropriate at 200 MeV (Fig. 2).

*NOTE ADDED BY THE AUTHOR OF THE REVIEW IN PROOF

The analysis of the polarization-transfer observables, presented in Figs. 5 and 6, was also performed for $D_{LL}$ data. The results of the analysis were described for a limited range of angles in our publication: A.V. Plavko, M.S. Onegin, V.I. Kudriashov // 60$^{th}$ Intern. Conf. on Nuclear

Physics – "Nucleus 2010". Book of Abstracts. St. Petersburg, 6–9 July, 2010, p. 283. For an extended range of angles the result remains practically unchanged from the point of view of experiment description. Thus, the DWBA 91, FL predictions are in excellent agreement with $D_{LL}$ measurements. There are also no conflicting conclusions resulting from the comparison with LEA, FL calculations. The role of direct and exchange contributions is likely to change at large scattering angles, where the calculations of these two types significantly differ, but the necessary measurements are not available. Therefore, the conclusions that were possible in the case of $D_{NN}$ and $D_{SS}$, but not for $D_{LL}$, remain uncertain until additional experiments are performed.

# SECTION 2

# A Comparison of Spin Observables in the $^{12}$C ($\vec{p}, \vec{p}'$) $^{12}$C Reaction at 100–500 MeV with Different Distorted-Wave Impulse Approximations

## Introduction to Background

As is known (see e.g. Refs. [1, 2]), polarization spin observables are essentially determined by the spin dependence of an effective nucleon-nucleon (NN) interaction. They include first of all such polarization transfer observables as triple $D_{NN}$, $D_{SS}$, $D_{LL}$, $D_{LS}$, and $D_{SL}$ scattering parameters (coefficients). All these coefficients are identical to the corresponding Wolfenstein parameters in the theory of triple scattering of unpolarized particles to unpolarized targets, where they have their own designations [3]. The indices L, N, and S refer to a longitudinal (beam direction), to normal (to the horizontal scattering plane), and to sideways ($\hat{N} \times \hat{L}$) directions in the laboratory coordinate system. It is natural that for the analyzed spin process, the functions of asymmetry always play an important role. Here $P(\theta)$ is a polarization function, and $A_y(\theta)$ is an analyzing power for the reaction.

All these seven polarization observables are parts of corresponding equations [4]. They can be easily combined into one general matrix equation [5] and defined through it. It will demonstrate the polarization of the scattered particles, $(P_S, P_N, P_L)$, in terms of the polarization of the incident beam, $(P_S^0, P_N^0, P_L^0)$. It is clear that in order to measure the polarization transfer observables, these three components of the incident beam polarization must be determined as exactly as possible, providing these components of the polarization of the scattered particles are available.

It is obvious from the structure of the matrix equation, including $(P_S, P_N, P_L)$ and $(P_S^0, P_N^0, P_L^0)$ values [5], that the main connecting link is the definite matrix. In it the $D_{SS}$, $D_{NN}$ and $D_{LL}$ coefficients are diagonal elements, and the $D_{LS}$ and $D_{SL}$ are off-diagonal ones. So the corresponding designations of the parameters of the polarization transfer or spin transfer follow from them.

Proton inelastic scattering with its spin observables can be used as a "filter" to examine particular pieces of the effective NN interaction. For example, transitions to unnatural parity states are driven mainly by the tensor and spin-orbit parts of the effective interaction [6]. One of the aims of the present review is to study these particular terms of the effective interaction. Complete sets of polarization transfer (PT) observables permit a comprehensive study of tensor and spin-orbit forces [6–8]. The induced polarization $P$, and the analyzing power $A_y$ traditionally play an important role. Differences between $P$ and $A_y$ are driven by nonlocal (momentum-dependent) terms in the effective interaction that arises due to knock-on exchange amplitudes [8–10].

The indicated experiments, involving incident proton beam polarizations oriented in-plane (along sideways and longitudinal directions), were followed by other important experiments, which allowed to obtain new results on the polarization transfer at 0° [11]. Recently, we have combined and analyzed all these experimental results for the excited $1^+$, $T = 0$ (12.71 MeV) state in $^{12}$C [12]. Presently we continue compiling and systematizing the data, combined with an appropriate set of ($\vec{p}, \vec{p}'$) observables. In doing so we conduct a comprehensive analysis of the results, covering not only the $1^+$, $T = 0$ state, but also $1^+$, $T = 1$ (15.11 MeV) level in $^{12}$C. Moreover, we also consider measured relative cross sections in $^{12}$C ($p, p'$) reactions at 0° in the region of $E_p = 100$–400 MeV for $1^+$, $T = 0$ and $1^+$, $T = 1$ states in $^{12}$C [13]. The $1^+$, $T = 1$ cross section at 0° is dominantly induced by the central part of the isovector spin dependent component $V_{\sigma\tau}$, which has been studied extensively in $(p, n)$ reactions. Finally, it is very important for us (see also Ref. [13]) that there are substantial contributions from the isovector tensor interaction $V_\tau$ through knock-on exchange processes for the 0° cross sections of the $1^+$, $T = 0$ state.

# Relative Cross-Sections in $^{12}$C ($p, p'$) $^{12}$C Reactions at 0º

Developing the Wolfenstein theory [3], Ohlsen showed in Ref. [4] that when invoking parity conservations and time reversal, the number of nonzero independent observables might amount to eight. We will consider this point later. Now let us note that according to the concept of Ref. [14], resulting from Ohlsen's formalism [4], the relationship between the polarization components of incident and outgoing nucleons may be expressed in a general matrix form (somewhat different than in Ref. [5]). It includes cross sections in addition to the seven observables discussed in the Introduction to the present review. If we replace the rather archaic form of the presentation of early theoretical studies [3, 4] by modern established symbols [2, 14, etc.], the scattering process may be represented as follows.

Thus, based on the nucleon-nucleon (*NN*) scattering amplitude $M(q)$, the nucleon-nucleus scattering amplitude $\bar{M}(q)$ can be formed. That will give us simple expressions for the scattering parameters of the Wolfenstein theory. As usual, we can define $D_{ij}$ polarization transfer coefficients in terms of a nucleon-nucleus (*NA*) scattering amplitude, i.e. $\bar{M}$. Then, according to Refs. [2–4, 14, etc.], we get:

$$\frac{d\sigma}{d\Omega} D_{ij} = \frac{1}{2} Tr\, [\bar{M}\sigma_i\, \bar{M}^+\sigma_j]. \qquad (1)$$

In Equation (1) and the other expressions in this section, the differential cross section has the traditional designation $\frac{d\sigma}{d\Omega}$. In the other sections of the review, we use, for the sake of simplicity (as other authors often do), the contracted designation $\sigma$ (without indexes). In Eq. (1), $\sigma_i$ ($\sigma_j$) is the Pauli spin matrix for the *i*th (*j*th) component of incident and scattering nucleon polarization, respectively. The traditional, though unusually represented, observable is

$$\frac{d\sigma}{d\Omega} = \frac{1}{2} Tr\, [\bar{M}\, \bar{M}^+], \qquad (2)$$

which is the usual unpolarized differential cross section.

Naturally, we also deal with the polarized cross section in polarization experiments, which we will mark by the index *p*. Then such a cross section will take the following form (see e.g. Refs. [4, 14]):

$$\frac{d\sigma_p}{d\Omega} = \frac{d\sigma}{d\Omega} (1 + D_{NO} P_N^0), \qquad (3)$$

where $D_{NO} \equiv A_y$. Here $D_{NO}$ is in fact scattered yield asymmetry due to the normally (*N*) polarized incident beam, i.e. the reaction analyzing power $A_y(\theta)$.

As is clear from Ref. [15], the cross sections for the ($p, p'$) excitation of the well-known $1^+$ states in $^{12}$C are peaking at $q_{min}$ (0º). The comparison of Formulae (2) and (3) shows that

$$\frac{d\sigma}{d\Omega}(0°) = \frac{d\sigma_p}{d\Omega}(0°), \qquad (4)$$

so the situation here gets ordinary.

In work [13] were performed zero-degree measurements of proton inelastic scattering for the $1^+$, $T = 0$ (12.7 MeV) and $1^+$, $T = 1$ (15.1 MeV) states in $^{12}$C at $E_p = 100, 200, 300,$ and 400 MeV. The total energy resolution was 200–300 keV. The yields for each state were obtained by the usual peak fitting procedure under the assumption of a Gaussian peak and the underlying continuum. The ratios of these cross sections,

$$R = \frac{d\sigma}{d\Omega}(1^+, T=0)\, /\, \frac{d\sigma}{d\Omega}(1^+, T=1), \qquad (5)$$

were determined from the peak yields. As a result, it appeared that the experimental ratios $R$ were almost constant ($R \approx 0.09$). All these results are shown in Fig. 1 by dots, with data uncertainties in the experiments taken into account. The cross sections for two $1^+$ states in $^{12}C$ were calculated by the authors of Ref. [13] with the program DWBA 91 from J. Raynal.

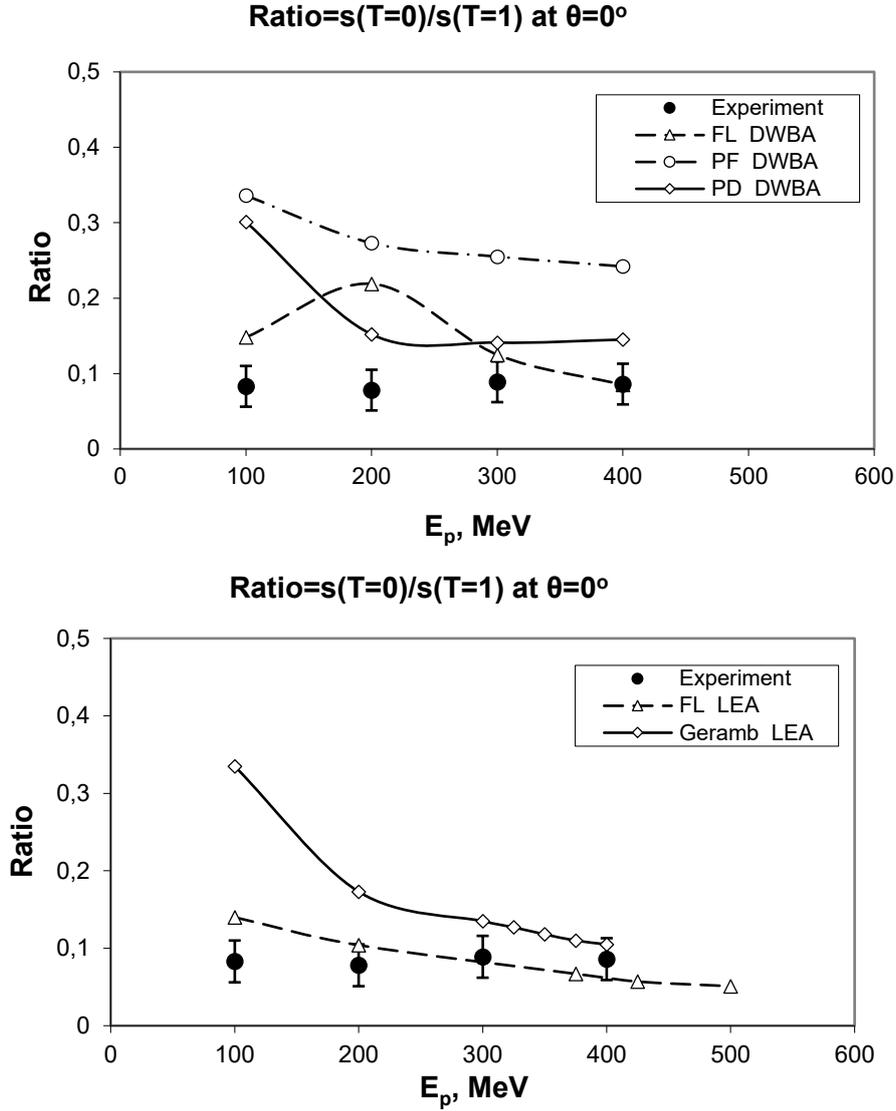

**Fig. 1**. The ratio of cross sections for the $1^+$, $T = 0$ and $1^+$, $T = 1$ states (the dark circles represent experiment [13], the white dots represent the calculated results of the present paper with curves smoothly plotted through them). The top panel represents the use of the program DWBA 91 [16], and the bottom panel represents the application of the program LEA [20]. The top and bottom panels show our calculations made using FL, the free *NN* *t*-matrix interaction (dashed lines) [17]. PD represents the Paris density-dependent interaction (solid lines) [18], and PF represents the Paris free interaction derived from the Paris potential (dot-dashed line) [18] in the top or bottom panels. Detailed designations are given in the boxes at the top of the figure.

We reproduced almost all the calculations made in Ref. [13], using a version of this code, kindly given to us by its designer, J. Raynal [16] (Fig.1, top panel). The overwhelming number of DWBA estimations were found to be larger than the experimental values (the same as given in Ref. [13]). Only the FL DWBA calculations agreed well with the experimental data in the region of 400 MeV. FL denotes the parameterization of the free *NN* *t*-matrix by Franey and Love [17]. The Paris density-dependent (or Geramb) interaction is denoted in Fig. 1 by PD. And PF stands for the Paris free interaction [18], derived from the Paris potential [19]*. The bottom panel of the figure shows all the calculations made using the computer program LEA from Kelly [20]. It is clear that our FL LEA calculations are the most satisfactory, if we consider all the studied region of $E_p$ (100–500 MeV). The PD (Geramb) LEA calculations provided similar results only at $E_p$ = 400 MeV.

The weakness of the program LEA is in the use of a zero-range integral. But the strength of this program is that it provides the best-fit transition density [21]. These positive qualities of LEA are apparent in the study of cross sections (Fig. 1, bottom panel)**. However, as we will see further on, for some other observables, where the treatment of exchange becomes a more serious matter, it is necessary to employ a finite-range program for the exchange part of the scattering, i.e. the DWBA 91 program.

An important feature of this study is that it allows us to describe quite satisfactorily, using just one program (LEA) and one interaction (FL), the character of $R$ in a wide range of $E_p$ at a zero-momentum transfer ($q = 0$). As many researchers indicate, the properties of certain forces become especially obvious in the energy region of $150 \leq E_p \leq 500$ MeV and give rise to the so-called energy window for nuclear structure studies. The incident-energy region around the indicated $E_p$ is well suited for nuclear structure studies for the reasons summarized in Ref. [22]. First, the distortion effects on the wave function are relatively small, because they are related to the strength of the central scalar-isoscalar interaction $V_o^C$, which is weak in this energy region. Second, a weak $V_o^C$ implies a suppression of multistep processes. This circumstance obviously makes the nuclear reaction mechanism simpler. Therefore, the inelastic nuclear excitation spectrum at not very large excitation energies is mainly the result of a one-step process that can be calculated within the so-called microscopic nuclear models.

## Diagonal Spin-Transfer Parameters

We want to prove here that the complete data on the spin-transfer parameters for normal ($N$), longitudinal ($L$), and sideways ($S$) polarized beams demonstrate quite clearly the sensitivity to individual terms in the $NN$ interaction. We consider not only diagonal ($D_{ii}$) spin-transfer parameters, which is more commonly done, but also off-diagonal ($D_{ij}$) ones as well, in accordance with the designations of Formula (1). At that, we use the most established modern designations for the indices in $D$:

$$(i, j) = N, L, S \text{ and } N', L', S', \quad (6)$$

where $i$ refers to incoming components, and $j$ represents outgoing spin ones. Here we omit, as it is commonly done, the prime on the outgoing (more often second) subscript.

It is important to determine to what extent FL LEA calculations, established for the parameter $R$ (Fig. 1), can be extended to spin-transfer observables measured in this or that part of the "energy window". As for the generalizations in Ref. [22], the force component $V_{\sigma\tau}^C$, which excites spin-isospin-flip transitions, is recognized to be especially large in the "energy window". At the same time, the isovector spin-independent component $V_\tau^C$ is strongly reduced. Thus, Ref. [22] represents the experimental and theoretical ratio $\left| V_{\sigma\tau}^C / V_\tau^C \right|^2$ at the zero-momentum transfer ($q = 0$) versus $E_p$. At $E_p = 200$–500 MeV, this ratio is found to be larger than 10. Such a predominance of $V_{\sigma\tau}^C$ over $V_\tau^C$ explains why nucleon-induced reactions at the given $E_p$ are an exceptionally valuable probe of isovector spin modes in nuclei.

It is well known that the FL (Franey and Love) interaction is tabulated for the range of energies from 100 to 800 MeV. If we consider a wider range of $E_p$ (from 25 to 800 MeV) and take into account the main factor in the determination of distorting potentials, it is worthwhile to single out the following three regions of $E_p$ [23]: $E_p < 100$ MeV, $100 < E_p < 400$ MeV, and $E_p > 400$ MeV. The present study is aimed at the second region, in which both the matter coupling and spin-dependent coupling vary slowly with $E_p$. In addition, as follows from the analysis in Ref. [23], spin excitations are seen with the greatest clarity. This and other facts allow us to extend the free nucleon-nucleon $t$-matrix from Franey and Love [17], used to describe the parameter $R$ (Fig.1), to the data $D_{ii}$ for $1^+$, $T = 0$ and $1^+$, $T = 1$ excited states in $^{12}$C, obtained in $(\vec{p}, \vec{p}')$ scattering (Fig. 2).

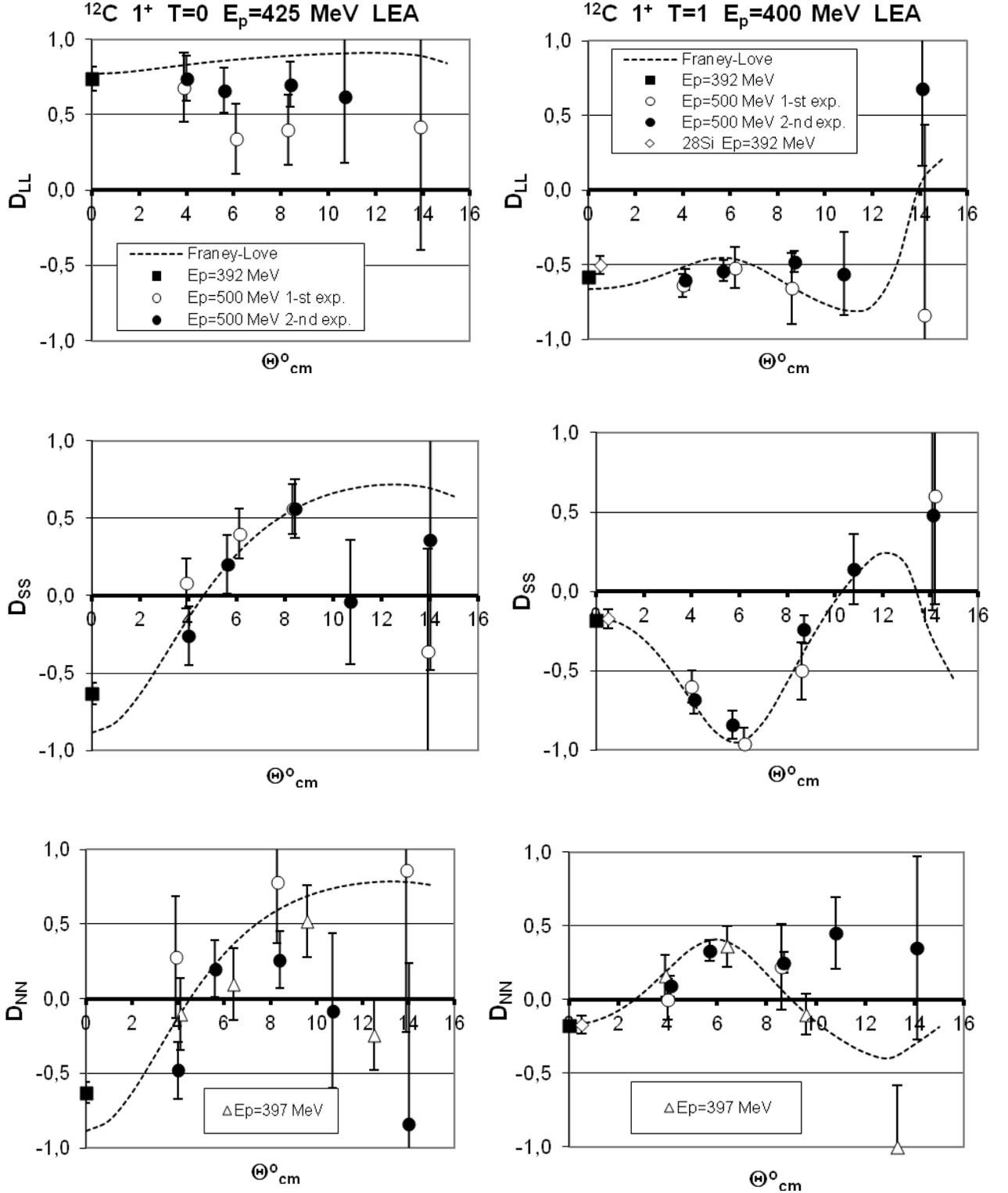

**Fig. 2.** The spin-transfer coefficients $D_{NN}$, $D_{SS}$ and $D_{LL}$ in $(\vec{p}, \vec{p}')$ scattering at the excitation level $1^+$, $T = 0$ in $^{12}$C (left) and $1^+$, $T = 1$ in $^{12}$C and $^{28}$Si (right). Experimental data in $^{12}$C for 0° at 392 MeV are represented by dark squares, and in $^{28}$Si by white rhombuses [24, 11]. In the case of $D_{SS}$ and $D_{NN}$ ($T = 1$), the corresponding data in $^{12}$C and $^{28}$Si at 0° are so similar that they overlap each other. Measurements at $E_p = 397$ MeV for $D_{NN}$ [25] are shown by white triangles. The black circles [8] (designated as 2nd exp.) and white circles [7] (1st exp.) represent experiments for 0.5 GeV. The curves demonstrate our FL LEA calculations made with the Franey and Love potentials for $E_p = 425$ MeV ($T = 0$) and $E_p = 400$ MeV ($T = 1$).

Let us consider the approach applied in an impulse approximation. In general, such an approximation should result in a form adopted for the effective interaction. The outcome is a local interaction at each energy, represented as a sum of potentials (the Yakawa type) in configuration space. The latter contains a sum of central, tensor, and spin-orbit parts [23]. Such a local effective potential allows for antisymmetrization between a projectile (proton) and a struck nucleon in the nucleus. This was taken into account by using a series of finite-range DWIA programs (like DWBA from Raynal) that explicitly evaluate exchange amplitudes. It is also possible to use another computer program that evaluates the exchange amplitudes approximately by adding a local exchange interaction and calculating without symmetrization [23]. The latter approximation is taken into account by using the program LEA from J. Kelly [20] for calculating the results shown in Figs. 1 (bottom part) and 2. Here exchange is treated in a zero-range approximation that adds this piece of the effective $NN$ interaction to the direct part [20, 21].

As is seen in Fig. 2, when using DWIA calculations, for which we employed the LEA program, the accuracy of the description of $D_{SS}$ and $D_{NN}$ at 0° is better in the case of $T = 1$, as compared with the similar data for $T = 0$, which is predictable. For the $1^+$, $T = 0$ state, the treatment of exchange generally requires the use of a program that can provide a more serious approach than the zero-range program LEA [12].

Indeed the isoscalar spin-dependent interaction term $V_\sigma^C$ is rather weak at all incident $E_p$, suggesting that the central part of the interaction is quite ineffective for the excitation of isoscalar spin modes [22]. Although the isoscalar tensor interaction is also quite weak for the isoscalar spin-transfer excitation of the $1^+$ state in $^{12}$C, there are substantial contributions from the isovector tensor interaction through knock-on exchange [26]. As a result, the tensor exchange process in this reaction is a dominant reaction mechanism, while the direct process (governed by $V_\sigma^C$) is suppressed [22, 25]. In our research [12] we were successfully guided by this notion.

In order to formulate the arising problem more clearly, first it was necessary to use lower values of $E_p$, and then choose such observables for which the role of exchange, and not direct processes, could be enhanced.

### $(P+A)\sigma$ and $(P-A)\sigma$ Spin Observables

From the fact that the descriptions of the experimental data $D_{SS}$ and $D_{NN}$ at 0° for $T = 0$ are not quite adequate, it can be derived that the LEA program for the exchange part of scattering is rather simplified. We took it into account when analyzing similar observables, $D_{SS}$ and $D_{NN}$, but at lower energies: $E_p = 200$ MeV in [12]. As is well known (e.g. from Ref. [23]), a better treatment of antisymmetrization is required at lower energies. In the present review, we want to prove that we are able not only to understand this process but also control it. Therefore, a similar procedure of combining observables for the $1^+$, $T = 1$ state in $^{12}$C [27] is reproduced for the excitation of the $1^+$, $T = 0$ state in $^{12}$C at $E_p = 200$ MeV. The result is shown in Fig. 3, where we demonstrate the total and difference functions $(P+A_y)\sigma$ and $(P-A_y)\sigma$, respectively. Here $P$ represents polarization, $A_y$ is the analyzing power, and $\sigma$ is an unpolarized differential cross section. As is shown in Refs. [10, 12, 27, etc.], it is common to consider that

$$(P + A_y)\sigma = 2(\sigma_{++} - \sigma_{--}),$$
$$(P - A_y)\sigma = 2(\sigma_{-+} - \sigma_{+-}), \qquad (7)$$

where $\sigma_{if}$ quantities represent a differential cross section for scattering from an initial ($i$) to a final ($f$) spin projection, as measured along the vector $\hat{N}$. Even such a simple isolation of the observables $(P+A_y)$ and $(P-A_y)\sigma$, as is shown in Eq. (7), makes for a better understanding of the processes of ($\vec{p}, p'$) scattering. Using this method, we will try to take forward the understanding of ($\vec{p}, p'$) processes.

As is demonstrated in Figs. 1 and 2, we used the DWIA program LEA since it incorporates a number of desirable features. When analyzing ($P+A$)$\sigma$ data, we do not find significant differences between the calculated results obtained with exact finite-range calculations made using the program DWBA 91 (or other finite-range DWIA programs), and calculations with the program LEA. Indeed a similar situation is observed when analyzing the data for natural-parity transitions [21] The point is, that here the role of spin-flip is very insignificant, or it is not seen at all (Fig. 3, e, f).

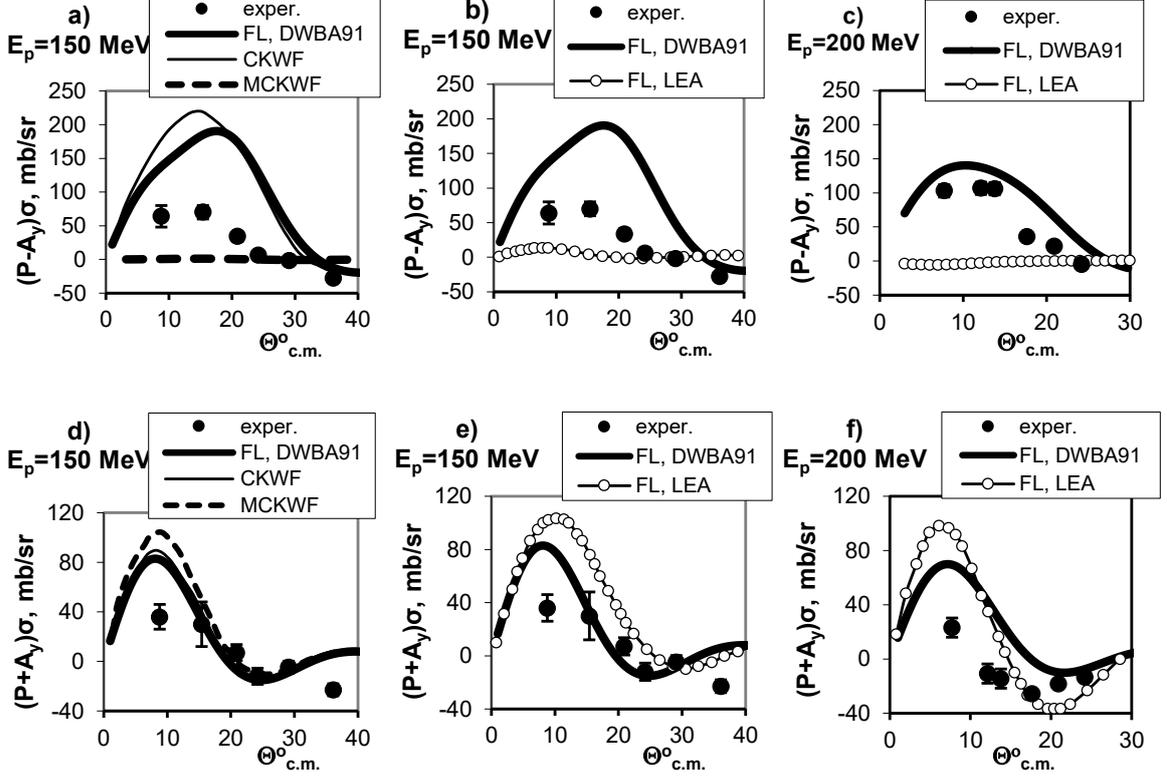

**Fig. 3**. The combinations $(P - A)\sigma$ are shown in the top part (a, b, c), and $(P + A)\sigma$ are shown in the bottom part (d, e, f) for the experiments (dark circles) and for the calculations (white circles and curves connecting them). The thick solid curves in all the cases represent FL DWBA 91 calculations, and the white circles with connecting curves represent FL LEA calculations. In addition, in a and d, the thin solid curves represent calculations with CKWF, and the dashed curve represents calculations with MCKWF (see the text). The experimental data at $E_p = 150$ MeV (a, b, d, e) are based on measurements from Ref. [9], which we complemented and corrected. The experimental results at $E_p = 200$ MeV (c, f) are derived from measurements of Ref. [14]. All the results refer to the $1^+$, $T = 0$ state in $^{12}$C. The corresponding designations are also given in the top boxes.

However, for the spin-flip observable $(P-A_y)\sigma$, similarity in the results in these two calculation methods is out of the question (Fig. 3, b, c). And the use of the LEA program is unacceptable here. If not the value, then at least the form of the $(P-A_y)$ data for the $1^+$, $T = 0$ state in $^{12}$C, requires exact finite-range calculations made using the program DWBA 91.

A different approach, represented in Fig. 3 (a, d), is also possible. It concerns the role of exchange processes. In this approach, one can also employ the same finite-range DWIA program with or without contributions from the strength assumed for the [$LSJ$] = 111 component of the transition density. It is well known (see e.g. Refs. [10, 26]) that measurements of the electromagnetic form factors for the $1^+$ ($T = 0$ and $T = 1$) excitations are totally insensitive to the [$LSJ$] = 111 part of the transition density. In transition density matrix elements for $LSJ$ transfer representations, the $A_{111}$ amplitude is relatively the largest [10]. However, due to an abnormality of the $1^+$ state, this amplitude cannot contribute to the direct matrix elements of ($\vec{p}$, $p'$) scattering. But it contributes to the knock-on exchange process. This, as we have already stated, is found to be driven primarily by the tensor-exchange contribution.

As is shown in Fig. 3 (a and d), at first we used complete DWIA calculations with wave functions of Cohen and Kurath (CKWF) [28], made in Ref. [10] (Fig. 3, a, d, thin solid lines, and symbols from CKWF). Then we performed our own similar calculations with the program DWBA 91 [16], using the same wave function and employing FL interactions (Fig. 3, a and d, thick solid lines). The results of these two types of calculations appeared to be almost identical for both total and difference functions. We then compared this block of data with the results of Ref. [10], using modified CKWF, i.e. MCKWF, constructed by deleting $A_{111}$ amplitudes (dashed curves in a, and d in Fig. 3). Thus we obtained practically the same effects as those shown in Fig. 3 (b, e, and c, f), resulting from the comparison of DWBA 91 and FL LEA calculations.

Consequently, though some calculations with finite-range DWIA programs or with complete wave functions ($A_{111} \neq 0$) are too large, the calculated $(P-A)\sigma$ and $(P+A)\sigma$ results have roughly correct shape and phase. By contrast, the zero-range calculations with LEA and the results obtained with MCKWF ($A_{111} = 0$) totally disagree with the $(P-A)\sigma$ data. Both methods in the case of the $1^+$, $T=0$ excited state in $^{12}$C clearly demonstrate that the spin-dependent part of the effective interaction, which plays here quite an important role, is significantly nonlocal in agreement with the conclusions made in Ref. [10].

### In-Plane Spin-Transfer Coefficients

We have already shown the comparison of the diagonal spin (polarization) transfer coefficients $D_{LL}$ and $D_{SS}$ for the $1^+$, $T=0$ (12.7 MeV) state in $^{12}$C (Fig. 2, left) with those for the $1^+$, $T=1$ (15.1 MeV) state in $^{12}$C, and for the $1^+$, $T=1$ (11.5 MeV) levels in $^{28}$Si (Fig. 2, right), at $E_p = 0.4$–$0.5$ GeV. The forward angle data on $D_{LL}$ reveal a very dissimilar character (in both experiments and calculations). Thus, the moduli of the diagonal observables $D_{LL}$ are roughly equal. Their values are large and positive for $T=0$, but negative for $T=1$. A rather comparable picture is seen for our plane-wave impulse approximation (PWIA) within the same FL LEA calculations (not shown).

Similar tendencies are demonstrated by simple calculations, based on defining the strength ratio of the central spin-dependent term to the knock-on exchange tensor term in the $NN$ scattering amplitude [11, 24]. Both $D_{LL}$ and $D_{SS}$ can become $-1/3$ if the central spin-dependent interaction is dominant, and $+1/3$ and $-2/3$, respectively, if the knock-on exchange tensor interaction is dominant.

The isovector ($\Delta T = 1$) $M1$ transitions from the $T_0 = 0$ ground state are mainly mediated by the isovector spin-dependent term ($V_{\sigma\tau}$) [29]. Due to this fact, $D_{LL} = -1/3$ for the $1^+$, $T=1$ exited states in $^{12}$C and $^{28}$Si. However, the isoscalar ($\Delta T = 0$) $M1$ transitions are mediated by the isovector tensor term ($V_\tau^T$) mainly through knock-on exchange [17]. This results in $D_{LL} = +1/3$ in the case of the $1^+$, $T=0$ state in $^{12}$C. Certainly, these are just rough approximations. However, the estimation of $D_{SS} (= D_{NN}) = -2/3$ at 0° is also in agreement with the experiment for $T=0$.

The DWIA program LEA that we applied using FL interaction gives excellent agreement with the measurements at 0° for $D_{LL}$ in the case of the $1^+$, $T=0$ excitation (Fig. 2), though similar calculations, made using PD and PF interactions, fail to describe the same $D_{LL}$ data [11, 30]. A comparable situation was observed earlier for DWIA calculations, made using the same forms of the effective interaction in the description of $D_{LL}$ for the same state at the bombarding energy of 200 MeV [31]. It was found there that the predictions of the FL interactions are in excellent agreement with the $D_{LL}$ and $D_{SS}$ data, whereas the free density-dependent Paris interactions are in a better agreement with the polarization transfer coefficients $D_{LS}$ and $D_{SL}$ for the inelastic proton excitation of the $1^+$, $T=0$ state in $^{12}$C at $E_p = 200$ MeV.

In order to expand the understanding of the role of different forms of interactions, we show in Fig. 4 our analysis of a whole set of $D_{LS}$ and $D_{SL}$ measurements at 500 MeV that were partially demonstrated in work [31]. There a striking similarity of such coefficients as $D_{LS}$ (or $D_{SL}$) at the two incident energies of 200 and 500 MeV (in momentum-transfer dependence) was particularly emphasized. The results of our study of $D_{LS}$ and $D_{SL}$, shown in Fig. 4, concern the case of the $1^+$, $T = 0$ excited state in $^{12}$C. We conducted a similar research for the $1^+$, $T = 1$ state in $^{12}$C for the same observables, consisting of two sets of measurements at the same energy $E_p = 500$ MeV [7, 8] with the application of the LEA program and FL interactions. The quality of the description of the measurements is identical (not shown). Here we want to stress that, unlike it was the case in [31], we did not have to apply different effective interactions for diagonal and off-diagonal elements of a full spin-transfer matrix. In the analysis shown in Fig. 4 we used the same effective FL interaction and similar conditions of the description of the scattering process as those employed for the description shown in Fig. 2.

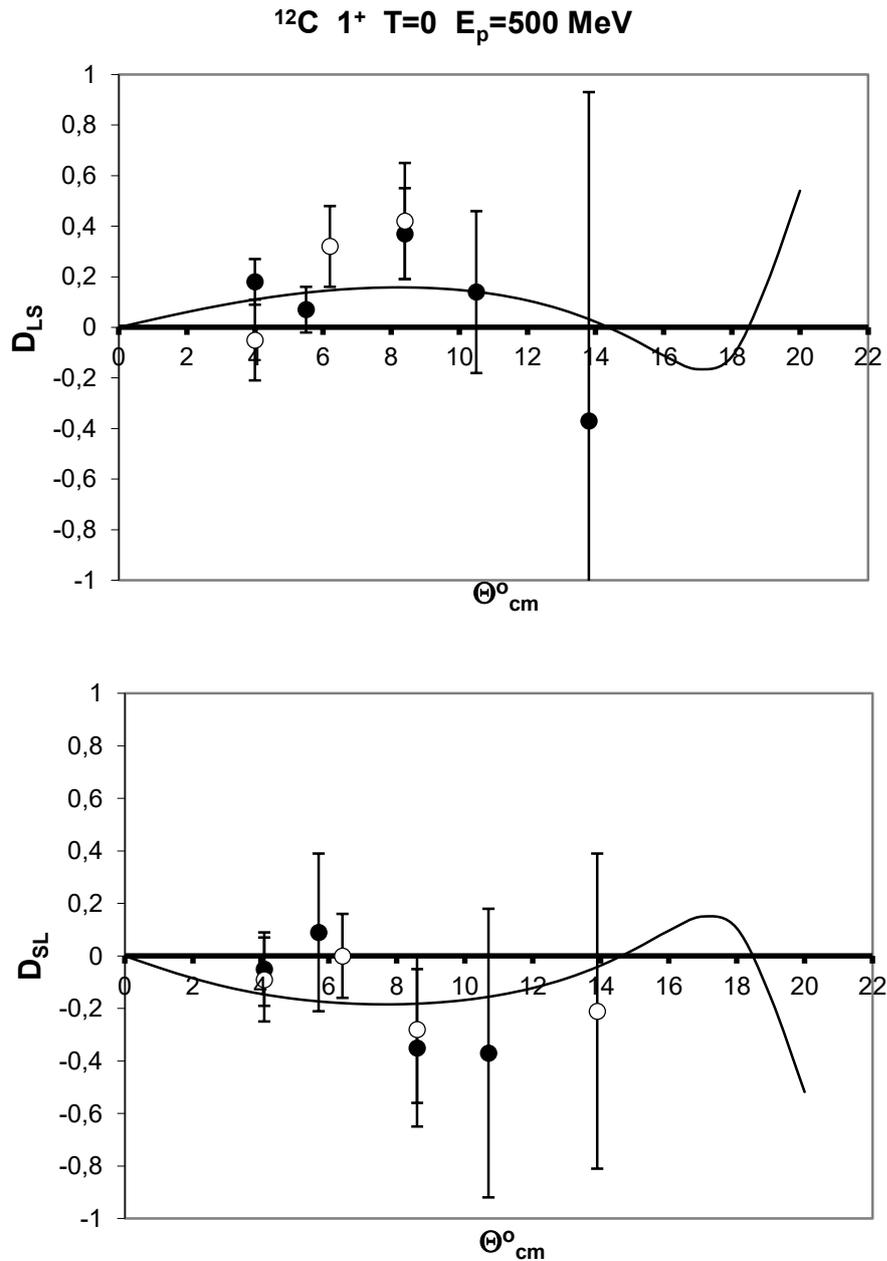

**Fig. 4.** The observables $D_{LS}$ and $D_{SL}$ versus c.m. scattering angles for the excitations of the $1^+$, $T = 0$ state in $^{12}$C at $E_p = 500$ MeV. The results of the experiment [8] are plotted by solid circles. The data [7] are shown by white circles. The curves are the results of our calculations made using the DWIA program LEA with the FL interaction at 425 MeV.

It is noteworthy that for the $D_{ij}$ magnitude the symmetry rules for elastic scattering are well known. One of them is as follows:

$$D_{LS} \approx -D_{SL}. \qquad (8)$$

For natural parity collective states, the rules for elastic scattering should work approximately in the same way. In particular, the effect of Rule (8) was confirmed in experiments in the case of $(\vec{p}, \vec{p}')$ for such states [32]. Besides, it follows from the theoretical considerations of [32] that Relation (8) is generally true for both natural and unnatural parity inelastic transitions, providing that the adiabatic approximation is valid. At the same time, a conclusion was derived from the analysis of the components of the collision matrix [32] that Relation (8) is more trivial for natural parity transitions than for unnatural parity ones. In the latter case, this rule cannot be implemented.

Ratio (8) was tested in work [33], in particular for the $6^-$, $T = 0$ (11.58) state of $^{28}$Si in a $(\vec{p}, \vec{p}')$ reaction at 500 MeV. However, the experiments were very limited: measurements of $D_{LS}$ and $D_{SL}$ were made at two angles only. On the other hand, the nonrelativistic distorted-wave impulse approximation and the Dirac relativistic impulse approximation were used at a rather wide range of angles. The bulk of the results confirms quite well the effectiveness of Rule (8) here. We believe that our calculated results at FL interaction, using the DWIA program LEA, combined with the data of Refs. [7, 8], clearly demonstrate that Rule (8) for the $1^+$, $T = 0$ state in $^{12}$C for the 500-MeV $(\vec{p}, \vec{p}')$ reaction works quite successfully (Fig. 4). However, similar results for the $1^+$, $T = 1$ state in $^{12}$C do not conform to this rule (not shown). In the case of $T = 1$, more complex conditions arise for the interference between the matrix elements of the collision matrix [32]. ***

**Energy Dependence of ($P$–$A_y$) Quantities for the $1^+$, $T = 1$ State in $^{12}$C**

The difference between the induced polarization $P$ and the reaction analyzing power $A_y$, ($P$–$A_y$), for $(\vec{p}, p')$ transitions is usually assumed to be especially sensitive to the tensor part of the effective $NN$ interaction as well as to the nonlocal or exchange nature of the interaction. After getting ($P$–$A_y$) data for the $1^+$, $T = 1$ state in $^{12}$C at 150 MeV [9], were obtained ($P$–$A_y$) quantities for the excitation of the same state in $^{12}$C by 400 MeV protons [34]. A significant decrease in the value of ($P$–$A_y$) data with increasing bombarding energies became evident. This spin-difference function revealed itself near zero or even negative quantities at $E_p = 500$ MeV [8]. At the same time, ($P$–$A_y$) measurements at 200 MeV [35, 14] generally followed the same trend as observed for these data at 150 MeV [9]. All these values for the ($P$–$A_y$) quantity, shown in Fig. 5, demonstrate an increase from small negative values at a small momentum transfer to positive and large quantities at higher $q$ at $E_p = 150$ and 200 MeV.

Indicated Fig. 5 represents systematization of ($P$–$A_y$) quantities for the $1^+$, $T = 1$ state in $^{12}$C at all the $E_p$ values specified above. We have already shown such systematization in our work [36]. The present review covers, along with the discussed ($P$–$A_y$) data for $E_p = 150, 200, 400,$ and 500 MeV, measurements of ($P$–$A_y$) at $E_p = 80$ MeV [37]. Moreover, the systematic presentation of the data in Fig. 5 includes measurements of the angular distributions $A_y$ and $d\sigma/d\Omega$ at $E_p \approx 400$ MeV [38].

As for the discussed dependences, their interpretations mainly concern the fact that such a behavior of ($P$–$A_y$) is consistent with the expected decrease in strength of nonlocal process at higher energies [34, 35, 14]. However, in the theoretical (calculation) process, these qualitative interpretations were explained in a systematic way only in our publication [36], and are now represented in Fig. 5.

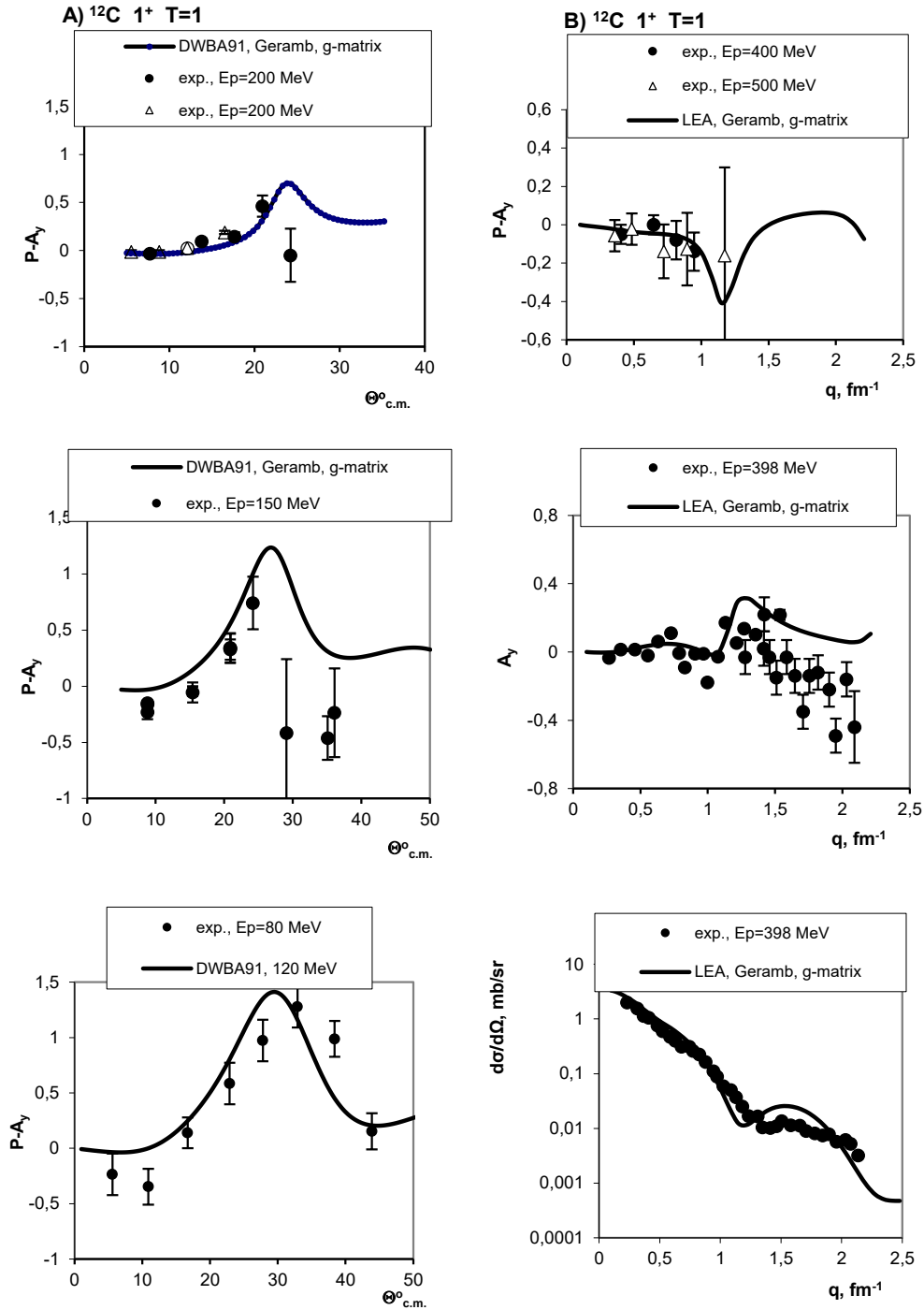

**Fig. 5**. The experimental data (dots) are systematized as follows. Part A (left): $E_p$ are 80 [37], 150 [9], 200 MeV (● from [14], Δ from [35]). Part B: for the ($P$–$A_y$) data, proton energies are ~400 [34], and 500 MeV [8]; for the $A_y$ and the $d\sigma/d\Omega$ data, $E_p = 398$ MeV [38]. Computer programs and the types of nuclear forces used are given in the boxes. In the analysis of the indicated ($P$–$A_y$) data at $E_p$ = 80 Mev we used the Geramb $G$ (or $g$)-matrix at $E_p = 120$ Mev. In Part A the program DWBA 91 is used, and in Part B the program LEA is employed, also with the Geramb $G(g)$- matrix.

Thus, similar to the case of the $1^+$, $T = 0$ state in $^{12}$C at $E_p = 400$–500 MeV [12], in order to describe the $1^+$, $T = 1$ state at the same $E_p$, it is worthwhile to employ a zero-range treatment of knock-on exchange in LEA program that approximately evaluates exchange amplitudes by adding a local exchange interaction. Then the agreement between the calculated and experimental data may be considered quite satisfactory (Fig. 5, right). However, the case of $E_p = 200$ MeV or lower, as we demonstrated in [12], requires the use of a finite-range program (of a DWBA 91 type) for the exchange part of the scattering. This is shown in the left-hand side of Fig. 5. Rather good agreement of the energy dependence of ($P$–$A_y$) observables in the region of $E_p$ = 80–200 MeV qualitatively confirms the assumption [14, etc.] that ($P$–$A_y$) is driven primarily by tensor exchange contributions.

Here, in the description of ($P$–$A_y$), we used the Paris density-dependent interaction [18], to some extent supporting work [39], where for the same reaction, a rather satisfactory description of ($P$–$A_y$) data was achieved at $E_p$ = 150 MeV for the $1^+$, $T = 1$ state (but not for the $1^+$, $T = 0$ level). We have expanded the same type of interaction to the description of the value $R$ at $E_p$ = 400 MeV, where, according to Fig. 1 (bottom part), PD and FL interactions are almost identical as regards their role.

## Summary and Conclusions

We have carried out a set of simultaneous calculations of ($\vec{p},\vec{p}'$) observables for the excitation of both $1^+$, $T = 0$ and $T = 1$ states in $^{12}$C at the incident proton intermediate energy $E_p \approx$ 100–500 MeV, where the impulse approximation is generally considered a valid reaction model. In this regime, we focused on the energy dependence of just one observable, the ($P-A_y$) difference of the $1^+$, $T = 1$ state in $^{12}$C, as a continuation of our analysis of energy dependences for the normal-component spin-transfer coefficients $D_{NN}$ of both $T = 0$ and $T = 1$ states [40, 41]. All these observables may be particularly sensitive to the nonlocal character (e.g., via exchange) of the $NN$ interaction, though such sensitivity can be different in each particular case.

As for the general character of the energy dependence of this or that observable, as it was justly pointed out in Ref. [42], within the spirit of impulse approximation, the changes in the experimental value of this observable at different $E_p$ should theoretically reflect changes in free $NN$ scattering at the corresponding $E_p$. Then possible density-dependent corrections, exchange features of the effective interaction and other effects may bring in significant corrections to the energy dependence.

In the present review, as well as in our works [36, 43], focus is given to the fact that in a wide range of $E_p$, the nonrelativistic impulse approximation (NRIA) allows to provide a rather good description of difference functions, based on ($P$–$A_y$), for the excitation of both $1^+$, $T = 0$ and $T = 1$ states in $^{12}$C. Papers [9, 39] fail to provide a satisfactory description for $T = 0$, and works [8, 14] – for $T = 1$ in $^{12}$C. Ref. [34] does not reproduce simultaneously the $T = 0$ and $T = 1$ data in the calculations. Moreover, in paper [35], the authors characterize the situation as an intriguing result obtained for this observable with relativistic calculations in the case of the $T = 1$ data at 200 MeV. They made full calculations, using a direct-plus-exchange parameterization of the $NN$ interaction (DREX) and calculations made using the direct-only (no exchange) $NN$ interaction (DRIA). It is curious that calculations assuming only direct contributions are able to describe experiments better than DREX calculations [35].

The NRIA analysis we carried out shows that exchange processes must contribute to this reaction, at least for some observables. Our study is aimed at analyzing the energy dependence of polarization data, including ($P$–$A_y$), in order to achieve a better understanding of the nonlocal/exchange nature of the scattering process, in particular for the $1^+$, $T = 1$ excited state in $^{12}$C (see Fig. 5). In the case of the $1^+$, $T = 0$ state in $^{12}$C, we employ another analytical method (Fig. 3), based on isolating the observables $(P-A_y)\sigma$ and $(P+A_y)\sigma$. It clearly demonstrates the way to uncover significant nonlocality in $(\vec{p},p')$ scattering and, thus in the spin-dependent part of the effective $NN$ interaction.

NOTE ADDED BY THE AUTHOR OF THE REVIEW IN PROOF

*A *G*-matrix representation of the effective *NN* interaction, derived from free *NN* scattering data, is given in Ref. [18]. We obtained these data in a clear tabulated form with detailed explanations curtesy of H.V. von Geramb.

**A more accurate description of the value $R$ (0°), using the LEA program (as compared with DWBA 91), has been achieved possibly due to the following facts. Some distinctive characteristics of LEA appear to be inferior to certain properties of DWBA 91 (e.g. accuracy in the description of exchange processes), but they remain not so important for calculations of differential cross sections in scattering at 0°. Generally speaking, small angles (low $q$) are appropriate for direct processes (consequently, exchange processes are not important), and higher momenta $q$ are suitable for exchange processes. In the description of $R$ (0°) in LEA, the best quality connected with the transition density may become apparent in the region where the role of exchange is minimal.

***Indeed, the bulk of the experimental results, similar to those shown in Fig. 4, but for the $1^+$, $T = 1$ state in $^{12}$C, demonstrates different relations between the $D_{LS}$ and $D_{SL}$ observables. We were able to describe all these data rather satisfactorily (allowing for experimental uncertainties) within the LEA framework. Approximately the same result was achieved in paper [8], where calculations were made using three different models (nonrelativistic and relativistic).

# SECTION 3
# Polarization Transfer Coefficients and Partial Differential Cross Sections for $(\vec{p}, \vec{p}')$ $^{16}$O ($4^-$, $T = 0$ and $T = 1$) Reactions at 200–350 MeV

## Introduction to Procedures

As is known (see e.g. Ref. [1]), the properties of the nucleus are often modelled through the use of the effective nucleon–nucleon (NN) interaction. In their turn, the instruments of the polarization transfer in $(\vec{p}, \vec{p}')$ reactions can provide new information on the NN force inside nuclei [2]. Taken together, studies of various polarization transfer processes in the above reactions are a rich source of information on both the effective NN interaction and nuclear structure [2].

In such studies, a particularly good choice present high-spin stretched transitions with unnatural parity. As is well known (see e.g. Ref. [3]), it is common for such cases that only one particle-hole configuration significantly contributes to the transition. This means that the wave functions necessary for the analysis may be constrained by (e,e') form-factor measurements.

Of special interest is the fact that for $(\vec{p}, \vec{p}')$ reactions these transitions ensure such polarization observables that allow comparing their combinations with the size of particular spin-dependent pieces of the effective NN interaction. Moreover, such a comparison can be made separately in both isoscalar and isovector channels [1–3]. As we will show further on, in single-collision approximations, functions related to polarization transfer observables may display dependence on a single spin-dependent amplitude of the NN interaction and on a certain experimental form factor.

## Conventional Polarization Phenomena for Unnatural-Parity Stretched States

Traditionally, in inelastic proton scattering from nuclei, especially with beams of polarized particles (protons), various observables can be measured, including those for the differential cross section $\sigma$, induced polarization $P$, and analyzing power $A_y$.

When contributions from nuclear currents are absent, such simple relationships between the observables as $P \approx A_y$ are to be expected. They are often explained by the fact that in the nonrelativistic model, differences between $P$ and $A_y$ in the effective NN interaction are driven by nonlocal (knock-on exchange) terms. Examples of such measurements are shown in Fig. 1. As is seen in the figure, the experimental measurements, as well as those in the range of the maximum of the differential cross section for the high-spin stretched transition to the $4^-$, $T = 1$ state in $^{16}$O, demonstrate a relatively small difference between $P$ and $A_y$. Here the transition form factor is calculated, assuming that the single particle-hole configuration, ($p_{3/2}^{-1} d_{5/2}$), is dominant [1–3], which follows in particular from pion-nucleus scattering [8]. However, it is obvious from Fig. 1 that the physical information extracted here is rather limited and, therefore, incomplete.

It is evident that three types of polarized proton beams should be used in full experiments: (1) a beam polarized in the direction perpendicular to the scattering plane (type $N$), (2) a beam polarized along the direction of motion of the accelerated particles (type $L$), and (3) a beam polarized in the scattering plane, but in a direction orthogonal to the axis of motion of the particles (type $S$) [1–4, 9–12].

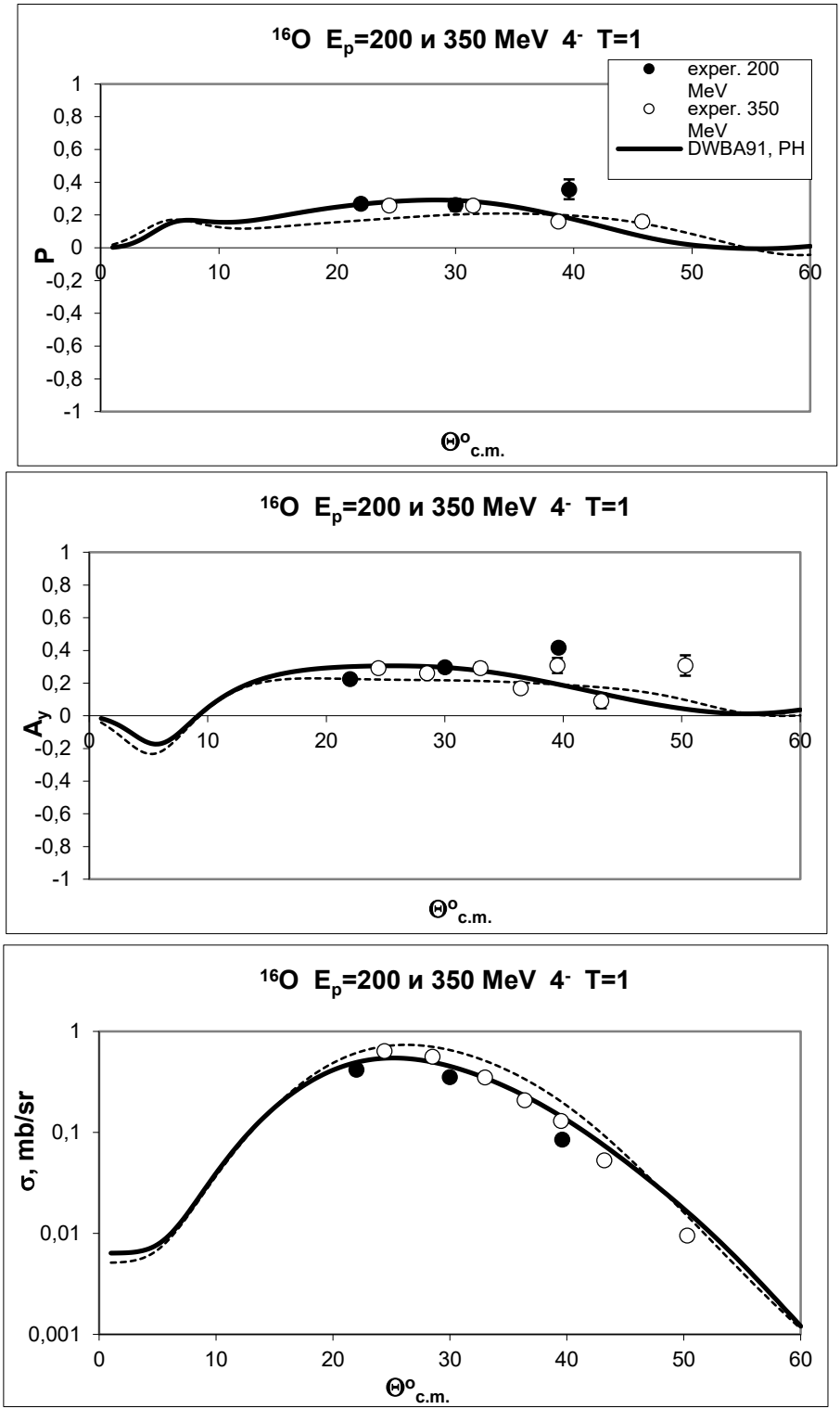

**Fig. 1**. A comparison of our calculated results at $E_p = 200$ MeV (curves) with the systematized experimental data at $E_p = 200$ MeV (solid points) from [1–3], and at $E_p = 350$ MeV (open points) from [4]. In order to compare the results at two energies, the values of the angles at $E_p = 350$ MeV were multiplied by the coefficient $k = (350/200)^{1/2} = 1.32$. We made calculations using the program DWBA 91 [5] and employed the density-dependent Geramb interaction [6], constructed by the Hamburg group with the use of the Paris potential, often labeled as the PH interaction (solid curves). Also shown (dashed curves) are some other calculated data we obtained using the effective interaction of Nakayama and Love (NL), based on a one-boson exchange model [7]. Systematized, measured and calculated observables for the $4^-$, $T = 1$ state at 18.98 MeV in $^{16}$O are represented.

The quantities $P$ and $A_y$ and the off-diagonal polarization transfer coefficients $D_{ij}$ (the latter were measured only in a full experiment) are sensitive to the interference between amplitudes. At the same time, $D_{ii}$ diagonal polarization transfer coefficients obtained from the full experiment are sensitive only to the absolute square of amplitudes [10]. The latter observables have been described more or less successfully in various models. However, data of an interference nature very often pose serious problems for description, which is not totally surprising [10].

As an example, shown in Fig. 2, let us examine the diagonal coefficients $D_{ii}$ for the same excitation as the one presented in Fig. 1.

**Forward-Angle Values of $D_{ii}$ Polarization Transfer coefficients and Their Combinations for the $^{16}O\ (\vec{p}, \vec{p}')\ ^{16}O\ (4^-, T = 1)$ Reaction, and some Common Properties of $D_{ii}$ for Stretched Levels**

Unnatural-parity transitions at extremely forward angles (at and near zero degrees) are characterized by the fact that the $D_{NN}$ value is practically equal to the $D_{SS}$ value. This may be due to the circumstance that in this case the $\hat{N}$ direction is basically identical to the $\hat{S}$ direction [13], owing to the symmetry around the scattering axis. Our calculations (Fig. 2) at $\theta_{c.m.} = 1°$ with the program DWBA 91 from Raynal with the Geramb $DD$ forces (PH[‡], solid curves) and the Nakayama–Love no$DD$ interaction (NL[§], dashed curves) confirm the above-mentioned fact for the stretched isovector $4^-$, $T = 1$ (19.98 MeV) transition in $^{16}O$. Indeed, in the case of the PH force, the calculations result in $D_{NN} = -0.73$, and $D_{SS} = -0.74$ for the indicated minimal scattering angle $\theta$ c.m. = 1°. As is seen from the figure, the ratio close to $D_{NN} = D_{SS}$ is obtained at extremely small angles and when the $NL$ interaction is used. All the calculated data can be considered to be a certain control test for the computer program DWBA 91.

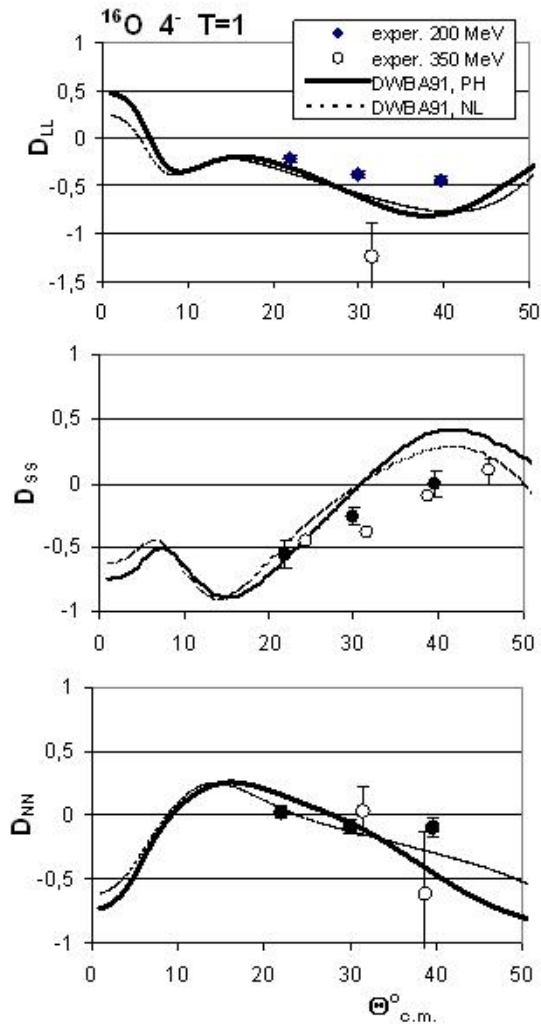

**Fig. 2.** Our calculations (curves) and systematized experimental data (points) are shown. Measurements at $E_p$ = 200 MeV (solid points) are taken from [3] – $D_{LL}$, $D_{SS}$, and from [2] – $D_{NN}$. Measurements at 350 MeV (open pints) are from [4]. Angles for 350 MeV were multiplied by the coefficient $k = (350/200)^{½} = 1.32$. All the calculations made at $E_p$ = 200 MeV are described in the text.

---

[‡] PH is the Hamburg $G$-matrix interaction based on the Paris potential (PH or GPH from H.V. von Geramb) [6].
[§] NL is the Nakayama–Love $G$-matrix based on the Bonn potential [7] (the tables were provided by W.G. Love and given to us courtesy of Prof. P. Schwandt).

The quantity $\Sigma$, as a linear combination of the polarization transfer coefficient $D_{ii}$ (called a total spin transfer in [13]), i.e. $\Sigma = [3 - (D_{NN} + D_{SS} + D_{LL})] / 4$, is equal to 1 for spin-flip ($\Delta S=1$) transitions, and to 0 for non-spin-flip ($\Delta S = 0$) transitions if the spin-orbit interaction is negligible. This may occur in the $(\vec{p}, \vec{p}')$ process at $\theta \approx 0°$. In our calculations at $\theta = 1°$, $\Sigma$ was equal to 1.00 for both PH and NL forces.

The relation
$$D_{NN}(0°) = \pm [1 + D_{LL}(0°)] / 2 \qquad (1)$$

is also well known [13]. The plus sign in it refers to natural-parity transitions, and the minus sign refers to unnatural-parity ones. In our calculations at $\theta = 1°$, this relation in a digital representation is as follows: $-0.730 \approx -0.734$ for the PH force, and $-0.614 \approx -0.618$ for the *NL* interaction.

Thus, all the calculated combinations of the polarization transfer coefficients $D_{ii}$ at and near zero degrees are in good agreement with the corresponding theoretical relations [13]. Moreover, the calculations made using DWBA 91 provide a satisfactory description of the experimental measurements $D_{ii}$ (Fig. 2), obtained in the region of maximal differential cross sections (see Fig.1).

For comparison we also performed a similar study of the $T = 1$ stretched $6^-$ state at 14.35 MeV in $^{28}$Si, using polarization transfer coefficients from $(\vec{p}, \vec{p}')$ measurements at 200 MeV [3] and 500 MeV [10]. Our analysis, using the program DWBA 91 and PH forces (not shown), revealed that $D_{NN}(0°) = D_{SS}(0°) = -0.52$. The quantity $\Sigma$ appears to be practically equal to 1 (0.98), and Equation (1), in a digital representation, gives the following: $-0.521 = -0.521$.

The main qualitative features of the measured and calculated polarization transfer coefficients for the $6^-$, $T = 1$ excitation (not shown) and those of the corresponding data for the $4^-$, $T = 1$ excitation (Fig. 2) in the region of maximal differential cross sections are principally of a similar character. However, the discrepancy between the calculated and experimental data is somewhat more noticeable for $D_{NN}$. Nevertheless, overall this is an important guide for unnatural-parity stretched transitions.

Therefore, we have confirmed the assumption [10, 14] that $D_{SS}$, $D_{LL}$ and $D_{NN}$ should resemble each other for all isovector stretched states, since the characteristics of $D_{ii}$ depend primarily on the isovector stretched-state assumption and the sampled properties of the force. Thus, for pure stretched states of high spin, the qualitative shapes of $D_{ii}$ should be almost independent of the nucleus and similar over a wide range of energies. Lastly, we would like to emphasize that, since $D_{SS}$, $D_{LL}$ and $D_{NN}$ are very insensitive to the type of distortion used (as they do not depend on the nucleus), all these common characteristics should become most apparent for scattering at and near $\theta_{c.m.} = 0°$ in the excitation of all $T = 1$ stretched states.

**Partial Differential Cross Sections for $(\vec{p}, \vec{p}')$ Transitions**

Subsequently, as we can see in Fig. 2, $D_{ij}$ polarization transfer coefficients, often labeled as spin transfer coefficients, were measured in a $(\vec{p}, \vec{p}')$ experiment. They are usually treated as components of the proton depolarization tensor, or simply depolarization parameters [15]. All of them are identical to the corresponding Wolfenstein parameters, i.e., triple $(p, p')$ scattering parameters [16]. Although these spin observables were established from experiments in a direct way, they are not related to the collision matrix of inelastic proton scattering in a transparent way [15]. Consequently, a different (alternative) parameterization of the polarization data was proposed in [15]. In was shown in works [15, 17, 12] that a complete set of $D_{ij}$ coefficients for a specific isolated state yielded information on individual components of the *NN* scattering amplitude.

Therefore, instead of $D_{ij}$ parameters were used their linear combinations $D_i$ [12, 15]. They appeared to be sensitive to individual terms in the *NN* interaction. Consequently, the spin-observable combinations $D_i$, when multiplied by the differential cross section $\sigma$, yielded the following quantities:

$$\sigma_i = \sigma D_i . \qquad (2)$$

They can be treated as partial differential cross sections or, according to the method of obtaining them, as the corresponding polarized cross sections.

In a plane wave approximation, such partial cross sections $\sigma_i$ (2) appear to be equal to the squares of products of individual amplitudes in the *NN* effective interaction and certain transition form factors. The form factor can be constrained, for example, by electron scattering data. In a formalized representation, Equation (2) can be expanded to the following ratios [2, 3, 10–12, 15]:

$$\begin{aligned}
\sigma_{ls} &\equiv \sigma D_{ls} = C^2 \chi_T^2, \\
\sigma_q &\equiv \sigma D_q = E^2 \chi_L^2, \\
\sigma_n &\equiv \sigma D_n = B^2 \chi_T^2, \\
\sigma_p &\equiv \sigma D_p = F^2 \chi_T^2.
\end{aligned} \qquad (3)$$

Thus, we observe that in a plane-wave approximation to inelastic scattering, each partial differential cross section is determined by the product of a square form factor (such as spin transverse $\chi_T$ or spin longitudinal $\chi_L$) and a square of one *NN* amplitude (*C*, *B*, *F* or *E*). Consequently, measurements and calculations of the partial cross section can be represented as spin-orbit ($\sigma_{ls}$), spin-longitudinal ($\sigma_q$), and spin-transverse ($\sigma_n$ and $\sigma_p$) components. The spin-observable combinations $D_i$ were established as follows [2, 3, 10–12, 15]:

$$\begin{aligned}
D_{ls} &= [1 + D_{NN} + (D_{SS} + D_{LL}) \cos\theta - (D_{LS} - D_{SL}) \sin\theta] / 4 \\
D_q &= [1 - D_{NN} + D_{SS} - D_{LL}] / 4 \\
D_n &= [1 + D_{NN} - (D_{SS} + D_{LL}) \cos\theta + (D_{LS} - D_{SL}) \sin\theta] / 4 \\
D_p &= [1 - D_{NN} - D_{SS} + D_{LL}] / 4
\end{aligned} \qquad (4)$$

Here $\theta$ is a center-of-mass scattering angle.

The expressions of $\sigma_i$ in a PWIA approximation (3) are applicable to unnatural parity states [10, 15]. The relative values of $\sigma_i$ point directly to the magnitudes of the *C*, *B*, and *F* terms in the *NN* interaction, since the same factor (the nuclear transfer form factor $\chi_T$) is included into each of them. As for $\sigma_q$, it is also directly comparable with the other $\sigma_i$ components, although a different, namely the longitudinal form factor $\chi_L$ enters into it. It is it due to the fact that here we deal with stretched states, and for them $\chi_T$ and $\chi_L$ are linked by a simple linear relationship [14].

Subsequently, the combinations of polarization transfer coefficients (4) were transformed into partial differential cross sections (3). The latter may reveal their selective dependence upon individual *NN* amplitudes (mostly spin-dependent), e.g., in KMT (Kerman, McManus, Thaler) [18], and some other representations, as is shown in Ref. [3]. Due to such a dependence, these partial cross sections $\sigma_i$ can be used, according to the authors of work [3], as an important diagnostic method. For the sake of clarity, let us remind the well-known KMT form of the *NN* interaction (*NN* scattering amplitude in the PWIA and in a nonrelativistic framework):

$$M(q) = A + B\, \sigma_{1n}\sigma_{2n} + C\,(\sigma_{1n} + \sigma_{2n}) + E\sigma_{1q}\sigma_{2q} + F\sigma_{1p}\sigma_{2p}. \qquad (5)$$

Here $\vec{\sigma}$ (or simply $\sigma$ with the appropriate index) is the Pauli spin operator for particle 1 (e.g., for the target nucleon), or 2 (the projectile nucleon), acting along the direction $\hat{n}$ (normal to the scattering plane), or along the direction $\hat{q}$ (the direction of the momentum transfer), and also along the direction $\hat{p}$, where $\hat{p} = \hat{q} \times \hat{n}$. The following ratio $(\sigma_{1n} + \sigma_{2n}) \equiv (\vec{\sigma}_1 + \vec{\sigma}_2) \cdot \hat{n}$ can serve as one of the additional clarifying examples.

Thus, provided the necessary conditions of the experiment ensure the required measurements of the relevant observables and form combinations of them (4), any differential cross sections for any state can be rearranged (2) into four separate components (partial cross sections). Then, according to Expression (3), such partial cross sections can be linked to separate amplitudes of a particular effective *NN* interaction.

### Three $4^-$ ($T = 0$ and $T = 1$) States in $^{16}$O, and $(\vec{p}, \vec{p}')$ Partial Differential Cross Sections

In the present review, we mainly use the well tested (primarily for the excitation of normal parity states) Paris density-dependent interaction, also known as PH *G*-matrix interaction [6], based on the Paris potential. One of our purposes was to compare our analysis with DBHF (Dirac–Brueckner Hartree–Fock) calculations and rather simplified BHF calculations with the effects from Brueckner theory only [3], performed for the same excitations $4^-$ with $T = 0$ and $T = 1$ in the $^{16}$O nucleus. It is worth noting that the difference between DBHF and BHF calculations is that the DBHF calculations added the effect of a strong relativistic mean field in the Dirac–Brueckner approach to nuclear matter. However, the description of partial (polarized) cross sections revealed that all the calculations (both BHF and DBHF) were similar [1–3].

The overall picture is that the comparison of our calculations with the PH interaction and DBHF calculations yielded almost the same results for all the values of $\sigma_i$ in the isovector channel under consideration. Herewith, the description of the experiment was good, as we will show it further on.

Moreover, in the isoscalar channel (two excitations of $4^-$ with $T = 0$ in $^{16}$O), our calculations made using the PH interaction ensured a good description of the maximal values of $\sigma_{ls}$ (see Figs. 3 and 4 below). The values of $\sigma_{ls}$ are sensitive to the response through the spin-orbit operator ($\sigma_{1n} + \sigma_{2n}$), what Expressions (3) and (5) reflect. The spin-orbit term in the isoscalar channel is dominant. In full compliance with that the partial cross section $\sigma_{1n}$, driven, as it was already noted, primarily by the strength of the spin-orbit amplitude, significantly surpasses all the other partial cross sections in both experiments and calculations (Figs. 3, 4). Moreover, our PH calculations are to some extent better than the DBHF calculations [2, 3]. We achieved a significantly better description for the $4^-$, $T = 0$ (19.81 MeV) level (Fig. 4). Thus, in comparison with [2], the improvement we achieved primarily referred to the predicted magnitude of $\sigma_{ls}$. Moreover, in our case the visible push of the calculated pick in $\sigma_{ls}$ out to larger angles was almost absent; however, it was rather noticeable in work [2], which was pointed out by its authors.

Inversely, for the isovector transition, the spin-orbit part of the effective interaction is small, as compared with the case in the isoscalar channel. It is evident from the small values of the observables $A_y$ and $P$ (see Fig. 1 and comments in Ref. [3]). Accordingly, the measured spin-orbit response $\sigma_{ls}$ is small for the $4^-$, $T = 1$ transition in Fig. 5 (the values of $\sigma_{ls}$ are the smallest, as compared with the other partial cross sections). It complies with our DWIA calculations made using the PH *G* matrix, also shown there. It should be stressed that despite the spin-orbit term $\sigma_{ls}$ being too small in this case, it can be described quite well, using the PH interaction. Moreover, the description is somewhat better than with the use of the DBHF interaction [2, 3].

While both PH and DBHF calculations generally describe $\sigma_{ls}$ data for isovector $4^-$ states satisfactorily, PH curves reflect the experiment much better than DBHF curves in the case of isoscalar $4^-$ states, especially for the isoscalar $4^-$, 19.81 MeV state in $^{16}$O. Possible reasons for such a defect in the case of DBHF calculations were analyzed in Ref. [2]. At the same time, in transitions where the isoscalar spin-orbit component is a dominant contribution (for two $4^-$, $T = 0$ transitions in $^{16}$O), the difference between experimental and calculated data is larger and more systematic for the normal spin-transverse cross section $\sigma_n$ in both DBHF and especially PH calculations (see Figs.

3, 4 and Refs. [2, 3]). In all these cases, the predictions for the value of $\sigma_n$ are too small. Especially dramatic is the discrepancy between calculations and experiments for the PH DD description in the case of the $4^-$, $T = 0$, $E_p = 19.81$ MeV (Fig. 4 in this section). However, this problem is almost absent for the $T = 1$ transition at any effective interaction analyzed here (Fig. 5 and Refs. [2, 3]).

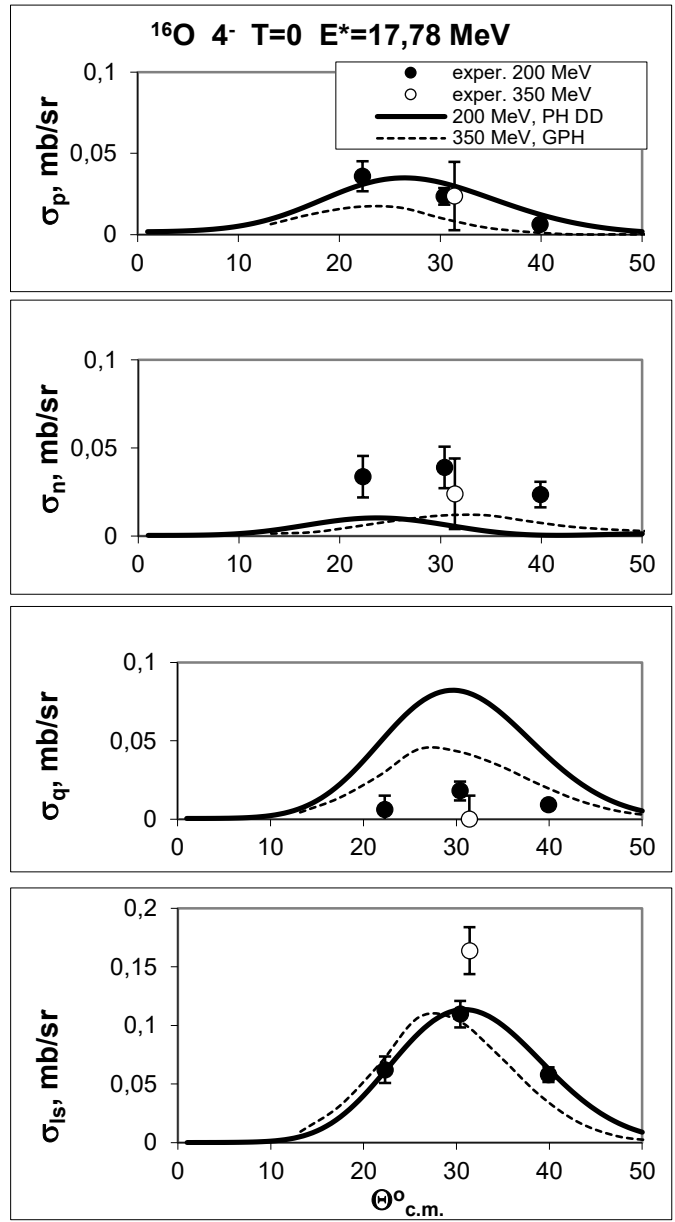

**Fig. 3.** The description within the framework of our PH *DD* calculations at 200 MeV (solid curves) of the experimental partial cross sections $\sigma_i$, measured at the same values of $E_p$ [2] (solid points). For obtaining $\sigma_i$, the $D_{NN}$ results [2] were combined with the previously measured in-plane spin-transfer coefficients [19]. The open points are similar experimental results, based on the corresponding measurements of $D_{ij}$ and $\sigma$ at $E_p = 350$ MeV [4]. Represented in the same way are PH (or GPH) calculations for partial cross sections (dashed lines), based on the previous calculations of $D_{ij}$ and $\sigma$ made at $E_p = 350$ MeV in [4]. The systematized angular distributions of the observables refer to the $4^-$, $T = 0$ (17.78 MeV) level in $^{16}$O.

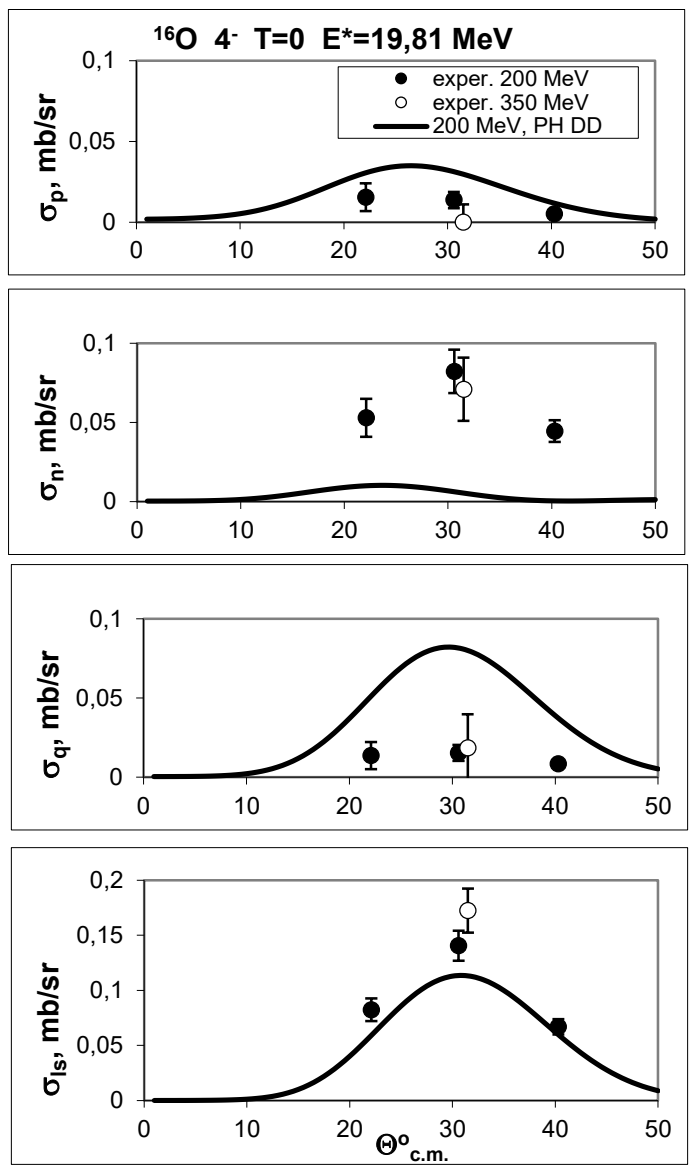

**Fig. 4.** The description within the framework of our PH *DD* calculations at $E_p = 200$ MeV (curves) of the experimental partial differential cross sections $\sigma_i$, measured at the same $E_p$ [2] (solid points). To obtain $\sigma_i$, the $D_{NN}$ results [2] were combined with the previously measured in-plane spin-transfer coefficients [19]. The open points are similar experimental results based on the corresponding measurements of $D_{ij}$ and $\sigma$ at $E_p = 350$ MeV [4]. The systematized angular distributions of the observables refer to the $4^-$, $T = 0$ (19.81 MeV) level in $^{16}$O.

On the one hand, at $T = 0$, $\sigma_n$ is underpredicted for all the interactions used (Figs. 3, 4 and Refs. [2, 3]), and on the other hand, also at $T = 0$, the spin-longitudinal component $\sigma_q$ appears to be overpredicted, mainly in PH calculations (Figs. 3, 4). It should be noted that the latter defects are characteristic of isoscalar channels. It is practically absent in the isovector channel, for both PH and DBHF calculations (Fig. 5 and [2, 3]). The description of the in-plane spin-transfer response $\sigma_p$, is a very important issue. This response is usually associated with tensor exchange amplitudes. In DBHF calculations and in our analysis, the PH calculations can be considered quite acceptable for all the three $4^-$ stretched states in $^{16}$O (Figs. 3–5).

The authors who performed the DBHF calculations in [3] believe that for transitions in which the isoscalar spin-orbit component is a dominant contribution (i.e., for $\Delta T = 0$) there is a clear signature for an increase in the tensor interaction associated with $\sigma_n$. Judging by the results represented in Figs. 3 and 4, the same problem concerns PH interactions, even to a larger extent.

In general, the present review confirms the conclusions made in Ref. [3] that the balance between spin-orbit and tensor components of effective *NN* interactions needs further modification, especially in the isoscalar channel.

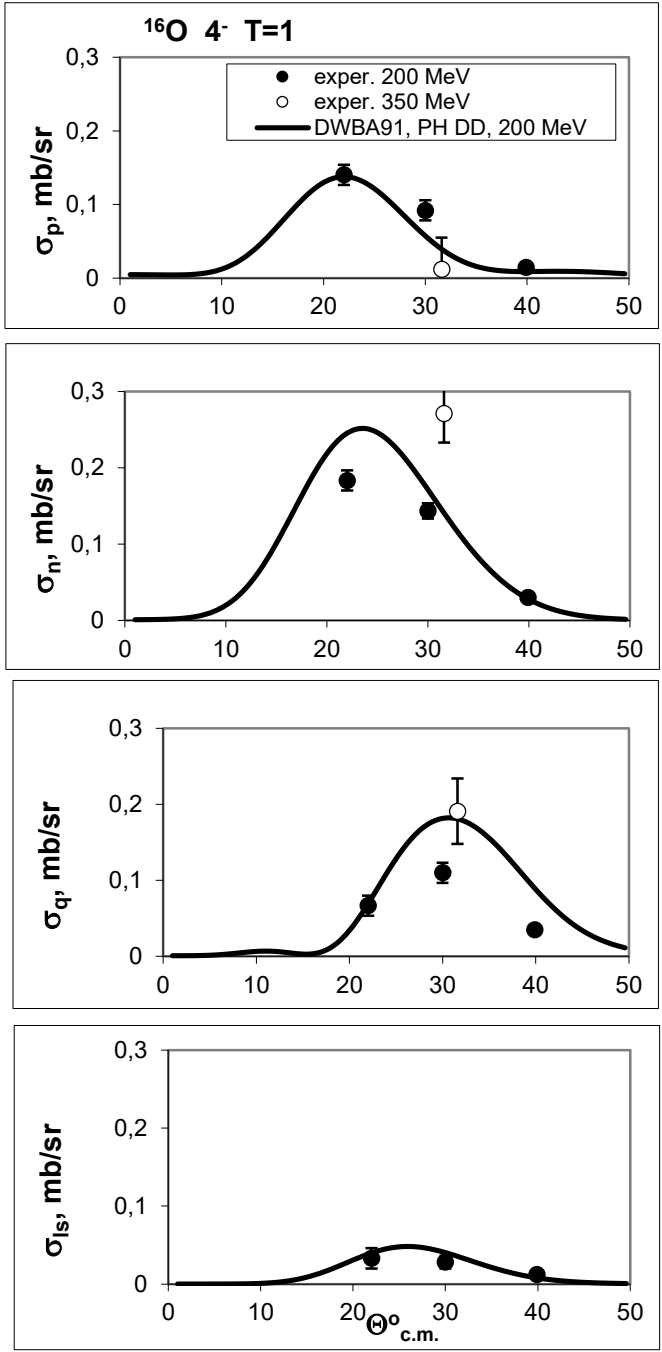

**Fig. 5.** The description within the framework of our PH DD calculations at $E_p = 200$ MeV (curves) of the experimental partial differential cross sections $\sigma_i$, measured at the same $E_p$ [2] (solid points). To obtain $\sigma_i$, the $D_{NN}$ results [2] were combined with previously measured in-plane spin-transfer coefficients [19]. The open points are similar experimental results based on the corresponding measurements of $D_{ij}$ and $\sigma$ at $E_p = 350$ MeV [4]. The systematized angular distributions of the observables refer to the $4^-$, $T = 1$ (18.98 MeV) state in $^{16}$O.

## Some Conclusions

The most important conclusion we arrived at in the present sections is that the method, developed here as a continuation of studies [1–3] of decomposing the full cross section into four partial differential cross sections, leads to a definite reexamination of the well-established effective *NN* interactions.

The use of the PH interaction here, as compared with the application of the DBHF interaction in Refs. [1–3], clearly reveals a common problem that existed even in [1–3] and is especially evident here in the analysis of individual terms in the effective PH interaction.

Consequently, the analysis of the results represented in the isoscalar channel (Figs. 3 and 4) unambiguously indicates the need for a smaller spin-longitudinal amplitude at the $\sigma_{1q}\sigma_{2q}$ spin-operator (i.e., $E$), and a larger transverse amplitude at the $\sigma_{1n}\sigma_{2n}$ spin-operator (i.e., $B$). As is clear from the present review, the supposition made in [1] for DBHF, and fully applicable for the PH interaction, suggests too much pion-like attraction and too little repulsion in particular parts of the tensor interaction.

It is especially worth noting that the spin-transfer response $\sigma_p$, which is usually associated with complex (tensor) knock-on exchange contributions, proved to be well predicted at both DBHF and PH interactions.

In some cases, the problems that arise for the isoscalar parts of the effective interaction are generally more significant than those for the isovector effective interaction. At the same time, it should be pointed out that the number of nuclear excitations, analyzed with the applied method, is still small, and the available studies are rather inconsistent. Therefore, the fact that the predictions for the $\sigma_n$ value in the isoscalar channel are too weak (see Fig. 4) for both DBHF and PH calculations may simply reflect, as suggested in Ref. [2], the uncertainty in the magnitude of the transition form factor.

An important achievement of the present review is establishing the overall ability of the PH interaction (Fig. 5), as it was in the case of the DBHF interaction [2, 3], to reproduce remarkably well all essential features of the isovector amplitude, and thus to describe complete data sets of partial differential cross sections for the $^{16}$O $(\vec{p}, \vec{p}')$ $^{16}$O $(4^-, T = 1)$ reaction at 200–350 MeV.*

Furthermore, all the discussed examples of partial (polarized) cross sections confirm that they can present a remarkable diagnostic instrument for examining effective *NN* interactions of different types and thus provide some important guidance for future studies.

### * NOTE ADDED BY THE AUTHOR OF THE REVIEW IN PROOF

In addition to the comparison of partial cross sections for high-spin stretched transitions of both $T = 0$ and $T = 1$ characters, we also compared them in the case of the excitation of the $6^-$, $T = 0$ (11.58) and $6^-$, $T = 1$ (14.35 MeV) stretched states in $^{28}$Si. The distinctive characteristic of our analysis was that we were the first to form experimental partial cross sections (based on measurements from [3]) and then to compare them with DWIA calculations for the $6^-$, $T = 0$ state in $^{28}$Si. As a result, we managed to contrast polarized cross sections for $6^-$, $T = 0$ and $6^-$, $T = 1$ states in $^{28}$Si, similar to the procedure performed for $4^-$, $T = 0$ and $4^-$, $T = 1$ states in $^{16}$O (Figs. 3–5). The calculations for the $^{28}$Si were made almost in the same manner as in the case of $^{16}$O.

Thus, with $T = 0$ and $T = 1$ in $^{28}$Si (similar to $^{16}$O), $\sigma_{ls}$ is more dominant in the region of the maximum of the cross section for $T = 0$, as compared with $T = 1$. It is due to the well-known fact that the isovector spin-orbit interaction is weak (by contrast with the isoscalar interaction) at intermediate energies with any set of nuclear forces. At the same time the differences between the data and calculations of $\sigma_{ls}$ (for the $T = 0$ and $T = 1$ transitions) are larger for $^{28}$Si than for $^{16}$O.

It is interesting to note that all the positive and negative features in the description of $\sigma_i$ (=$\sigma_K$) for the $6^-$, $T = 1$ state in $^{28}$Si with the use of the DBHF effective interaction [3] are almost accurately reproduced in our calculations made using other (well-known and often employed) nonrelativistic nuclear forces.

On the other hand, an exceptionally significant underrating of the $\sigma_n$ value in the description of the $4^-$, $T = 0$ (19.81 MeV) state in $^{16}$O (Fig. 4) is fully present in our analysis of the $\sigma_n$ for the $6^-$, $T = 0$ state in $^{28}$Si. This testifies, in particular, of the existence of certain systematic problems regarding stretched transitions (especially in the case of $T = 0$).

# SECTION 4
# Unnatural-Parity $(\vec{p}, \vec{p}')$ Reactions in a Factorized Impulse-Approximation Model for the Polarization Transfer (PT) and Spin Responses[*]

## A Schematic Formalism

This section is aimed at further study of the parameterization of complete spin-transfer measurements for inelastic polarized-proton scattering on the basis of $D_K$ polarization observables, introduced by Bleszynski et al. [1] and expressed in terms of conventional Wolfenstein parameters [2].

Complete measurements of $D_{ij}$ spin-transfer coefficients require beams of protons with initial polarization, longitudinal ($L$), normal to the reaction plane ($N$), and sideways ($S = N \times L$). The spin (polarization)-transfer observable $D_{ij}$ relates to the component $i$ of the incident (initial) proton polarization, and to the component $j$ of the outgoing (final) proton polarization [3, 4]. The components $D_{ij}$ (alternate parameters suggested by Wolfenstein [2]) are experimental measurements.

Let us remind that in theory, according to Refs. [1, 2], the spin-transfer coefficients $D_{ij}$ can be defined in terms of the *NA* (nucleon–nucleus) scattering amplitude $\bar{M}(q)$ and nucleon spins:

$$D_{ij} = Tr\ [\bar{M}\sigma_i\ \bar{M}^\dagger \sigma_j]/2I_0. \qquad (1)$$

Here $\sigma_i$ and $\sigma_j$ are the Pauli components of the proton (spin) matrix with respect to the $i$ ($L, N, S$) and $j$ ($L', N', S'$) axes. The prime on the second (outgoing) subscript is normally omitted. $I_0$ is the usual unpolarized differential cross section.

The *NA* scattering amplitude $\bar{M}(q)$ or, in other words, the collision operator for the $(\vec{p}, \vec{p}')$ transition in the PWIA can be expressed in terms of the free *NN* scattering amplitude operator in the nonrelativistic framework $M(q)$ [1–6].

The use of the polarization-transfer observables $D_{ij}$ in (1), and especially of the combinations of these observables $D_K$, obtained, according to Bleszynski et al [1], in inelastic proton scattering, allows to employ this scattering (see Ref. [7]) as a filter to examine particular pieces of the effective *NN* scattering.

Model calculations for $\bar{M}(q)$ are represented in detail in Ref. [5]. It is worth noting that in the case of spin-flip transitions at the angle of 0°, the *NN* scattering amplitude in a PWIA approximation becomes significantly simplified [6].

The authors of paper [1] believed that the Wolfenstein parameters [2] were not related in a transparent way to the $(\vec{p}, \vec{p}')$ collision matrix. Bleszynski et al. [1] introduced a particular set of observables $D_K$ that were expressed in terms of linear combinations of $D_{ij}$ parameters. Each of these observables $D_K$ (where $K = ls, q, n, p$), according to the authors of work [1], depend separately on the specific tensor components of the $(\vec{p}, \vec{p}')$ collision matrix. Following these model representations, a different parameterization should be also applied to experimental results, alternative to the conventional Wolfenstein parameters.

Consequently, the $D_K$ observables can be defined in terms of components of the general collision matrix and also in those of experimental $D_{ij}$ parameters. The authors of [1] managed to obtain, in a single collision approximation, expressions for the observables $D_K$ that displayed dependence of an individual $D_K$ on particular components of the *NN* amplitude and on particular nuclear form factors.

Such an approach to the case of nucleon-nucleus scattering was highly appraised in Refs. [3, 4, 15, 16, etc.], since it demonstrated that spin observables could be used to isolate specific parts of the nuclear response. In [15, 16], especially highly rated was the fact that this approach opened up the possibility to separate the transverse- and longitudinal-spin-response functions of the nuclei by measuring spin (polarization)-transfer observables using polarized beams.

**Parameterization of Complete Spin-Transfer Measurements in ($\vec{p}, \vec{p}'$) Scattering for Unnatural-Parity States in $^{12}$C, $^{16}$O, and in $^{28}$Si Nuclei**

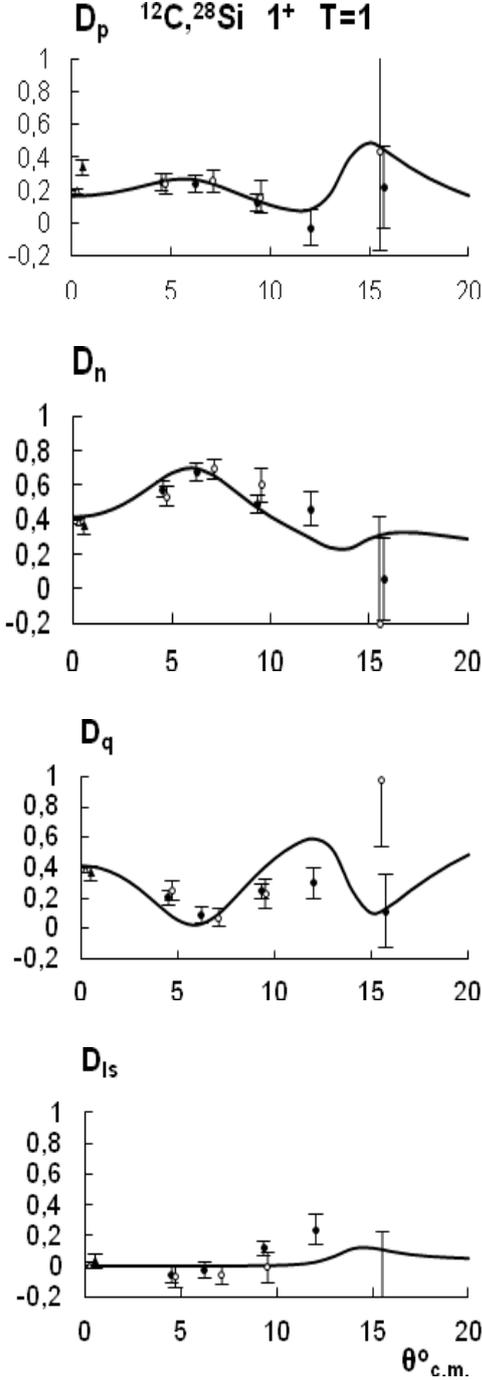

As noted above, four polarization observables $D_K$ (in previous Sec. 3 designated as $D_i$), introduced by Bleszynski et al (see Ref. [1] in this section), can be expressed in terms of linear combinations of the Wolfenstein parameters [2], or analogous complete PT coefficients ($D_{ij}$) for the ($\vec{p}, \vec{p}'$) reaction. As we have already noted, in the PWIA with an optimal factorization approximation, the combinations $D_K$ are given as:

$$D_{ls} = [1 + D_{NN} + D_{SS} + D_{LL}\ \cos\theta - \delta\sin\theta]/4, \text{ etc.} \quad (2)$$

where $\theta \equiv \theta_{c.m.}$ (deg.) is a scattering angle, and $\delta = D_{LS} - D_{SL}$.

Expressions (2), in their full form, are given in Equations (4) in Sec. 3.

**Fig. 1.** Systematized angular distributions of the spin-transfer probabilities $D_K$ for the isovector $1^+$ (15.11 MeV) level in $^{12}$C, and the isovector $1^+$ (11.5 MeV) level in $^{28}$Si based on measured ($\vec{p}, \vec{p}'$) quantities at 400–500 MeV. At 500 MeV filled circles [3], and open circles [4] – $^{12}$C; and at ~400 MeV [6] open triangles – $^{12}$C, filled triangles – $^{28}$Si. Our calculations (curves) are described in the text. At 0° the corresponding values $D_K$ for $^{12}$C (open triangles) and $^{28}$Si (filled triangles) overlap and become visually almost undistinguishable in the figure.

$D_{ls}$ is associated with the spin-orbit term in the *NN* effective interaction. The other three $D_K$ are associated with the tensor terms for each axis as $D_q$ with the momentum transfer $\vec{q} = \vec{k}_{in} - \vec{k}_{out}$, $D_n$ – with normal $\vec{n}$ to the reaction plane, and $D_p$ – with $\hat{p} = \hat{n} \times \hat{q}$.

As an example, Fig. 1 shows experimental and calculated dependences for all four $D_K$ combinations for the $1^+$, $T = 1$ level in $^{12}$C and $^{28}$Si.

The novelty of the present review is that the range of the measured spin-observable combinations $D_K$ has been extended to 0°, which allowed for the first time to evaluate the validity of analytical results at extremely forward angles (at and near zero degrees). Besides, the measured $D_K$ for the $1^+$, $T = 1$ state in $^{12}$C at 0° were completed by similar data for the $1^+$, $T = 1$ level in $^{28}$Si (Fig. 1).

The above-mentioned combinations $D_K$ become simpler at small scattering angles since $\delta \approx 0$, $\cos\theta \approx 1$ and $\sin\theta \approx 0$. According to Ref. [6], in the case of 0° for the isovector $(\Delta T = 1)$ $M1$ transition we have $D_{SS} = D_{NN} \approx D_{LL} \approx -1/3$. If these quantities are inserted into the simplified expressions $D_K$, we easily get the following set: $D_{ls} = 0$, and $D_q = D_n \approx D_p \approx 0.3$. Furthermore, these quantities appear to be approximately equal to the measured values and to our DWIA calculations using the FL (Franey and Love) interaction and the program LEA from Kelly. This is quite similar to what is shown in Fig. 1.

As for the description of $D_K$ data at other angles, the program LEA, used here for the first time, appeared to be more effective overall than some other programs, previously employed for such purposes [3], which proceeded from nonrelativistic and relativistic calculations.

As is seen in Fig. 1 for $T = 1$, the angular distributions of the transverse- and longitudinal spin-transfer probabilities, $D_n$ and $D_q$, respectively, are considerably different in shape, as they are in the case of a lower energy ($E_p = 200$ MeV), which we will discuss further on. This phenomenon can be explained by a significant difference in the momentum dependence of transverse- and longitudinal axial form factors.

At the same time the angular distributions $D_n$ and $D_q$ in calculations and experiments, have similar smooth shapes at both $E_p = 400$–500 MeV and 200 MeV for the isoscalar $1^+$ (12.71 MeV) state in the $^{12}$C$(\vec{p}, \vec{p}')$ $^{12}$C reaction [8]. This is in good agreement with the fact that the moduli of the isoscalar interaction components of the free nucleon–nucleon $t$-matrix for 140-Mev (and 800-Mev) nucleons from Franey and Love, responsible for transverse and longitudinal transitions, respectively, are similar in value and are rather flat as functions of the quantity $q$ [5].

**Particular Combinations of $D_K$, and Specific Amplitudes of the Effective Interaction**

According to the model of Bleszynski et al. [1], certain combinations of $D_{ij}$ should demonstrate a particular selectivity to specific components of the $(\vec{p}, \vec{p}')$ collision matrix. Basically, it comes to the fact that the four functions $D_K$ (2) appear to be sensitive primarily to individual terms in the $NN$ interaction. In this case, the $NN$ interaction is normally the $t$ matrix in the KMT (Kerman–MacManus–Thaler) representation [9].

The three polarization observables $D_K$ are predominantly sensitive to the response of the nucleus through tensor operators that act along three axes: $\sigma_{1q}\sigma_{2q}$, $\sigma_{1n}\sigma_{2n}$, and $\sigma_{1p}\sigma_{2p}$. Here, as was indicated above, $\hat{n}$ is normal to the scattering plane, $\hat{q}$ lies along the momentum transfer to the projectile, and $\hat{p} = \hat{n} \times \hat{q}$. The observable $D_{ls}$ is sensitive to the response through the spin-orbit operator $(\sigma_{1n} + \sigma_{2n})$. Then are used all four spin-dependent KMT amplitudes: $E$, $B$, $F$, and $C$ for $D_q$, $D_n$, $D_p$, and $D_{ls}$, respectively (see e.g. Ref. [10])[**]. The amplitudes $E$, etc., are complex and have isospin dependence, with the isospin index (0 or 1) being omitted.

It is worth reminding (see Refs. [1] and [4]) that particular combinations of $D_K$ (2) in simplified assumptions can be related to specific amplitudes of the effective interaction and nuclear-structure-dependent components via

$$D_{ls} \approx |C|^2 \, \chi_T^{\,2} / I_0 \qquad D_n \approx |B|^2 \, \chi_T^{\,2} / I_0$$
$$D_q \approx |E|^2 \, \chi_L^{\,2} / I_0 \qquad D_p \approx |F|^2 \, \chi_T^{\,2} / I_0 \,. \tag{3}$$

(The above is just a different layout of Expressions (3) in Sec. 3.)

---

[**] The KMT form of the $NN$ interaction with detailed decomposition is represented by Eq. 5 in Sec. 3.

Here $\chi_T$ and $\chi_L$ are transverse and longitudinal form factors, and $I_0$ is an unpolarized cross section.

We employed Equations (3) for the case of $T = 0$ in [8], and in this review we extended it to the excitation of states with $T = 1$. We also tested it at $\theta = 0°$, where $M(q)$ was so simplified that the two-body spin-orbit term could be neglected ($C = 0$), as well as the direct tensor interaction. In the case of no tensor interaction, we got the ratio $B = E$ (see e.g. Ref. [11]).

In general, $\chi_T$ and $\chi_L$ are coherent sums of matrix elements for the transfers of different $\Delta L$. In the case of a dominant single $\Delta L$, the ratio $\chi_T^2/\chi_L^2$ can be simply replaced by a number. In the present review, we analyze such particular cases.

For $M1$ transitions ($\Delta J = 1$) at very forward angles, including $0°$, a contribution with $\Delta L = \Delta J - 1$ dominates. For the excitation of $1^+$, $T = 0$ and $1^+$, $T = 1$ states, dominates $\Delta L = 0$ (a $\Delta L = \Delta J + 1$ contribution is negligible) [6]. This ensures that in this case

$$\chi_T^2 = \chi_L^2. \qquad (4)$$

Taking all the above characteristics of Eq. (3) into account, we get the following results at $0°$:

$$D_{ls} \approx 0, \quad D_q \approx D_n. \qquad (5)$$

It is this particular phenomenon that we observe in Fig. 1 in both experiments and our DWIA calculations for $T = 1$. We established Ratio (5) in [8] for the excitation of the $1^+$, $T = 0$ state in $^{12}C$. The only difference was that at the $\Delta T = 0$ transition the coefficients $D_q$ (and $D_n$) acquired different values, which is natural due to the dependence of all members of $M(q)$ on the isospin. In this way, Ratio (3) at $0°$ can be reliably tested and appear to be quite adequate.

### Isospin and Angular Dependences of $D_{ls}$ Spin-Transfer Probabilities in Unnatural-Parity ($\vec{p}, \vec{p}'$) Reactions at Intermediate Energies

In this section, we demonstrate that $D_{ls}$ observables, based on a set of polarization-transfer measurements, and their analysis within the framework of the model of Bleszinski et al., are applicable for a systematic evaluation of the role of spin-orbit interactions in nucleon inelastic scattering on a number of light-weight nuclei.

Accordingly, numerous research data (see e.g. Refs. [3, 13, 14]) prove that the isovector spin-orbit interaction is consistently weak at intermediate energies. On the other hand, the isoscalar spin-orbit component of the effective interaction is large. It is in agreement with the fact that the corresponding spin-transfer probabilities, $D_{ls}$, are primarily driven by the strengths of spin-orbit amplitudes (Fig. 2).

In the case of $^{12}C$ we employed DWIA calculations (solid curves) with the DBHF interaction ([15, 16] and a private communication from Stephenson). For $^{16}O$, we performed calculations using the DWBA 91 program from Raynal with the $G$-matrix from Geramb (solid curve). For $^{28}Si$, all our calculations were made using the same DWBA 91 program with three types of interactions: the $G$-matrix from Geramb (solid curve) and two versions of the Idaho interaction [18] for $T = 1$ (almost undistinguishable in Fig. 2). In the case of $T = 0$ for $^{28}Si$ we used one of the versions of the Idaho interaction and LEA FL at 500 MeV (thin solid and dashed curves, respectively).

In full agreement with experimental and calculated results for $D_{ls}$ for the $1^+$ ($T = 1$) state in $^{12}C$ (see Fig. 2) are the corresponding data for the same excitation at higher energies of 400–500 MeV (Fig. 1). Indeed, the experimental and calculated values of $D_{ls}$ for the isovector excitation are still very small.

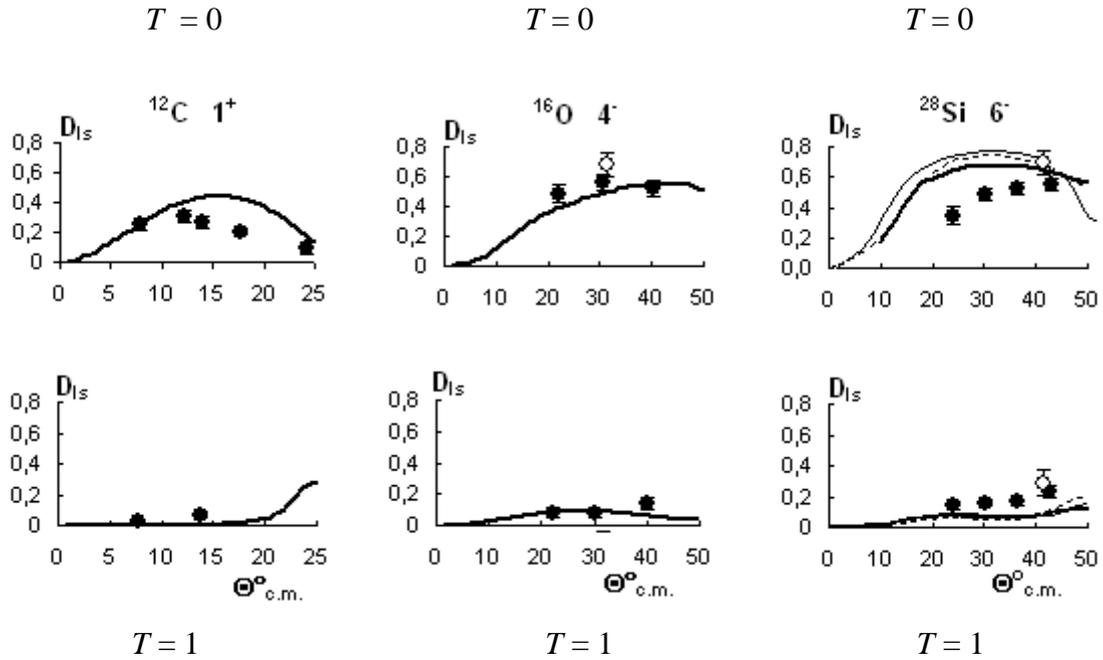

**Fig. 2.** The systematized experimental (points) and predicted (curves) data $D_{ls}$ for $(\vec{p}, \vec{p}')$ scattering on $^{12}$C (1$^+$), $^{16}$O (4$^-$) and $^{28}$Si (6$^-$) to the $T = 0$ (top) and $T = 1$ (bottom) excitations. We based the represented experimental results $D_{ls}$ on $(\vec{p}, \vec{p}')$ spin-observable measurements at $E_p = 200$ MeV [15, 16] (solid points) and at $E_p = 350$ MeV [7] (open point – $^{16}$O), with the added data $D_{ls}$ at $E_p = 500$ MeV [17] (open points – $^{28}$Si). Due to the difference in energies $E_p$ we introduced certain kinematic corrections. The calculated data (curves), described in the text, refer to $E_p = 200$ MeV, apart from one incident for $^{28}$Si (500 MeV).

The isospin representation for the moduli of spin-orbit components in the effective *NN* interaction (parameterization of the free *NN* t-matrix from Love and Franey) at $E_p = 140$ and 800 MeV is in qualitative agreement with the analyzed results. Undeniably, these isoscalar and isovector moduli, displayed as a function of *q*, have similar shapes. However, at all *q* quantities, isoscalar (as against isovector) "interaction" components are typically dominant in value [5].

**Transverse- and Longitudinal- Spin-Transfer Probabilities for Unnatural-Parity $(\vec{p}, \vec{p}')$ Reactions at Intermediate Energies**

In this section, we present systematized angular distributions of two spin (polarization)-transfer probabilities $D_K$ (Fig. 3), based on the available complete $(\vec{p}, \vec{p}')$ measurements at $E_p = 200$ MeV (solid points), at $E_p = 350$ MeV, and at $E_p = 500$ MeV (open points for $^{16}$O and $^{28}$Si, respectively) for the unnatural-parity $T = 1$ levels in a set of light-weight nuclei. Instead of measured and calculated polarization-transfer coefficients, Fig. 3 shows their combinations, which, as probabilities $D_K$, are associated in the PWIA prediction with the squares of particular amplitudes in the *NN* effective interaction. The given $D_K$ are polarization observables introduced in Ref. [1].

As noted above, a distinctive feature of $T = 1$ excitations is that the isovector spin-orbit interaction is weak at intermediate energies. The smallest quantity of $D_{ls}$ in both experiments and our calculations confirms this fact (Fig. 2). As a rule, normal spin-transverse $D_n$ quantities exceed $D_{ls}$ values (Fig. 3).

Consequently, the spin-observable combinations $D_n$ that depend on the isovector spin-spin interaction and the transverse spin-matrix element are generally well described by calculations for the excitations represented in Fig. 3: 1$^+$ (15.11 MeV) in $^{12}$C, 4$^-$ (18.98 MeV) in $^{16}$O, and 6$^-$ (14.35 MeV) in $^{28}$Si.

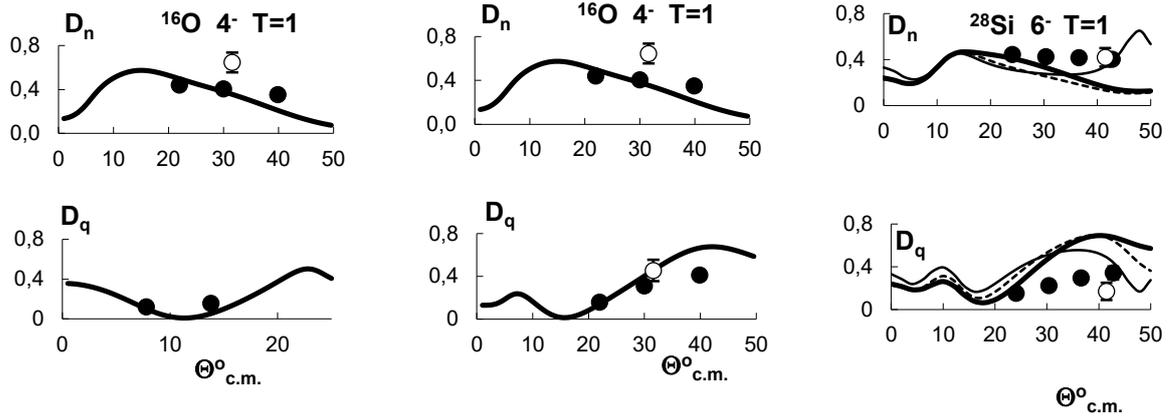

**Fig. 3.** Systematized angular distributions of the spin-transfer probabilities $D_n$ and $D_q$ for the indicated isovector levels in $^{12}$C, $^{16}$O and $^{28}$Si, based on the measured ($\vec{p}, \vec{p}'$) quantities at 200 MeV [15, 16], as well as at 350 and 500 MeV. Measurements at 350 MeV ($^{16}$O) were performed in [7], and at 500 MeV ($^{28}$Si) – in [17]. Calculations made at 200 MeV were as follows. In the case of $^{12}$C, we used DWIA calculations based on the DBHF interaction [15, 16]. For $^{16}$O, we employed the DWBA 91 program from Raynal and G-matrix (DD) from Geramb. For $^{28}$Si, all our calculations were made using the DWBA 91 program with three types of interactions: the $G$-matrix from Geramb (thick solid curve), and two variations of the Idaho interaction (thin solid and dashed curves) [18].

The contrasting term $D_q$ is a spin-longitudinal component of the spin-transfer probability. $D_q$ values depend on the spin-spin interaction, namely on the pion-dominated isovector $(\sigma_1 \cdot \hat{q}\; \sigma_2 \cdot \hat{q})$ piece of the *NN* interaction, as well as on the longitudinal spin-matrix element.

Different $q$ dependences in $D_n$ and $D_q$ can be explained by the fact that transverse and longitudinal axial form factors have a different $q$ dependence. The qualitative features of such a relative behavior of $D_n$ and $D_q$ can be understood from Fig. 4, which represents isovector transverse ($V_T$) and longitudinal ($V_L$) parts of the *t*-matrix interaction at 210 MeV (see Ref. [19]). When $\theta_{cm}$ in Fig. 3 varies from 20° to 40°, this corresponds to $q$, changing from 1 to 2 fm$^{-1}$ (Fig. 4).

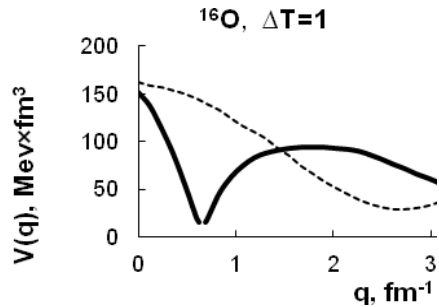

**Fig. 4.** Transverse and longitudinal isovector parts of the t-matrix interaction at $E_p = 210$ MeV ($V_T$ and $V_L$, respectively) – from Franey and Love. $V_L$ is shown by a solid curve. $V_T$ is represented by a dashed curve.

Let us apply these results to the $4^-$, $T = 1$ state in $^{16}$O. Thus, near 1.5 fm$^{-1}$, where $V_T \sim V_L$ (Fig. 4), $D_n$ and $D_q$ have similar quantities (Fig. 3). The transverse and longitudinal transition densities are also comparable for the stretched $4^-$, $T = 1$ excitations [19]. Near 0.7 fm, where $V_L$ is very small, $D_q$ becomes roughly equal to 0. The $D_n$, however, acquires its maximum value, since here $V_T$ is large and dominant. At larger $q$, where $V_L$ becomes bigger than $V_T$, $D_q$ acquires larger values, as compared with $D_n$. At 0° ($q = 0$). When $V_T$ and $V_L$ are practically the same (Fig. 4), we also observe similarity of $D_n$ and $D_q$ in our calculated data (Fig. 3).

Although isoscalar spin-dependent forces present a different picture, they can be analyzed in a similar manner. Let us note that the relation between isoscalar $D_n$ and $D_q$ probabilities is entirely different for $T = 0$ states (as compared with $T = 1$). We will show this relation in detail further on. Now we will only point out a unique sensitivity of the nucleon to the longitudinal spin response of the nucleus, which cannot be detected in $e$- and $\pi$-nucleus interactions [1].

## The Exchange Part of Unnatural-Parity Transitions in (p,p′) Scattering

As is shown above, $(\vec{p}, \vec{p}')$ polarization-transfer measurements allow to establish the fact that the three spin-transfer probabilities $D_K$ are predominantly sensitive to the response of the nucleus through tensor operators ($D_q$, $D_n$, and $D_p$), while the observable $D_{ls}$ is driven primarily by the spin-orbit operator.

As we observed in Fig. 1, the value and shape of the component $D_p$ can be reasonably predicted. At the same time, it is known that this characteristic is not simple. Indeed, the in-plane spin-transverse response $D_p$ is usually associated with tensor exchange amplitudes (see e.g. Ref. [10]), therefore this specific quality of $D_p$ appears to be difficult to study.

One of the analytical methods is based on a simultaneous description of the $(\vec{p}, \vec{p}')$ observable combinations and their $(e,e')$ counterparts, as is shown in Ref. [3]. In principle, the obtained information should allow to extricate from the $(\vec{p}, \vec{p}')$ data possible contributions of nuclei currents (convection and/or composite). The key element of such a study in a nonrelativistic DWBA treatment (similar to the one we employed) is that the convection current arises exclusively through knock-on exchange and other nonlocalities [3, 13. 14].

A combined study of $(\vec{p}, \vec{p}')$ and $(e,e')$, using both relativistic and nonrelativistic analyses [3], shows that $(\vec{p}, \vec{p}')$ quantities can be relatively insensitive to the knock-on exchange. Therefore, convection-composite currents are negligible at the $1^+$, $T = 1$ (15.11 MeV) excitation in $^{12}$C.

However, as is noted in Ref. [3], for the $1^+$, $T = 0$ (12.71 MeV) isoscalar transition in $^{12}$C, the analysis indicates that the convection-composite-current contributions are relatively much more important than they are for the $1^+$, $T = 1$ $(\vec{p}, \vec{p}')$ isovector transition. Based on this characteristic, especially prominent are the measured $T = 0$, $(\vec{p}, \vec{p}')$ quantities, related to the response $I_0 D_p$ and $I_0 D_{ls}$, where the factor $I_0$ represents an ordinary unpolarized cross section.

In a number of our works in order to evaluate direct and exchange pieces of the effective interaction, we compared the results of the analysis of $(\vec{p}, p')$ or $(\vec{p}, \vec{p}')$ spin observables obtained with the use of two calculation models. One of these models, represented in the LEA program from Kelly, is based on a zero-range treatment of knock-on exchange. To oppose this approach, we employed another, a more complex model that involves the use of a finite-range treatment for the exchange part of the scattering. In this case, it was necessary to solve a more serious problem of establishing the exact nucleon-nucleus kinematics.

To make such exact calculations, we used the program DWBA 91 from Raynal, and in some cases, we checked them against our calculations based on the data from Stephenson. In the latter case, the DWBA 86 code was used as a program with a finite-range DWIA (based on the works by Schaeffer and Raynal).

Having compared the experimental data and the results of the analysis with the zero-range calculations of LEA and those performed with the finite-range DWBA 91 program, we successfully obtained two sets of polarization transfer coefficients $D_{ij}$ for $(\vec{p}, \vec{p}')$ reactions, leading to the unnatural-parity $1^+$, $T = 0$ and $T = 1$ states in $^{12}$C [20].

It follows from the analysis of the combinations $D_K$ for the $1^+$, $T = 0$ and $T = 1$ states in $^{12}$C, using both programs (DWBA 91 and LEA), that for these purposes the approach that does not represent the full *NN* amplitude cannot be used in certain cases. This applies to the program LEA in which the exchange approximation is such that the exchange interactions are independent of the momentum transfer and are reduced to delta functions [21].

The most striking discrepancy between experimental and calculated data within the framework of LEA formalism is established by us in the case of the $1^+$, $T = 0$ excited state in $^{12}$C for the combination $D_p$ (Fig. 5, dashed line) and $D_{ls}$ (not shown) at $E_p = 200$ MeV.

However, there is no such discrepancy between similar calculated and experimental data in the case of the $1^+$, $T = 1$ excited state in the same nucleus, $^{12}$C, for the combination $D_p$ (Fig. 5, dashed line) and $D_{ls}$ (not shown) at the same $E_p$ value.

All the results of the analysis are in good agreement with the above-mentioned studies [3] aimed at extracting information on $M1$spin responses from $(e,e')$ measurements and the analogous $(\vec{p},\vec{p}')$ quantities at $E_p = 500$ MeV, respectively. Based on the combined analysis of the indicated data, the conclusion about the corresponding convection-current contributions, and, therefore, about the role of knock-on exchange was made in [3]. It is worth noting that in the $I_0D_p$ and $I_0D_{ls}$ data for the $1^+$, $T = 0$ state in the $^{12}$C excitation, the indicated contributions turned out to be relatively more important than they are for the $1^+$, $T = 1$ transition. This, in particular, is in agreement with our analysis discussed above and represented in Fig. 5.

Moreover, the research results, systematized in Ref. [22], clearly indicate that for the inelastic excitation of the $1^+$, $T = 0$ state in the $^{12}$C by inelastic proton scattering, the tensor exchange process is a dominant reaction mechanism, while the direct process is suppressed. In this connection it is clear that DWBA 91 calculations describe $D_p$ for the $1^+$, $T = 0$ state (Fig. 5, solid curves) and $D_{ls}$ for the $1^+$, $T = 0$ state (not shown) fairly well. Moreover, antisymmetrization introduces additional nonlocalities in the form of knock-on exchange amplitudes, as it occurs in the case of nucleons [5].

Nevertheless, it is also obvious (Fig. 5) that for a number of the observable combinations $D_K$ a zero-range treatment of knock-on exchange in LEA is not inferior, as regards to the quality of the description of experiments, towards the exact finite-range calculations made with the use of the DWBA 91 or DWBA 86 programs.

To ensure a better understanding of the analytical picture represented in Fig. 5, let us consider the previous results of the analysis of $(\vec{p},\vec{p}')$ polarization transfer measurements (polarized cross sections) for the $6^-$, $T = 1$ transition in $^{28}$Si $(\vec{p},\vec{p}')^{28}$Si [10]. There experimental data are compared with two types of calculations. One of them is based on the original DWIA program, LEA [21], the same one as we used (Fig. 5). The other type of calculations is based on its modification that agrees well with the exact finite-range treatment, i.e., basically, the same that we employed for two types of analysis represented in Fig. 5, using the DWBA 91 and DWIA 86 programs. If we consider the analytical data, represented in Fig. 5, from the angle of those studies, we will observe the following similarity. Thus, the spin-transfer response $\sigma_p \equiv I_0D_p$ for the $6^-$, $T = 1$ transition in $^{28}$Si [10], and spin-transverse component $D_p$ for the $1^+$, $T = 1$ excited state in $^{12}$C (Fig. 5) are in good agreement with the calculations of both types indicated above.

Besides, it should be noted that there also exist relatively small differences between the data and calculations made using both $NN$ matrixes, considered above, in the case of the response $I_0D_n$ [10], and the component $D_n$ (Fig. 5) for the same isovector ($\Delta T = 1$) transitions, respectively.

However, in the case of the $I_0D_q$ response and the component $D_q$ for the same transitions and in the same nuclei, the situation is very different. Therefore, despite the fact that $D_q$ data still agree equally well with the calculations of both types, for the $I_0Dq$ response (in the case of $6^-$, $T = 1$ in $^{28}$Si, not shown) only finite-range calculations are necessary [10]. Besides, our analysis of the spin-transfer responses $D_n$ and $D_q$ for the $6^-$, $T = 1$ state in $^{28}$Si shows that the calculated results (Fig. 3) underestimate $D_n$ and overestimate the experimental data for $D_q$, which largely corresponds to the estimations from [16].

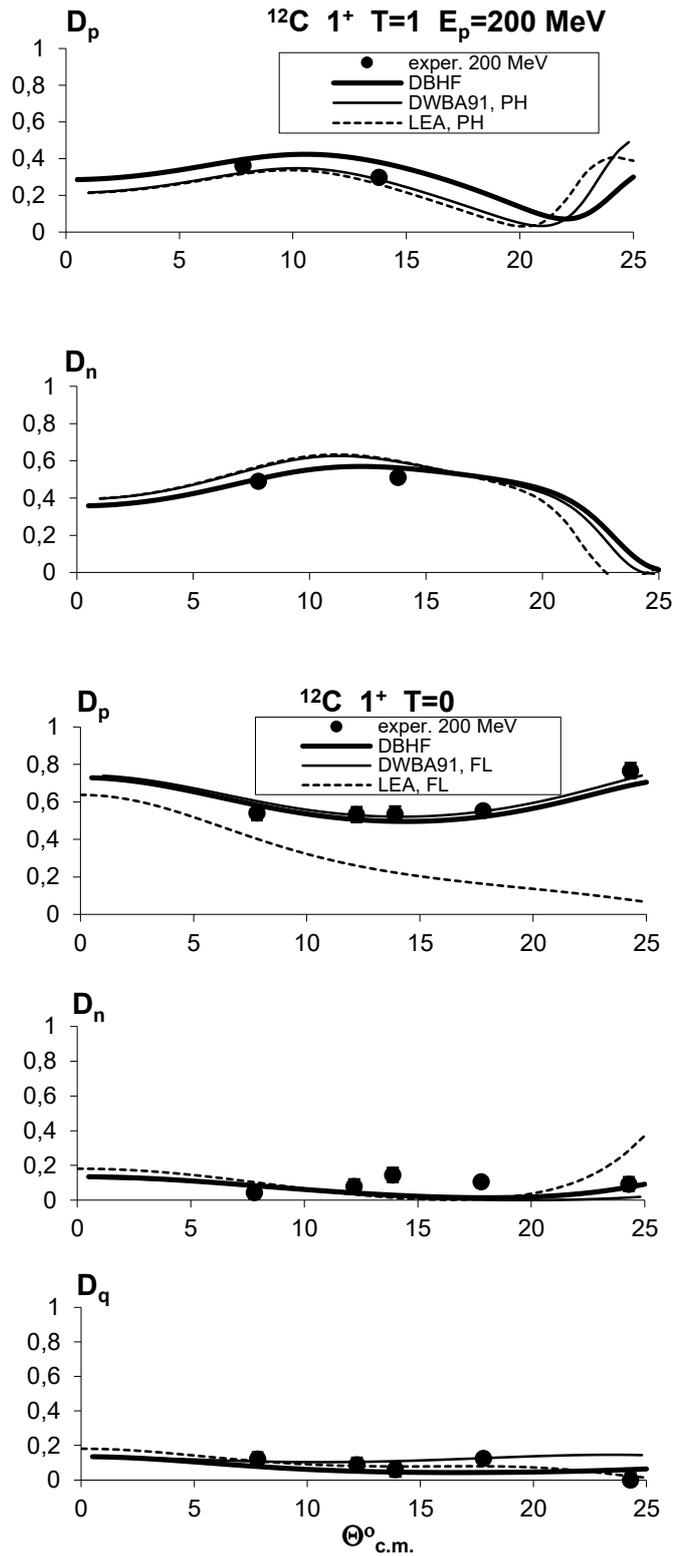

**Fig. 5.** The sensitivity of $D_K$ to direct contributions and the expected role of nonlocal processes. All solid lines (both thick and thin) represent calculations made with exact finite-range amplitudes (two versions of the DWBA programs based on the work by Schaeffer and Raynal); the dashed curves are zero-range calculations made using LEA. The lower part represents observables for the $1^+$, $T = 0$ state, and the upper part represents observables for the $1^+$, $T = 1$ state in $^{12}$C. The thick solid curves are DWBA 86 calculations made using the DBHF interaction of Sammarruca and Stephenson, while the thin solid curves use DWBA 91 calculations with the free interaction of Franey and Love ($T = 0$) and the effective PH interaction of Geramb and his Hamburg associates, based upon the Paris potential ($T = 1$). This potential is used in all corresponding calculations made using LEA. The combination of the experimental data is based on measurements from Ref. [15].

As we noted above, in a nonrelativistic framework, unnatural-parity transitions appear to be sensitive to the coupling of the nucleon spin to the bound current, with these couplings arising through exchange processes.

It is known now that $(p,p')$, $(\vec{p}, p')$ and $(\vec{p}, \vec{p}')$ observables and some of their combinations are sensitive, to a different extent, to the nonlocal and exchange nature of the scattering process. Moreover, there may be several sources of nonlocality in spin-dependent parts of the effective *NN* interaction.

According to Ref. [14], although a quantitative understanding of the sources of nonlocality in spin-dependent parts of the *NN* interaction is rather difficult, it is important to find a signature of their presence. Such examples, as those represented in Fig. 5, help to discriminate between possible *NN* effective interactions with different nonlocal behavior.

## Summary of the Section

As we established it for the spin-transfer response $D_K$, a part of which is shown in Fig. 5, in some cases exchange processes become especially important for transitions with $\Delta T = 0$ in the $^{12}$C nucleus. At the same time, for other quantities (particularly for transitions with $\Delta T = 1$ in the same nucleus) the role of exchange processes is much weaker, judging by the fact that there the zero-range treatment of knock-on exchange in LEA works as satisfactorily as exact finite-range calculations in DWBA 91.

Such a result for the $\Delta T = 1$ transition in $^{12}$C is in good agreement with the conclusions of work [3] in the case of 500 MeV. The authors of this paper showed that in the case of $\Delta T = 1$ the $(\vec{p}, \vec{p}')$ quantities were relatively insensitive to the knock-on exchange, and the convection-composite currents were confirmed to be very small.

We conducted our analysis at noticeably smaller energies ($E_p = 200$ MeV) because with such energies exchange (nonlocal) processes are expected to be more significant in strength as the incident proton energy decreases (see specifically Ref. [23]).

In fact such expectations proved to be true, although most clearly only for the $\Delta T = 0$ transition. Here our nonrelativistic calculations with exchange showed that in some cases experiments could be carried out only with $t$ (or $G$) matrix used to represent the full *NN* amplitude. In any case we did not get anywhere near the "curious situations" at any $\Delta T$ values as those observed in Ref. [23] where relativistic codes were used, so, to achieve a better description of the data, the authors had to exclude knock-on exchange contributions altogether.

The main approach we used in this work was that in the framework of the model represented in Ref. [1], according to works [5, 16], complete $(\vec{p}, \vec{p}')$ spin-transfer observables can be used to isolate specific parts of the nuclear response. This is supported by the fact that the combinations of polarization-transfer coefficients selectively (separately) depend on individual (particular) components of the *NN* amplitude.

Criticism regarding this method [24] primarily concerns the point that certain combinations of $(\vec{p}, \vec{p}')$ spin-transfer observables and their relation to individual terms in the *NN* interaction are linked to a direct plane-wave impulse approximation. Therefore, according to Ref. [24], the inclusion of distortion of projectile waves and knock-on exchange blurs the physical interpretation of corresponding nuclear response functions.

However, as was shown above, such doubts can largely be overcome. Thus, as is suggested in Ref. [5], the considered method allows obtaining significant information about transverse and longitudinal spin-response functions of the nucleus. Indeed, the comparison of the corresponding quantities $D_n$ and $D_q$ for the $1^+$, $T = 0$ (Fig. 5) and $1^+$, $T = 1$ states (Figs. 1 and 3) demonstrates their obvious isospin separation.

Good agreement between two DWBA calculations shown in Fig. 5 by thick and thin lines, respectively, should also be pointed out. The thin solid lines were obtained within several variations of the approach that formulated the proton-nucleus scattering problem in the nonrelativistic impulse approximation. The second approach (thick solid lines) represents the DBHF technique based on the Dirac–Brueckner theory. There the authors followed closely the relativistic approach to nuclear matter [15, 16]. As is seen in Fig. 5, there is no significant discrepancy between these two types of calculations.

In paper [8] we described a number of difficulties in coordinating calculations and experiments for the transverse and longitudinal responses of the stretched states, $4^-$, $T = 0$ in $^{16}$O and $6^-$, $T = 0$ in $^{28}$Si. It is evident that the theoretical picture presented there was too simplified in comparison with the experiment.

Nevertheless, at the excitation of states with the same spins and parities in the same nuclei, but for $\Delta T = 1$ transitions, such responses can be described rather adequately (Fig. 3). Besides, well known is the fact that for the unnatural-parity transition, the spin-orbit part of the effective interaction is small for isovector transitions, and the spin-orbit amplitude is large in the isoscalar channel, which is effectively confirmed in Fig. 2. Therefore, we quite agree with the assertion of the authors of work [16] that such representations can be used as a powerful diagnostic tool. Nevertheless, many unsolved problems remain, and a more comprehensive investigation is needed in the study of polarization-transfer processes in $(\vec{p}, \vec{p}')$ reactions.

### *NOTE ADDED BY THE AUTHOR OF THE REVIEW IN PROOF

In Fig. 5, we present several treatments of calculations. The most important calculations for us are those that differentiate between treatments of exchange. As a result, contrasted are series with exact finite-range calculations (solid curves of two types in the figure) with other series with zero-range approximations of knock-on exchange (dashed curves). In the latter case, substantial differences may appear between the data and a calculation. However, in the case of a series of finite-range DWIA programs all data are reasonably well predicted.

It should be noted that the exact finite-range calculations were made using competing interaction models, among them are the density-independent $t$ matrix of Franey and Love (FL), the density-dependent Paris–Hamburg $G$-matrix, and density-dependent DBHF (Dirac–Brueckner Hartree–Fock) interaction from Stephenson and Sammarruca. The DBHF effective interaction includes all conventional medium effects and certain additional relativistic corrections, arising from an attractive scalar and a repulsive vector nuclear mean fields.

We formed theoretical angular distributions of the probabilities $D_K$ for the $1^+$, $T = 0$ and $T = 1$ levels in $^{12}$C in the case of the DBHF interaction (thick solid curves in Fig. 5), based on the spin observables $D_{ij}$ calculated by Ed Stephenson who kindly provided us with their tabulated form. That ensured a better understanding of the whole bulk of results, represented in Fig. 5. Moreover, it helped to assess the influence of tensor exchange contributions upon the spin-observable combinations $D_K$. In particular, our conclusions are in agreement with those of K. Nakayama and W.G. Love (Phys. Rev. C 1988, v. 38, p. 51) for other data, namely for isoscalar $S = 1$ excitations, that the exchange contributions associated with the tensor force are large.

Especially satisfactorily the results of the $D_p$ study for $T = 0$ and $T = 1$ in Fig. 5 comply with the assertions of W.G. Love [W.G. Love. Conf. on the (p,n) reaction and the nucleon-nucleon force . N.Y.: Plenum Press, 1980, p. 23]. Thus, $\Delta S = \Delta T = 1$ transitions are sensitive to the isovector part of the tensor force (large) that acts in direct amplitudes. At the same time transitions, resulting in $\Delta S = 1$ and $\Delta T = 0$, are quite sensitive to knock-on exchange terms, which are normally associated with the tensor force.

*The present section is a development of the following extended abstracts by A.V. Plavko, M.S. Onegin, V.I. Kudriashov, presented at the 65th Intern. Confer. "Nucleus 2015", Book of Abstracts, June 29–July 3, 2015, St. Petersburg, Russia:

1. Unnatural-Parity $(\vec{p}, \vec{p}')$ Reactions in a Factorized Impulse-Approximation Model for Polarization Transfer (PT) and Spin Responses (p. 182).
2. Isospin and Angular Dependences of Spin-Transfer Probabilities $D_{ls}$ in Unnatural-Parity $(\vec{p}, \vec{p}')$ Reactions at Intermediate Energies (p. 179).
3. Transverse- and Longitudinal-Spin-Transfer Probabilities for Unnatural-Parity $(\vec{p}, \vec{p}')$ Reactions at Intermediate Energies (p. 180).

# SECTION 5
# Specific Examples and General Conclusions

## Full Spin-Transfer Probabilities for the 1$^+$, $T = 0$ and $T = 1$ Levels in $^{12}$C

As noted above, Equation 5 in Sec. 3 represents basic scattering amplitudes. This equation is of course also isospin dependent. Strictly speaking, $M, A, B,$ etc., should be written as $M_\tau, A_\tau, B_\tau,$ etc., where $\tau$ is the isospin index (0 or 1).

Hence, the combinations $D_i$ of the $D_{ij}$ in Equations 4 (Sec. 3) should be written as follows: $D_{ls}^\tau$, $D_q^\tau$, etc. In Equations 3 (Sec. 3), the transverse and longitudinal form factors should be indicated as $\chi_T^\tau$ and $\chi_L^\tau$. Such particularization is also applicable to Equations 2 and 3 in Sec. 4.

The combinations $D_i$ (4 in Sec. 3) or, which is the same, $D_k$ (2 in Sec. 4), made up of the spin-transfer parameter $D_{ij}$, relate to unnatural parity states in the PWIA. For natural parity states in the PWIA, the corresponding expressions change. However, we do not discuss them here, as we do not analyze the corresponding polarization-transfer observables. Let us just note that for natural parity states we get $D_q^\tau \approx 0$, which has been confirmed in both calculations and experiments.

It should be emphasized that with the inclusion of distortions, certain $D_i$ ($D_k$) combinations are still sensitive primarily to the same indicated terms. Accordingly, as we showed above (see Fig. 5 in Sec. 4), specific combinations of polarization transfer coefficients, i.e. $D_i$ ($D_k$), may be especially sensitive to the piece of a nonlocal or exchange nature of the scattering process. The use of DWBA 91 assists to that, as this program allows for exact finite-range DWIA for exchange contributions. Such an effect becomes evident in comparison with the application of the DWIA program LEA, in which the exchange is handled in a zero-range approximation.

With reference to the above, we will further demonstrate that some specific cases require the use of a finite-range program for the exchange part of the scattering. For this purpose, we will represent Fig. 5 (Sec. 4) in a more detailed form and show in full all the four spin observables $D_K$, respectively, for both the 1$^+$, $T = 1$ (Fig. 1 in this section) and the 1$^+$, $T = 0$ (Fig. 2 in this section) excited states in $^{12}$C.

In order to clarify the picture, we have chosen the appropriate data at the lowest available energy ($E_p$ = 200 MeV) because exchange processes, as is quite well known, become more important as the incident energy $E_p$ decreases.

The spin-transfer observables ($\vec{p}, \vec{p}'$) for the 0$^+$ →1$^+$ inelastic transitions in $^{12}$C are known to be selectively sensitive to isovector ($T = 1$) and isoscalar ($T = 0$) spin-dependent components of the $NN$ interaction. Figures 1 and 2 in this section clearly confirm that. Yet the impact of exchange components on isoscalar and isovector transitions varies. For isovector transitions, the zero-range treatment of knock-on exchange in LEA is mostly satisfactory for the description of the experimental data $D_K$ (Fig. 1, Sec. 5). However, this approximation is in fact unacceptable in the case of $D_p$ and $D_{ls}$ for isoscalar transitions (see the dashed curves in Fig. 2 in Sec. 5). In this case, the tensor exchange process is a dominant reaction mechanism. Since the zero-range treatment of exchange is approximate, it is natural that it leads to the biggest deviations from the experiment.

The explanation of such deviations seems clear enough. Indeed, the observables $D_K$ (both experimental and calculated) are based on Eqs. 4 (Sec. 3), or, which is the same, on Eqs. 2 (Sec. 4). In the expression for the observable $D_p$, the polarization-transfer coefficients $D_{NN}$ and $D_{SS}$ are given with negative signs, but in the expression for $D_{ls}$, their signs are positive. At the same time, as follows from Figs. 5 and 6 (Sec. 1), the calculations made using mainly LEA overrate the calculations $D_{NN}$ and $D_{SS}$ in comparison with the experiment. Therefore, the resultant LEA calculations significantly underrate the experimental data for $D_p$, but the LEA calculations for $D_{ls}$, on the contrary, overrate the respective experimental data for $D_{ls}$. As for the expressions with the observables $D_q$ and $D_n$, the coefficients $D_{NN}$ and $D_{SS}$ in them have different signs, thus canceling each other, which is also true in the case of experimental data. Consequently, the approximate character of LEA calculations for $D_{NN}$ and $D_{SS}$ eventually does not surface in analytical LEA results for the combinations $D_q$ and $D_n$ (see Fig. 2 in this section).

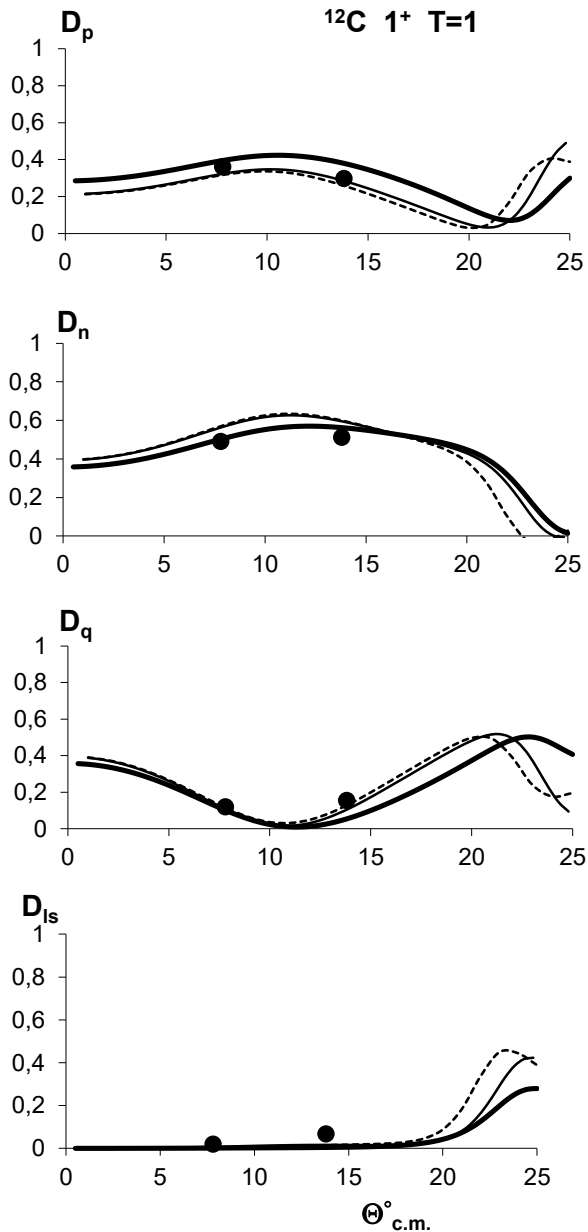

**Fig. 1**. The same as Fig. 5 in Sec. 4 (top) for $^{12}$C, $1^+$, $T = 1$, but for all the four observables $D_K$.

The presented explanations deal with simplifying assumptions, e.g. neglecting distortions. However, DWBA calculations (Fig. 2 in this section), with distortions of projectile waves included, do not alter much the simple interpretation given above.

It should be noted that the programs DWBA 91 and DWBA 86, employed in the case of Fig. 5 (Sec. 4), and Figs. 1 and 2 in this section, do not use a zero-range treatment of knock-on exchange (as opposed to the precise program LEA), but include both direct and knock-on exchange terms exactly. When we follow the latter nucleon-nucleus kinematics, the difference from various interactions is minimal. In a number of cases, it is relatively small in comparison with the dependence on the treatment of exchange (finite range or zero-range).

In the case of the $1^+$, $T = 1$ excited state in $^{12}$C at 200 MeV, the differences between calculations made using either DWBA or LEA programs do not seem critical. Both provide quite a good description of all experimental data for all the four probabilities $D_K$. Besides, for two interactions (DBHF and FL) the differences between angular distributions within the analyzed angles are rather insignificant in DWBA calculations. All this may signal that the direct process is dominant here and the exchange one is minimal.

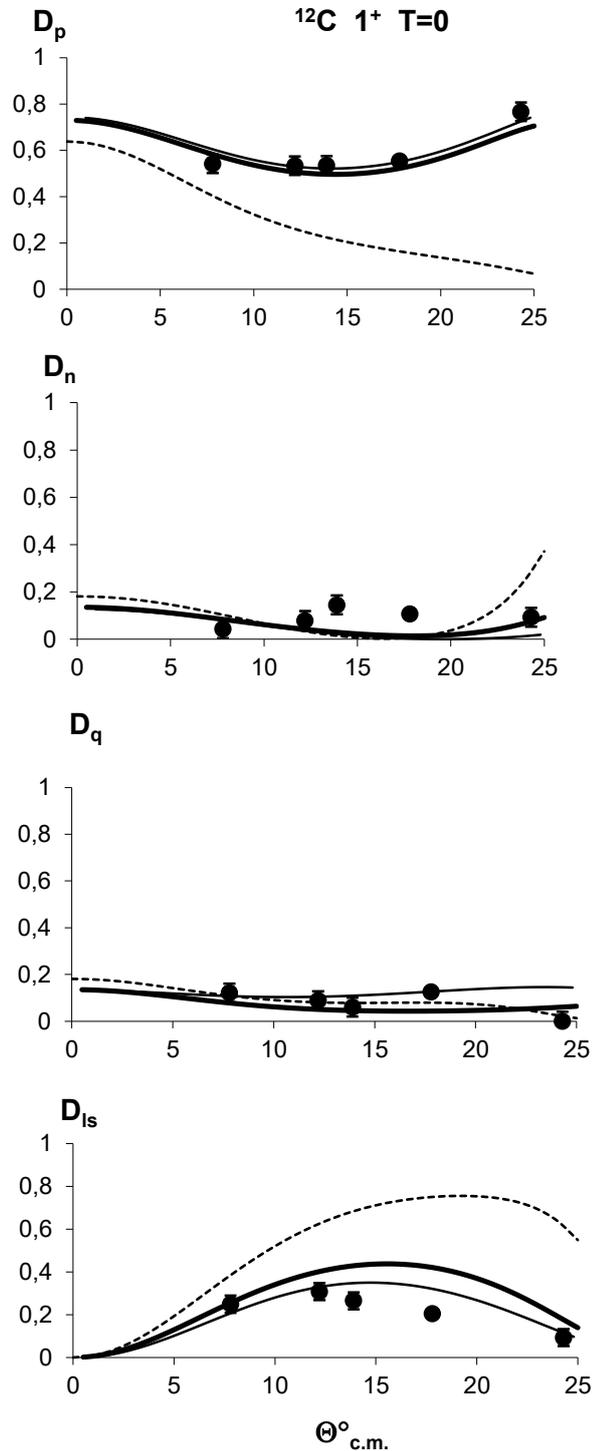

**Fig. 2**. The same as Fig. 5 in Sec. 4 (bottom) for $^{12}$C, $1^+$, $T = 0$, but for all the four observables $D_K$.

We should emphasize that the thin solid curves in Figs. 1 and 2 in this section are calculations representing the DBHF (Dirac–Brueckner Hartree–Fock) effective interaction. It means that these theoretical results are based on special density-dependent $G$-matrix calculations. They include all established effects of other interactions and, in addition, they take into account strong relativistic mean-field potentials (in the Dirac–Brueckner approach to nuclear matter). In DBHF cases, considered in both the present and previous sections, in the calculation scheme, the mean nuclear potential is indeed constructed in terms of a large (partially canceling) competing scalar and vector field (see, e.g., F. Sammarruca, E.J. Stephenson, K. Jian. Phys. Rev. C, 1999. V. 60, p. 064610).

As follows from Figs. 1 and 2 in this section, calculations within DBHF describe the observable combinations $D_K$ for the $1^+$, $T = 0$ and $1^+$, $T = 1$ states in $^{12}$C extremely well. What is surprising is the fact that an almost accurate description provides basically traditional (simpler) effective forces, also shown there. The important thing is to ensure that the appropriate calculation programs allow for an exact finite range DWIA for exchange contributions, especially in the case of the $1^+$, $T = 0$ state in nuclei.

Eventually, if we suppose that convection-composite currents arise exclusively through knock-on exchange in a nonrelativistic DWBA treatment, such currents for the $1^+$, $T = 1$ state in $^{12}$C are confirmed to be very small (Fig. 1 in this section). Within the framework of this understanding, in the case of the excitation of the $1^+$, $T = 0$ state in $^{12}$C at 200 MeV (see Fig. 2 in this section), the isoscalar convection current is most prominent.

At the same time, in the ($\vec{p}, \vec{p}'$) reaction at $E_p = 500$ MeV for off-diagonal $D_{ij}$ coefficients in the case of the $1^+$, $T = 0$ excited state in $^{12}$C (for both experiments and LEA calculations), the ratio $D_{SL} = -D_{LS}$ is observed (see Fig. 4 in Sec. 2). It is analogous in its meaning to the relationship $A_y = P$, as all the four indicated observables are theoretically the result of the interference between spin-orbit and tensor contributions to the reaction amplitude. The fact that the LEA treatment provides a satisfactory description of the ratio $D_{SL} = -D_{LS}$ (Fig. 4, Sec. 2) allows us to dismiss the exact finite-range treatment. Therefore, this circumstance indicates that convection-composite-current contributions are relatively less important here than they are for 200 MeV. This conclusion is qualitatively consistent with the expected decrease in strength of nonlocal processes at higher $E_p$ (see for example the top graphs of Fig. 2 in Sec. 1). In addition, according to Love and Comfort, with a noticeable decrease of $E_p$, the nonlocality (or velocity dependence) of the effective coupling between projectile and target nucleons must hold in nonrelativistic calculations, which we generally observed.

Indeed, in a relativistic treatment, the nuclear current operator appears automatically, but it is neglected in a nonrelativistic impulse approximation. Composite spin-convection current amplitudes of the relativistic theory in nonrelativistic treatments normally arise exclusively from nonlocal processes like knock-on exchange. Generally speaking, the zero-range treatment of knock-on exchange in LEA can be modified (following nucleon-nucleus kinematics) to exact finite-range calculations in DWBA 91. As a result, the method we use merely amounts to the comparison of different exchange effects of the effective interaction.

It should be noted that the nonlocality, present in the exchange terms, corresponds to different coordinates for incident and scattering nucleons. This nonlocality arises from the finite (nonzero) range of the *NN* effective interaction.

## A Comparison of Combinations of ($\vec{p}, p'$) and ($\vec{p}, \vec{p}'$) Observables, and the Normal-Component Coincident Analyzing Power

Figures 1 and 2 in this section represent the combinations of observables, defined via Expressions (4) in Sec. 3, or via Expressions (2) in Sec. 4. These combinations include both out-of-plane and in-plane polarization observables (a set of polarization-transfer coefficients). Our next step will be constructing other combinations, made up exclusively from the normal-components spin observables $D_{NN}$, $D_{0N}$ (also designated as $D_{0n}$), and $D_{N0}$ (also designated as $D_{n0}$). Traditionally, the spin observables $D_{NN}$ and $D_{n0}$ ($A_y$) are measured using an incident proton beam in which the polarization vector is normal to the scattering (often horizontal) plane. A different single-spin asymmetry, polarization $D_{0n}$ (P), can be induced by scattering unpolarized protons. The aim of the proposed step remains the same: to construct a different, independent combination of observables that would have a definite physical meaning and be sensitive enough to the coupling of the nucleon spin to the bound nucleon current. Such a combination can be expressed in the following form:

$$-(P - A_y)/(1 - D_{NN}) = A_y(\hat{n}). \qquad (1)$$

The left-hand side part of Eq. (1) represents a combination of $(p, p')$ and $(\vec{p}, \vec{p}')$ observables, shown in different form in Figs 1, 2 and 5 in Sec. 1, and in Fig. 5 in Sec. 2. These combinations are shown in Fig. 3 below by ark circles.

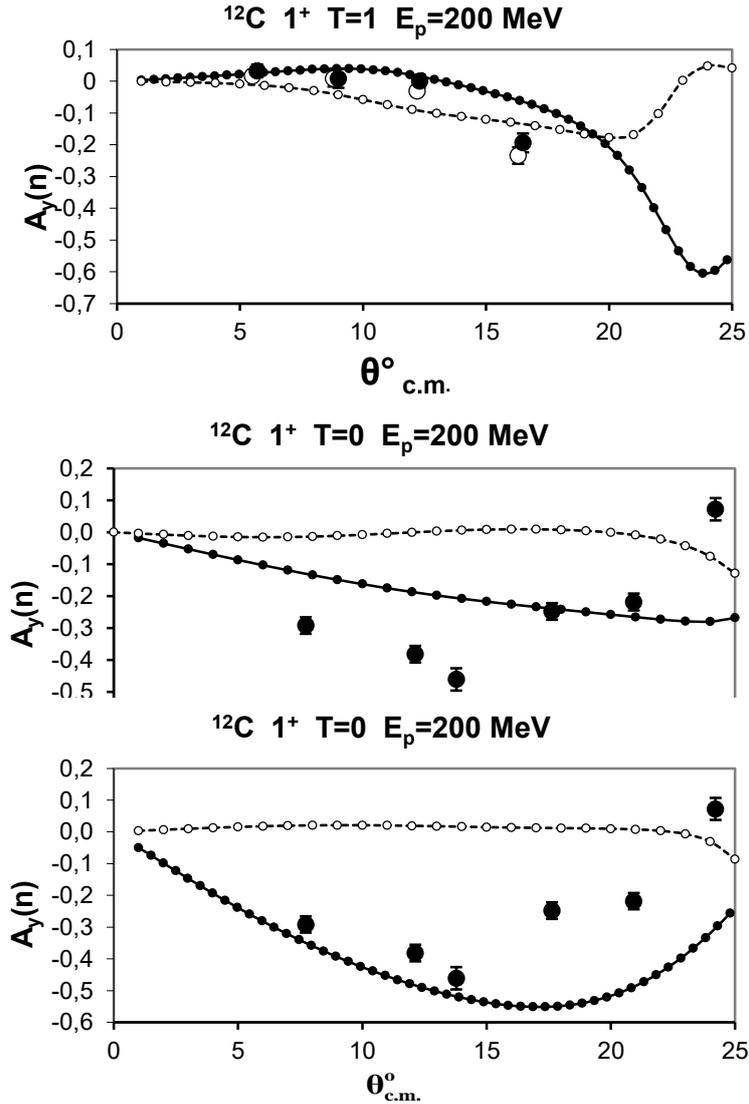

**Fig. 3.** Calculations and experiments of the observables, expressed in Eq. (1) in this section: all data for $1^+$, $T = 1$ in $^{12}C$ (top panel), and for $1^+$, $T = 0$ in $^{12}C$ (middle and bottom panels). In the case of the experiment, the combinations obtained with Eq. (1, left part) – dark circles. The open circles represent the quantities of the experimental normal-component coincident analyzing power $A_y(\hat{n})$ – Eq. (1, right part). Calculations with LEA – open dots and dashed line, and those with DWBA 91 – filled dots and solid lines. For all the calculations (DWBA 91 and LEA), in the top and middle panels of this figure, PH forces are used; and for the bottom panel, FL forces are employed.

As for the right-hand side part of Eq. (1), its corresponding values, represented by clear circles in Fig.3, were obtained independently in a reaction of a $(\vec{p}, p'\gamma)$ type. In order to maintain

Eq. (1), one makes the assumption that the $(\vec{p}, p'\gamma)$ reaction is a two-step process. It results in the fact that the total transition amplitude (excitation + decay) can be presented as a product of strong and electromagnetic amplitudes. Then coincident $(\vec{p}, p'\gamma)$ observables, measured when the photon is emitted normal $(\hat{n})$ to the scattering plane, reproduce basically the same information that is contained in normal-component $(\vec{p}, \vec{p}')$ single observables (see e.g. Ref. 35 in Sec. 2). As a result, the right-hand side of Eq. (1) represents the quantities of the measured $(\vec{p}, p'\gamma)$ coincident analyzing power $A_y(\hat{n})$. The same as the left-hand side of Eq. (1), it is expressed as a function of the center-of-mass proton scattering angle.

Fig. 3 shows combinations given in the left-hand side of Eq. (1) for the $1^+$, $T = 0$ state in $^{12}$C (two bottom panels), and for the $1^+$, $T = 1$ state in $^{12}$C (top panel). In the latter case, we also demonstrate a comparison with the normal-component coincident analyzing power $A_y(\hat{n})$, as is indicated in the right-hand side part of Eq. (1). Here we observe very good agreement between two completely independent measurements of practically the same physical quantity in the case of the $1^+$, $T = 1$ state in $^{12}$C. For the $1^+$, $T = 0$ level in $^{12}$C, there are still no available direct measurements of $A_y(\hat{n})$ values, so the corresponding designations on the axis of ordinates are hypothetical. They just represent a theoretically equal value, expressed by the combination in the left-hand side of Formula (1) in this section. In Fig. 3 we present a full-scale picture of the corresponding experimental dependencies for both $1^+$ states ($T = 0$ and $T = 1$) in $^{12}$C. We have established it, based on the available measurements, which were described, for example, in Refs. [15, 23, and 24] in Sec. 4. All these data we analyzed in the same way as the combinations shown in Figs. 1 and 2 in this section.

The conclusions derived from the evaluation of experimental and calculated data in all the three figures appear to be essentially comparable. Thus, in the angular region near the cross section peak (see Fig. 3 in Sec. 1) we can observe a common phenomenon. As before (see, for example, Ref. [42] in Sec. 2), it is characterized by the fact that large isoscalar unnatural-parity amplitudes for the 12.71-MeV transition to the $1^+$, $T = 0$ state in $^{12}$C produce significant contributions through an exchange coupling to the isovector tensor interaction. Indeed, the 12.71-MeV transition (for which most of the direct contributions are intrinsically weak) is rather strongly influenced by exchange couplings (for details see Ref. [26] in Sec. 2). That is why exact finite-range calculations made using the program DWBA are more important here than a zero-range treatment in LEA, which is clearly demonstrated in Fig. 3 in this section.

The transitions for all the discussed observables in proton scattering in $^{12}$C were described using the Cohen–Kurath wave function (see Sec. 2 and Ref. [28] there), including the corresponding radial parts of the transition density (see Ref. [13] in Sec. 1, [23] in Sec. 2, and [25, 26] in Sec. 4). For distorting potentials, we used either classical optical model parameters (fits to $p + ^{12}$C elastic scattering), or such potentials were generated "self-consistently".

**A Model of Nonlocal Knock-on Exchange Contribution to the $(\vec{p}, \vec{p}')$ Reaction**

Therefore, certain combinations of $(\vec{p}, \vec{p}')$ observables (Figs. 2 and 3 in this section) show sensitivity to composite spin-convection current amplitudes (naturally known in a relativistic formalism), which appear in nonrelativistic treatments of proton-nucleus scattering only through an explicit treatment of exchange processes. The indicated amplitudes, or the corresponding composite-current matrix elements, essentially involve the composite-current operators $(\sigma \cdot j)$ and $(\sigma \times j)$, where $j$ is a nucleon convection current operator and $\sigma$ is a projectile spin-operator.

F. Petrovich and W.G. Love, in particular, developed a consistent description of the structure and theory of these phenomena (see Ref. [27] in Sec. 4). They suggested, for instance, that exact estimates of tensor exchange terms, based on realistic interactions, showed to be most important in isoscalar unnatural parity transitions. The primary aim of the present review is to find and present evidence for such an assumption.

Next, we would like to emphasize that we tried, where it was feasible, to achieve a separation of the scattering problem into two parts: a nuclear reaction part and a nuclear structure part. Therefore, in unnatural-parity ($\vec{p}, p'$) and ($\vec{p}, \vec{p}'$) reactions we tried to evaluate the "strength" of nonlocal processes. Exchange terms are supposed to be one of the main source of nonlocal coupling, though other sources of nonlocality in the spin-dependent parts of a nonrelativistic NN interaction, often discussed in publications, can also be important. Thus, a nonlocal spin-dependent coupling in the effective NN interaction was shown to probe current $\otimes$ spin correlations for inelastic nuclear excitations in Love and Comfort's classical paper (Ref. [13] in Sec. 1). By the time this study was completed it had already been known (Refs. [9, 42] in Sec. 2) that contributes to the knock-on exchange processes were driven primarily by the tensor-exchange interaction.

Consequently, exchange terms, arising from the tensor force, must be a more important source of $(P–A) \neq 0$ than are those associated with the central force. Love and Comfort claimed that they aimed at a more simple purpose, i.e., that of establishing trends and exploring sensitivities of DWIA calculations to the experimental data. According to these authors, experimental polarization-analyzing-power differences are explicitly illustrated by schematic calculations. So they singled out the most important features of the exchange mechanism and considered only the central force. In their model, the nonlocality arises from the finite (nonzero) range of the central parts ($V_{ip}^C$) of the complex interaction ($V_{ip}$), where $i(p)$ denotes a target (projectile) nucleon. In that case for the interaction $V_{ip}$, was applied a widely used version of an impulse approximation that consists of representing the NN t matrix (see Ref. [24] in Sec. 1). For $0^+ \rightarrow 1^+$ (I→F) transitions in $^{12}$C the direct part of the NN t matrix is associated with the $\vec{\sigma}_i \cdot \vec{\sigma}_p$ part of $V_{ip}^C$. Other particularities of such an approach can also be found in publication [26] in Sec. 4. To make things clearer, let us specify that in Love and Comfort's model the inclusion of a zero-range pseudo potential to approximate exchange effects simply adds a constant to $V_I^C(q)$. Here $V_I^C(q)$ in the PWIA is the Fourier transform of the $\vec{\sigma}_i \cdot \vec{\sigma}_p$ part of $V_{ip}^C$.

In the above model, a source of $(P–A)$ arises from interference between two time-reversal operators ($\vec{B}_{LS}$ and $\vec{B}_S$). Then, in schematic considerations, the quantity $(P–A)\sigma$ can be defined with the precision up to coefficients by the imaginary part of $\langle B_{LS}\rangle\langle B_S\rangle^*$, where both $\vec{B}_{LS}$ and $\vec{B}_S$ operators arise from the central part of the force. Let us note that $\langle B_S \rangle \equiv \vec{B}_S$, etc., and $\vec{B}_S$, $\vec{B}_{LS}$, and $\vec{B}_L$ denote target transition "spin" densities, analogous to $\vec{\sigma}_p$ for the projectile.

The transition matrix can be characterized, in the terms of transferred angular momenta (LSJ), by the sum of the following constituents: $\vec{B}_S \cdot \vec{\sigma}_p \leftrightarrow (011)$, $(\vec{B}_{LS} \times \hat{n}) \cdot \vec{\sigma}_p \leftrightarrow (111)$, and $\vec{B}_L \cdot \hat{n} \leftrightarrow (101)$

Within the present model, it is easy to carry out the following procedure. As is shown in Sec. 2, it is possible to make calculations for the indicated transitions using the modified Cohen–Kurath wave function (CKWF), i.e. MCKWF, constructed by deleting the (1–p)-shell spectroscopic amplitudes $A_{LSJ} \equiv A_{111}$. In the absence of (LSJ ≡ 111) terms, $\vec{B}_{LS} = 0$, and thus the quantity $(P–A)$ will approach zero if there is basically only one source of $(P–A) \neq 0$.

In accord with the above schematic model, it is easy to establish that the calculated quantities $(P+A)\sigma$ are negligibly different when MCKWF (instead of CKWF) are used. In fact, according to this model, the quantities $(P+A)\sigma$ are proportional to the interference between two time-reversal odd operators ($B_L$ and $B_S$). In terms of transferred momenta (LSJ), the corresponding components of the transitions matrix are as follows: $\vec{B}_L \cdot \hat{n} \leftrightarrow (101)$ and $\vec{B}_S \cdot \vec{\sigma}_p \leftrightarrow (011)$. Thus, the operator $B_{LS}$ is not included in the real part of $\langle B_L \rangle \langle B_S \rangle^*$, which defines $(P+A)\sigma$, and, therefore, the quantity $(P+A)\sigma$ is not sensitive to the inclusion of LSJ = 111 ($\leftrightarrow B_{LS}$) terms.

The above description clearly, although only qualitatively, explains the picture shown in the left column of Fig. 3 in Sec. 2. It is also evident that the DWIA calculations include analogs, especially for a very large $B_{LS}$–type amplitude. This criticality factor persists in our calculations, too, with certain quantities reevaluated, based on comparing the results that were obtained with the use of two programs: DWBA 91 and LEA (a middle column of Fig. 3 in Sec. 2).

For comparing these results with those shown in Figs. 1–3 of this section, let us single out an appropriate fragment from Fig. 3 in Sec. 2 and represent it separately in Fig. 4 below.

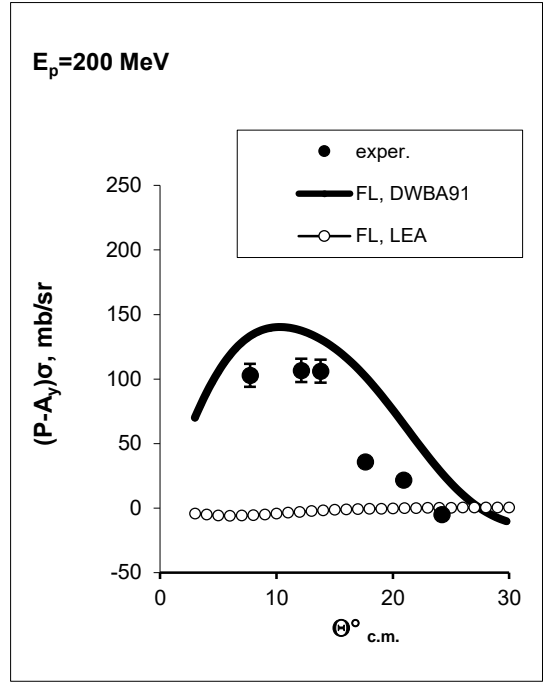

**Fig. 4.** The same as Fig. 3 in Sec. 2, but for the fragments related to the difference function $(P–A)\sigma$ at 200 MeV.

A definite improvement becomes obvious in a quantitative description of the $(P–A)\sigma$ dependence for the $1^+$, $T = 0$ in $^{12}C$ state at $E_p = 200$ MeV. We may hope that this improvement can be expanded to other cases.

### Discussions, Summary, and Conclusions

For the sake of clarity, it is necessary to emphasize it once again that the structure of the CKWF for both $1^+$ states in $^{12}C$ is such that the abnormal-parity terms with $L = S = 1$ are the largest components in *LS* representations of the transition amplitudes. A specific feature of the $0^+ \to 1^+$, $T = 0$ transition is that most of the direct contributions are intrinsically weak. Therefore, it is strongly influenced by exchange couplings. That is why it is not surprising that here we observe the most evident disparity between calculations with DWBA and LEA, demonstrating differing nonlocal behaviors (through knock-on exchange terms), which is shown in Fig. 4 in this section. Earlier (see Fig. 3 in Sec. 2) we have already demonstrated a qualitatively similar difference between DWIA calculations made with the use of CKWF and MCKWF in. It is for the latter WF that a removal of abnormal-parity transition amplitudes with $(LSJ) = (111)$ occurs. This term, according to numerous works by Love, Comfort et al., cannot be tested by other probes, and it contributes to the $(\vec{p}, p')$ reaction only through knock-on exchange. As demonstrated above, especially in Figs. 2, 3 and 4 in this section, very often there may be no motivation for eliminating exchange amplitudes, apart from $(\vec{p}, p')$ and $(\vec{p}, \vec{p}')$ results.

For further advancement in understanding the numerous measurements of $(p, p')$, $(\vec{p}, p')$, $(\vec{p}, \vec{p}')$, and $(\vec{p}, p'\gamma)$ observables for the excitation of the $1^+$, $T = 0$ and $T = 1$ states in $^{12}C$, we can proceed from the fact that different components of the transition operator demonstrate a dissimilar nonlocal behavior.

Accordingly, it has different manifestations in calculations of both isoscalar and isovector excitations. In particular, the exchange tensor amplitude proves to be extremely important in $1^+$, $T = 0$ transitions. On the other hand, in the case of the $1^+$, $T = 1$ transition in the calculations, one of the major contributions comes from the central terms (see e.g. Ref. [7] in Sec. 3, and [28] in Sec. 4). For instance, the cross sections, corresponding to central, central plus spin orbit (LS), and central plus LS plus tensor parts of the interaction, are clearly shown for the $T = 0$ and $T = 1$ states in $^{12}$C excitation at 200 MeV in Ref. [7] in Sec. 3. Thus, the isovector central part appears to be a dominant factor for the $T = 1$ excitation. However, for the $T = 0$ central part, it is also rather insignificant, even for the cross section. Advancement in calculating their quantities is impossible without taking into account tensor contributions.

Thus, here we mainly concentrate on two models. The first one, the popular nonrelativistic model of Love and Comfort, predicts nonzero values of the function $(P–A_y)$, based on single-spin asymmetries of $P$ and $A_y$. This model indicates the importance of spin $\otimes$ current terms in the effective *NN* interaction.

The second model is that of E. Bleszynski, M. Bleszynski, and C.A. Witten. According to this model, in the PWIA, a certain combination of $(\vec{p}, \vec{p}')$ polarization-transfer coefficients is directly related to individual terms in the effective *NN* interaction.

Some authors consider these models to be the most important diagnostic tools. Others, however, criticize this assumption for being a physically simple interpretation. All these approaches deserve attention, and the author of the review continues to study them.

The coupling of the nucleon spin to the bound nucleon current of the Love and Comfort model, in our case, is expressed through the tensor-exchange piece of the effective *NN* interaction in nonrelativistic treatments.

The specific feature of the present study is that for certain unnatural-parity transitions (in order for DWIA calculations to agree with measured quantities) we established a common exchange approach for both $(P–A_y)$, or $(P–A_y)\sigma$, data and sets of $(\vec{p}, \vec{p}')$ spin-transfer observables, or physically plausible spin-observable combinations.

A possible satisfactory description of the characteristics for structurally different transitions (the $T = 0$ and $T = 1$, $1^+$ levels in $^{12}$C), shown in Fig. 2 (Sec. 2), together with the description of their energy dependence (see also Figs. 5 and 6 in Sec. 1), give evidence of the efficacy of the formalism we applied (Eqs. 1 and 2 in Sec. 2). An important conclusion derived from that is that the transition amplitude can be factored into nuclear-structure-dependent and interaction-dependent compositions, as it was supposed in Ref. [14] in Sec. 2. In addition, the results presented in Figs. 1–6 in Sec. 1 demonstrate a wide variation among competing models of interaction at various proton energies.

From this standpoint, the use of a zero-range treatment of knock-on exchange in LEA may be quite acceptable in certain cases. Though it was unacceptable for the description of certain $D_K$ data in the case of the $1^+$, $T = 0$ excited state in $^{12}$C at 200 MeV (Fig. 2 in this section), now, at much higher energies $E_p$ (in the range of 400–500 MeV), using LEA appears to be rather appropriate. This particular situation is presented in Fig. 5 in this section. The comparison of this figure with Fig. 2 in this section clearly demonstrates that the substantial difference existing between data and calculations made using LEA at 200 MeV significantly decreased at 400–500 MeV. At such high energies, the quality of description using LEA appears to be much better. It becomes almost the same as exact finite-range calculations made using the program DWBA 91 at 200 MeV.

Thus, the described situation can be considered quite an adequate assumption (Refs. [14 and 23] in Sec. 4) that at higher energies either tensor-exchange contributions become less important or (what is basically the same) at these higher energies occurs decrease in strength of nonlocal processes.

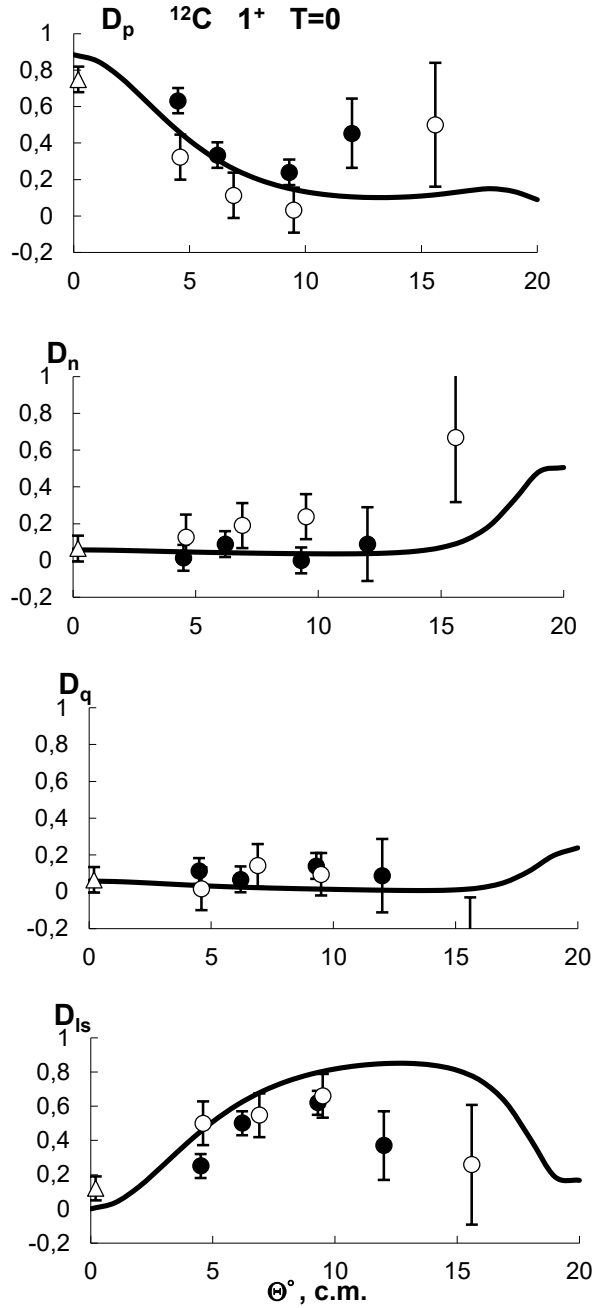

**Fig. 5.** The same as Fig. 1 in Sec. 4, but for the isoscalar $1^+$ state at 12.71 MeV in $^{12}$C.

To make it clearer, let us note that in a zero-range approximation (LEA), exchange interactions are independent of the momentum transfer and are reduced to delta functions (Ref. [21] in Sec. 4). If such a simple approach employed in LEA is unacceptable, it often becomes necessary to modify nucleon-nucleus kinematics. In the LEA program, it is assumed that at the exchange of a scattered nucleon with a nucleon of the nucleus, the first nucleon (projectile) stops and loses all its momentum. Otherwise, in a zero-range approximation, the overlap (for the exchange amplitude) is calculated as it would be if the outgoing nucleon were the projectile. Then in the treatment of LEA, the kinematics for the exchange amplitude use a momentum transfer calculation that is consistent with the assumption that a struck nucleon (a nucleon in the target) is in motion prior to collision. Such procedures were undertaken in Ref. [21] in Sec. 2 for the purpose of developing LEA, and also in a series of finite-range DWIA programs from Raynal and Schaeffer (see Refs. [4, 25] in Sec. 1).

We have shown in the present review, using numerous examples, the application of the Bleszynski et al. model. The deficiencies of this model are well known to us. Let us remind that according to the model, certain combinations of $(\vec{p}, \vec{p}')$ spin-transfer observables, such as $D_K$ or $\sigma_K$, are related to individual terms in the effective $NN$ interaction. The opponents of this approach, for example the authors of Ref. [24] (Sec. 4), point to the model's weakness of which we are aware quite well. This model is known to be based on a plain-wave impulse approximation. According to the mentioned opponents, when knock-on exchange is included, it becomes less clear what the combinations $D_K$ or $\sigma_K$ represent physically. For our part, let us mention that it is not the only difficulty. Often uncertainties result from calculated effects of the nuclear medium. That is why we paid special attention to the extent of taking into account exchange contributions. In fact, it turns out that in many cases, at certain $E_p$ values, a simplified LEA treatment, in which the exchange momentum transfer is a constant, appears to be sufficient. Certainly, quite a large part of the difference between model and scattering data arises from the accuracy of taking into account the density dependence in the interaction. We compared the results obtained using medium-modified forces with the results acquired with the use of free interactions of the same or different group.

It might be worth reminding in the final section that we used constructions that not just follow from procedures of mathematical physics, but acquire forms allowed by parity conservation.

In order to emphasize the correctness (as we understand it) of the starting positions, let us consider the evaluation operations of DWIA calculated data shown in Fig. 5 of the present section. We will apply the same assessment procedures as in Sec. 4 (Fig. 1) to ensure quality control of DWIA calculations for the combinations $D_K$ in the case of the $\Delta T = 1$ transition to $1^+$ levels in $^{12}$C and $^{28}$Si at 0°. But this time we will discuss the $\Delta T = 0$ transition for the $1^+$ level in $^{12}$C and also at 0°. As it was for the isovector ($\Delta T = 1$) $M1$ transition, we will be using the same $D_K$ expressions (Eq. 2 in Sec. 4) and the predicted $D_{ii}$, obtained in Ref. [6] (Sec. 4), and in [11] (Sec. 2) for the isoscalar ($\Delta T = 0$) $M1$ transition. In this case, the transitions are mainly mediated by the isovector tensor term through knock-on exchange. Then at 0° the indicated simple predictions give $D_{SS} = D_{NN} = -2/3$, and $D_{LL} = +1/3$ quantities.

If these quantities are inserted into the expressions $D_K$ (Eqs. 2 in Sec. 4), we can easily get the quantities:

$$D_{ls} = 0, \qquad D_q \approx D_n \approx 0.17, \qquad D_p \approx 0.7$$

These quantities, as is seen in Fig. 5 (and in Fig. 2) in this section, even in such a rough approximation, are rather close to our DWIA calculations at 0°, as well as to the experiment itself. Such results are quite inspiring; they indicate that we have chosen the right and rather innovative direction in our research.

In conclusion, our brief communication[††] should be also mentioned. There we pointed out that in the study of the $^{12}$C $(\vec{p}, p')$ $^{12}$C reaction, the Cohent-Kuratch shell-model wave functions, adjusted to reproduce the available electromagnetic data, were employed for all our microscopic calculations of $(\vec{p}, p')$ - scattering. This made it possible to extract effective radial dependences of particle, spin, spin-orbit, and current transition densities for the $1^+$, $T = 0$ and $T = 1$ states in $^{12}$C. This part of our studies is still in progress.

We believe that the present review provides explanation and gives details of certain positions regarding spin and isospin excitations. The author can only share the opinion expressed by Dr. F. Osterfeld in his review (Ref. [4] in Sec. 2) that this field of nuclear physics "will continue to be challenging and exciting".

---

[††] A.V. Plavko, M.S. Onegin, V.I. Kudriashov. 62nd Int. Conf. "Nucleus 2012". Abstracts. Voronezh, 2012, p. 205–206.

## Acknowledgements

The author of the present review is deeply indebted to Dr. M.S. Onegin and Dr. V.I. Kudriashov, his co-authors of many publications, for their participation and help in preparing many of the scientific results used here. The author also wishes to express his gratitude to Prof. E.J. Stephenson who provided him with unique tabulated DWIA calculations (with DBHF formalism), which helped to make established analytical results more reliable and conclusions more consistent. The author acknowledges that the present study would not have been possible without employing the computer program DWBA 91 from Prof. J. Raynal and the computer code LEA from Prof. J. Kelly.